\documentclass[12pt]{article}
\usepackage{amsmath,amssymb,mathtools}
\usepackage{graphicx,enumitem,color,threeparttable,booktabs,multirow,titlesec,subcaption}
\usepackage[title]{appendix}
\usepackage[hyphens]{url}
\usepackage{hyperref}

\titlespacing{\section}{0pt}{8pt}{5pt}
\titlespacing{\subsection}{0pt}{8pt}{5pt}
\usepackage[round]{natbib}
\allowdisplaybreaks[4]
\newtheorem{theorem}{Theorem}

\newtheorem{proposition}{Proposition}
\newtheorem{lemma}{Lemma}
\newtheorem{definition}{Definition}
\newtheorem{assumption}{Assumption}

\newtheorem{algorithm}{Algorithm}
\newtheorem{remark}{Remark}
\newcommand{\indep}{\perp\!\!\!\perp}
\newcommand{\E}{\mathrm{E}}
\newcommand{\Var}{\mathrm{Var}}

\renewcommand{\P}{\mathrm{P}}

\newcommand{\bX}{X}

\newcommand{\anon}{1}

\begin{document}

\def\spacingset#1{\renewcommand{\baselinestretch}%
{#1}\small\normalsize} \spacingset{1}


\if1\anon
{
  \title{\bf Causal Mediation Analysis for Panel Data with Double/Debiased Machine Learning}
  \author{Peikai Wu \hspace{.2cm}\\
    Department of Statistics and Data Science, Fudan University, China\\
    and \\
    Zhiguo Xiao \\
    Department of Statistics and Data Science, Fudan University, China}
  \maketitle
} \fi

\if0\anon
{
  \bigskip
  \bigskip
  \bigskip
  \begin{center}
    {\LARGE\bf Causal Mediation Analysis for Panel Data with Double/Debiased Machine Learning}
\end{center}
  \medskip
} \fi

\bigskip
\begin{abstract}
We develop a double/debiased machine learning framework for causal mediation analysis in panel data with two-way fixed effects. Binary, multivalued, or continuous treatments and continuous mediators affect outcomes via histories. We formulate partially linear models for potential outcomes and mediators, define history-specific direct, indirect, and total effects, and identify them under a sequential conditional independence assumption given unit effects, allowing arbitrary time-invariant confounding. Estimation combines first-differencing, cross-fitting sample splitting that removes period effects, and machine-learned confounding functions within a GMM framework whose score is Neyman orthogonal in all nuisance components. We establish asymptotic normality of the estimators for the structural parameters and causal estimands and provide consistent variance estimators. Simulations show small bias and near-nominal coverage under nonlinear confounding, whereas a DML-free benchmark fails severely for indirect effects. An application to U.S. county data illustrates that a stable total effect can conceal substantial sensitivity in its direct-indirect decomposition.
\end{abstract}

\noindent%
\vfill

\newpage
\spacingset{1.8} 

\section{Introduction}
\label{sec:introduction}
Many empirical questions concern not only whether an intervention changes an outcome, but also how that change is produced. A policy may affect income through employment and wages, a medical treatment may operate through an intermediate biomarker, and an educational program may alter later outcomes through a sequence of behavioral responses. Causal mediation analysis formalizes such questions by decomposing a total effect into direct and indirect components associated with a mediator. The potential-outcome literature has developed identification and semiparametric theory for controlled, natural direct, and natural indirect effects in cross-sectional settings \citep{intro_robins1992,intro_imai2010,intro_tchetgen2012}. In panel data, however, both treatment and mediator may vary over time, current outcomes may depend on their histories, and the relevant counterfactual is therefore nested over an entire mediator path rather than a single intermediate variable.

Longitudinal extensions of mediation analysis, by allowing exposures and mediators to vary over time, have clarified how dynamic pathways can be defined and identified \citep{vanderweele2017mediation,intro_zheng2017,intro_diaz2023}. Yet repeated observations also create a problem that is central in many panel applications: persistent unobserved heterogeneity may jointly affect treatment, mediator, and outcome in every period. Conditioning on a rich observed history need not remove this confounding, whereas a conventional fixed-effects regression, by itself, does not identify the nested counterfactuals required for natural effects. At the same time, observed confounding may be high-dimensional and nonlinear, making a fully parametric adjustment fragile. A useful panel mediation method must therefore exploit repeated measurements to accommodate time-invariant unobservables, retain sufficient flexibility for observed confounders, and still deliver valid inference for a decomposition that depends on more than one structural equation.

Double/debiased machine learning (DML) provides a general route to inference on low-dimensional parameters in the presence of complex nuisance functions \citep{chernozhukov2018double}. Within causal mediation analysis, \citet{intro_farbmacher2022} construct cross-sectional DML estimators for direct and indirect effects with a binary treatment, and \citet{intro_zenati2025} study a mediated response curve with a continuous treatment. More recently, \citet{wei2025debiased} develop cross-sectional debiased ML for ultra-high-dimensional mediation, combining orthogonal scores and cross-fitting with screening and regularization for mediator selection and inference. Two contemporaneous contributions develop mediation analysis for difference-in-differences designs: \citet{intro_huber2026} combine DML with conditional parallel-trends restrictions, while \citet{intro_deng2026} derive semiparametric estimators under mediator-adjusted parallel trends. Unlike \citet{intro_huber2026}, whose mediation effects are identified through conditional parallel trends in a DiD design, our framework targets history-specific mediation effects in a longitudinal panel system and uses sequential conditional independence together with fixed effects to accommodate time-invariant unobserved confounding.

A separate literature extends machine learning and orthogonalization to sequential and panel treatment-effect problems. \citet{intro_lewis2021} develop DML for dynamic treatment sequences, \citet{intro_semenova2023} study heterogeneous effects in high-dimensional dynamic panels under weak dependence, and \citet{intro_clarke2026} construct DML procedures for static fixed-effects models. Most closely related is \citet{intro_wu2026}, who develop DML--GMM inference for treatment effects in dynamic panels with two-way fixed effects, general treatment scales, and history-dependent potential outcomes. Their auxiliary-fold construction for eliminating period effects is particularly relevant here. However, moving from treatment effects to mediation effects is not a mechanical stacking of two panel DML regressions. The mediator is both an intermediate response and a regressor in the outcome equation; its estimated innovation is also needed to construct instruments for that equation. Natural effects involve nested counterfactual mediator histories and depend jointly on the parameters of both structural equations and on their estimation covariance. These features create identification, orthogonalization, and inference problems absent from a single-equation treatment-effect analysis.

We develop DML-based causal mediation analysis in panel data with two-way fixed effects; however, this generalization requires overcoming several difficulties. First, the identifying argument must recover nested potential outcomes in a two-equation history-dependent system rather than a single potential-outcome regression. Second, sequential exogeneity does not generally validate contemporaneous instruments after fixed-effect removal. We therefore construct GMM moments from lagged residualized treatment and mediator innovations. The outcome instrument contains an estimated mediator innovation and hence depends on the mediator coefficient. We introduce a nuisance copy of that coefficient, learned outside the score-evaluation fold, so that mediator innovations can be used without making the final GMM score nonlinear in the target parameter. Third, eliminating two-way fixed effects while preserving the independence required by cross-fitting calls for a three-way use of the sample. First differences remove unit effects, and period effects are removed with cross-sectional means computed from an auxiliary fold that is distinct from both the score-evaluation fold and the nuisance-training sample. The resulting mean-subtraction terms induce additional dependence that must be shown to be asymptotically negligible. Fourth, inference for natural effects cannot be obtained by applying a delta method to the structural coefficients alone: the estimators also average individual counterfactual contributions, and the mediator and outcome coefficients are estimated jointly. The influence function must retain both sources of variation and their covariance.

The paper makes three contributions. First, it provides a unified causal framework for history-specific mediation effects in panels. We formulate partially linear models for the potential outcome and potential mediator in which researcher-chosen dictionaries describe dynamic and heterogeneous causal components, while unknown functions flexibly absorb nonlinear observed confounding. Treatments may be binary, multivalued, or continuous, mediators are continuous, and we define history-specific controlled direct, natural direct, natural indirect, and total effects together with their derivative counterparts. Identification relies on sequential conditional independence given the observed history, the relevant unit effect, and the period effects, which permits arbitrary time-invariant confounding and translates the potential-outcome system into observed-data equations governed by the same structural parameters. A distinctive feature of this paper is that two-way fixed effects are allowed in the outcome, mediator, and treatment-assignment models, thereby exploiting repeated observations throughout the treatment--mediator--outcome system. To our knowledge, no existing procedure combines history-specific natural effects, two-way fixed effects in all three equations, sequentially valid residual instruments, and DML-based joint inference.

Second, we develop a DML--GMM procedure tailored to the two-equation mediation system. Sequential exogeneity motivates moments based on lagged residualized treatment and mediator innovations rather than contemporaneous instruments. First differences remove unit effects, while period effects are removed using cross-sectional means from an auxiliary fold that is distinct from both the nuisance-training and score-evaluation samples. Because the outcome instrument contains an estimated mediator innovation, we introduce a fold-specific nuisance copy of the mediator coefficient; this preserves an affine score and a quadratic GMM problem. Jointly stacking the two equations and estimating their full moment covariance retains the cross-equation dependence required for inference on natural direct and indirect effects.

Third, we establish an oracle representation and root-$N$ asymptotic normality for the joint estimator of the outcome and mediator parameters under high-level mean-square and product-rate conditions on the first stages. Auxiliary-fold demeaning is shown not to affect the first-order distribution, and a consistent sandwich variance estimator is provided. For any fixed smooth collection of level or derivative effects, we derive an asymptotic linear representation that combines sampling variation in individual counterfactual contrasts with estimation error in both parameter blocks. The resulting theory supports joint Wald inference for direct, indirect, and total effects, including collections whose exact decompositions make the covariance matrix singular.

In simulations with nonlinear time-varying confounding, correlated fixed effects, equation-specific period effects, and sequentially endogenous differenced regressors, the proposed estimator has small bias and near-nominal coverage, whereas a benchmark using the same first-stage learner can fail sharply for the mediator equation and indirect effects. In the U.S. county application, allowing treatment--covariate interactions leaves the estimated total effect essentially unchanged but reduces the estimated mediated share from about $65\%$ to $41\%$, showing that stability of the total effect alone need not validate its direct--indirect decomposition.

The remainder of the paper is organized as follows. Section \ref{sec:setup} introduces the causal estimands, the panel mediation model, and the identification conditions. Section \ref{secest} develops the DML--GMM algorithm and the plug-in estimators of the causal effects. Section \ref{secas} establishes the asymptotic theory. Sections \ref{sec:simulation} and \ref{sec:empirical} report the simulation study and empirical application, respectively, and Section \ref{sec:conclusion} concludes. Extension to multi- and high-dimensional mediator cases, additional theoretical and numerical results, and proofs are provided in the online supplementary material.

\section{Problem setup}
\label{sec:setup}
\subsection{Notation and framework}
We consider a balanced panel comprising $N$ units observed over $T$ periods, with units indexed by $i=1,\ldots,N$ and periods by $t=1,\ldots,T$. For unit $i$ in period $t$, let $Y_{it}\in\mathbb{R}$, $M_{it}\in\mathbb{R}$, $D_{it}\in\mathbb{R}$, and $X_{it}\coloneq(X_{it1},\ldots,X_{itL_X})'\in\mathbb{R}^{L_X}$ denote, respectively, the outcome, the mediator, the treatment, and a $L_X$-dimensional vector of observed covariates. In Appendix A, we discuss how to extend our framework to vector-valued or high-dimensional mediators. The framework accommodates continuous mediators and binary, multivalued, or continuous treatments. Within each period, $X_{it}$ is realized before $D_{it}$, followed by $M_{it}$ and then $Y_{it}$.

We begin by stating the standard causal assumptions required for our analysis. The first is a mediation-specific version of the stable unit treatment value assumption \citep[SUTVA;][]{imbens2015causal}.
\begin{assumption}[Stable unit treatment value assumption]
\label{assut}
For each unit, the potential outcome depends only on that unit's treatment and mediator assignments. Moreover, each treatment and mediator level has a unique version. Equivalently, there is neither interference across units nor hidden variation in treatment or mediator assignments.
\end{assumption}
We further impose a no-anticipation condition, under which future treatment and mediator assignments cannot affect current or past outcomes.
\begin{assumption}[No anticipation]
For every $t$, a unit's potential outcome in period $t$ depends only on treatment and mediator assignments through period $t$ and is invariant to assignments in periods after $t$. 
\end{assumption}
For any time-indexed variable, such as $V_{it}$ or $W_t$, define $\overline{V}_{it}\coloneq(V_{i1},\ldots,V_{it})'$, $\overline{W}_t\coloneq(W_1,\ldots,W_t)'$, and $\overline{W}_{j:k}\coloneq(W_{j},\ldots,W_{k})'$ for $j\leqslant k$. Following the potential-outcomes notation of \citet{rubin1974estimating}, let $Y_{it}(\overline{d}_{t},\overline{m}_{t})$ denote the potential outcome for unit $i$ in period $t$ under treatment history $\overline{d}_{t}$ and mediator history $\overline{m}_{t}$, where $\overline{d}_{t}$ and $\overline{m}_{t}$ are possible realizations of $\overline{D}_{it}$ and $\overline{M}_{it}$, respectively. Likewise, let $M_{it}(\overline{d}_{t})$ denote the potential mediator for unit $i$ in period $t$ under treatment history $\overline{d}_{t}$, and define $\overline{M}_{it}(\overline{d}_t^*)\coloneq(M_{i1}(d_1^*),\ldots,M_{it}(\overline{d}_t^*))'$.

To link these potential variables to their observed counterparts, we impose the standard consistency assumption.
\begin{assumption}[Consistency]
\label{asscon}
For every unit $i$ and period $t$, the observed outcome and mediator coincide with their potential counterparts under the realized treatment and mediator histories: $Y_{it}=Y_{it}(\overline{D}_{it},\overline{M}_{it}) \quad\text{and} \quad M_{it}=M_{it}(\overline{D}_{it}).$
\end{assumption}

\subsection{Causal estimands}
\label{secce}
For two given treatment histories $\overline{d}_t\coloneq(d_1,\ldots,d_t)'$ and $\overline{d}_t^*\coloneq(d_1^*,\ldots,d_t^*)'$, define the nested potential outcome $Y_{it}(\overline{d}_t,\overline{M}_{it}(\overline{d}_t^*))$, which sets the treatment history to $\overline{d}_t$ while setting the entire mediator history to its natural value under $\overline{d}_t^*$. The following estimands extend the conventional controlled direct, natural direct, and natural indirect effects to treatment and mediator histories; related longitudinal formulations are discussed by \citet{vanderweele2017mediation}.

\begin{definition}[History-specific causal effects]
For two treatment histories $\overline{d}_t$ and $\overline{d}_t^*$, define the following causal effects at time $t$: \\
1. For a prespecified mediator history $\overline{m}_t\coloneq(m_1,\ldots,m_t)'$, the history-specific controlled direct effect is $
\operatorname{CDE}_t(\overline{d}_t,\overline{d}_t^*;\overline{m}_t)\coloneq\E\bigl[Y_{it}(\overline{d}_t,\overline{m}_t)-Y_{it}(\overline{d}_t^*,\overline{m}_t)\bigr]$. \\
2. The history-specific natural direct effect is $\operatorname{NDE}_t(\overline{d}_t,\overline{d}_t^*)\coloneq\E\bigl[Y_{it}\bigl(\overline{d}_t,\overline{M}_{it}(\overline{d}_t^*)\bigr)-Y_{it}\bigl(\overline{d}_t^*,\overline{M}_{it}(\overline{d}_t^*)\bigr)\bigr]$. \\
3. The history-specific natural indirect effect is $\operatorname{NIE}_t(\overline{d}_t,\overline{d}_t^*)\coloneq\E\bigl[Y_{it}\bigl(\overline{d}_t,\overline{M}_{it}(\overline{d}_t)\bigr)-Y_{it}\bigl(\overline{d}_t,\overline{M}_{it}(\overline{d}_t^*)\bigr)\bigr]$.
\end{definition}

The history-specific total effect comparing $\overline{d}_t$ with $\overline{d}_t^*$ is defined by $\operatorname{TE}_t(\overline{d}_t,\overline{d}_t^*)\coloneq\E\bigl[Y_{it}\bigl(\overline{d}_t,\overline{M}_{it}(\overline{d}_t)\bigr)-Y_{it}\bigl(\overline{d}_t^*,\overline{M}_{it}(\overline{d}_t^*)\bigr)\bigr]$. By construction, this effect admits the decomposition $\operatorname{TE}_t(\overline{d}_t,\overline{d}_t^*)=\operatorname{NDE}_t(\overline{d}_t,\overline{d}_t^*)+\operatorname{NIE}_t(\overline{d}_t,\overline{d}_t^*)$. Unlike the natural direct effect, the controlled direct effect is indexed by a prespecified mediator history and does not generally enter this two-way decomposition.

When both the treatment and mediator are continuous, the history-specific effects defined above can be complemented by local derivative effects that characterize the impact of an infinitesimal change in the treatment at a particular period. Define the nested mean response surface $\Psi_t(\overline{a}_t,\overline{b}_t)\coloneq\E\bigl[Y_{it}\bigl(\overline{a}_t,\overline{M}_{it}(\overline{b}_t)\bigr)\bigr]$.

\begin{definition}[History-specific derivative effects]
For $j\leqslant t$, define the following derivative effects at time $t$: \\
1. For a prespecified mediator history $\overline{m}_t$, the controlled direct derivative is $\displaystyle \operatorname{CDD}_{t,j}(\overline{d}_t,\overline{m}_t)\coloneq\frac{\partial}{\partial d_j}\E\bigl[Y_{it}(\overline{d}_t,\overline{m}_t)\bigr]$. \\
2. The natural direct derivative is $\displaystyle \operatorname{NDD}_{t,j}(\overline{d}_t)\coloneq\frac{\partial}{\partial a_j}\Psi_t(\overline{a}_t,\overline{b}_t)\big|_{\overline{a}_t=\overline{b}_t=\overline{d}_t}$. \\
3. The natural indirect derivative is $\displaystyle \operatorname{NID}_{t,j}(\overline{d}_t)\coloneq\frac{\partial}{\partial b_j}\Psi_t(\overline{a}_t,\overline{b}_t)\big|_{\overline{a}_t=\overline{b}_t=\overline{d}_t}$.
\end{definition}

The corresponding total derivative is $\operatorname{TD}_{t,j}(\overline{d}_t)\coloneq \partial\E\bigl[Y_{it}\bigl(\overline{d}_t,\overline{M}_{it}(\overline{d}_t)\bigr)\bigr]/\partial d_j$, which satisfies $\operatorname{TD}_{t,j}(\overline{d}_t)=\operatorname{NDD}_{t,j}(\overline{d}_t)+\operatorname{NID}_{t,j}(\overline{d}_t)$ by the chain rule.

\subsection{Mediation model for the potential outcomes}
\label{secmod}
Since the causal estimands are defined in terms of the potential outcomes, to estimate them, we should build models for the potential outcomes instead of the observed outcomes. In this paper, we consider the following partially linear panel mediation model with two-way fixed effects:
\[
\begin{gathered}
Y_{it}(\overline{d}_t,\overline{m}_t) = \tau_Y(\overline{d}_{t-L:t},\overline{m}_{t-L:t},\overline{\bX}_{i,t-L:t})'\theta_0 + f_0(\overline{\bX}_{i,t-L:t}) + \alpha_i^Y + \lambda_t^Y + \varepsilon_{it}^Y,\\
\E(\varepsilon_{it}^Y \mid \overline{\bX}_{it}, \alpha_i^Y, \overline{\lambda}_t^Y) = 0,\\
M_{it}(\overline{d}_t) = \tau_M(\overline{d}_{t-L:t}, \overline{\bX}_{i,t-L:t})'\gamma_0 + g_0(\overline{\bX}_{i,t-L:t}) + \alpha_i^M + \lambda_t^M + \varepsilon_{it}^M,\\
\E(\varepsilon_{it}^M \mid \overline{\bX}_{it}, \alpha_i^M, \overline{\lambda}_t^M) = 0, 
\end{gathered}
\]
where $L$ is a fixed nonnegative integer that serves as a common upper bound on all lag lengths in the model, and variables before period 1 are treated as 0. $\tau_Y(\overline{d}_{t-L:t},\overline{m}_{t-L:t},\overline{\bX}_{i,t-L:t})\in\mathbb{R}^{s_Y}$ is a researcher-specified outcome dictionary that describes the finite-dimensional dependence of the potential outcome on the recent treatment history, mediator history, and their interactions with each other and with the covariate history. 
For example, an outcome dictionary containing contemporaneous and lagged treatment and mediator terms, nonlinear transformations, treatment--mediator interactions, and interactions with covariates is $\tau_Y(\overline{d}_{t-L:t},\overline{m}_{t-L:t},\overline{\bX}_{i,t-L:t})=(d_t,d_t^2,d_{t-1},d_td_{t-1},d_tX_{it1},m_t,m_{t-1},d_tm_t)'$.
$\tau_M(\overline{d}_{t-L:t},\overline{\bX}_{i,t-L:t})\in\mathbb{R}^{s_M}$ is a researcher-specified mediation dictionary that describes the effect of the recent treatment history and treatment-covariate interaction on the mediator. Terms depending only on the covariate history need not be included in the dictionaries because they are absorbed by the nuisance functions $f_0$ and $g_0$. The finite-dimensional parameters satisfy $\theta_0\in\mathbb{R}^{s_Y}$ and $\gamma_0\in\mathbb{R}^{s_M}$. The unknown functions $f_0$ and $g_0$ may be nonlinear and high-dimensional; $\alpha_i^Y$ and $\alpha_i^M$ are unit fixed effects; $\lambda_t^Y$ and $\lambda_t^M$ are time fixed effects; and $\varepsilon_{it}^Y$ and $\varepsilon_{it}^M$ are the outcome and mediator innovations, respectively.

The common lag length $L$ is an upper bound rather than a requirement that every lag have a nonzero effect. Hence, shorter lag structures are nested in the model by setting the corresponding coefficients or dictionary terms equal to zero. 

Under the above model specification, after specifying appropriate dictionaries for $\tau_Y$ and $\tau_M$, all causal estimands introduced in Section \ref{secce} can be expressed in terms of the model variables and the parameters $\theta_0$ and $\gamma_0$. An explicit calculation illustrating these formulas is provided in Appendix B.

\subsection{Identification}
The causal estimands of Section \ref{secce} compare potential outcomes of the same unit under different treatment and mediator histories, whereas the consistency condition in Assumption \ref{asscon} reveals only the trajectory that was actually realized. The remaining potential quantities must be recovered from units with different realized histories, which requires the treatment and the mediator to behave as if they were randomly assigned in each period, given the information accumulated up to that period.

\begin{assumption}[Sequential conditional independence (SCIA)]
\label{assscia}
For every full treatment history $\overline{d}_T$, every full mediator history $\overline{m}_T$, and every $t=1,\ldots,T$, 
{\small
\begin{gather}
\big\{\big(\bX_{ij},Y_{ij}(\overline{d}_j,\overline{m}_j),\lambda_j^Y\big)_{j=t+1}^T, Y_{it}(\overline{d}_t,\overline{m}_t)\big\} \indep D_{it} \mid \overline{\bX}_{it}, \overline{D}_{i,t-1} = \overline{d}_{t-1}, \overline{M}_{i,t-1} = \overline{m}_{t-1}, \alpha_i^Y, \overline{\lambda}^Y_t, \label{eqsciad} \\
\big\{\big(\bX_{ij},Y_{ij}(\overline{d}_j,\overline{m}_j),\lambda_j^Y\big)_{j=t+1}^T, Y_{it}(\overline{d}_t,\overline{m}_t)\big\} \indep M_{it} \mid \overline{\bX}_{it}, \overline{D}_{it} = \overline{d}_t, \overline{M}_{i,t-1} = \overline{m}_{t-1}, \alpha_i^Y,\overline{\lambda}_t^Y, \label{eqsciam} \\
\big\{\big(\bX_{ij},M_{ij}(\overline{d}_j),\lambda^M_j\big)_{j=t+1}^T, M_{it}(\overline{d}_t)\big\} \indep D_{it} \mid \overline{\bX}_{it},\overline{D}_{i,t-1}=\overline{d}_{t-1},\alpha^M_i,\overline{\lambda}^M_t. \label{eqsciam2}
\end{gather}
}
\end{assumption}

Condition \eqref{eqsciad} requires that, given the covariate history, the past treatment and mediator histories, and the unobservables of the outcome process, the current treatment be independent of the current potential outcome and of future covariates, time effects, and potential outcomes. Condition \eqref{eqsciam} imposes the same requirement on the current mediator, with the contemporaneous treatment added to the conditioning set, and thus rules out mediator--outcome confounding left unexplained by the observed history and the unit effect. Condition \eqref{eqsciam2} is the analogue of \eqref{eqsciad} for the mediator process. 

Because $\alpha_i^Y$ and $\alpha_i^M$ enter the conditioning sets, Assumption \ref{assscia} tolerates arbitrary time-invariant confounding among the treatment, the mediator, and the outcome, as well as arbitrary dependence between the two unit effects; only the variation left after the observed history and the unit effects have been accounted for must be ignorable. This is precisely what the panel structure buys, and it is substantially weaker than the corresponding cross-sectional requirement. Because the assumption restricts the joint behavior of unobserved potential variables, its plausibility in a given application may be examined with single-world intervention graphs \citep{richardson2013single, intro_wu2026}.

The mediation model of Section \ref{secmod} is formulated in terms of potential outcomes, whereas the data contain realized outcomes and mediators only. The following theorem translates the potential-outcome system into observed-data restrictions.

\begin{theorem}[Identification]
\label{th1}
Under Assumptions \ref{assut} to \ref{assscia}, for every $t$ and every $(\overline{d}_t,\overline{m}_t)$ in the relevant conditional support,
\begin{align*}
&\E(Y_{it}\mid\overline{\bX}_{it},\overline{D}_{it}=\overline{d}_t,\overline{M}_{it}=\overline{m}_t,\alpha_i^Y,\overline{\lambda}_t^Y)
=\E\{Y_{it}(\overline{d}_t,\overline{m}_t)\mid\overline{\bX}_{it},\alpha_i^Y,\overline{\lambda}_t^Y\},\\
&\E(M_{it}\mid\overline{\bX}_{it},\overline{D}_{it}=\overline{d}_t,\alpha_i^M,\overline{\lambda}_t^M)
=\E\{M_{it}(\overline{d}_t)\mid\overline{\bX}_{it},\alpha_i^M,\overline{\lambda}_t^M\}.
\end{align*}
If the mediation model for the potential outcomes also holds, the observed data satisfy
\begin{equation}
\label{eq:observed-mediation-model}
\begin{gathered}
Y_{it}=\tau_Y(\overline{D}_{i,t-L:t},\overline{M}_{i,t-L:t},\overline{\bX}_{i,t-L:t})^\prime\theta_0+f_0(\overline{\bX}_{i,t-L:t})+\alpha_i^Y+\lambda_t^Y+\varepsilon_{it}^Y,\\
\E(\varepsilon_{it}^Y\mid\overline{\bX}_{it},\overline{D}_{it},\overline{M}_{it},\alpha_i^Y,\overline{\lambda}_t^Y)=0,\\
M_{it}=\tau_M(\overline{D}_{i,t-L:t},\overline{\bX}_{i,t-L:t})^\prime\gamma_0+g_0(\overline{\bX}_{i,t-L:t})+\alpha_i^M+\lambda_t^M+\varepsilon_{it}^M,\\
\E(\varepsilon_{it}^M\mid\overline{\bX}_{it},\overline{D}_{it},\alpha_i^M,\overline{\lambda}_t^M)=0.
\end{gathered}
\end{equation}
\end{theorem}

All variables in \eqref{eq:observed-mediation-model} are observed and carry the same $\theta_0$ and $\gamma_0$ that govern the potential outcomes. Hence every estimand in Section \ref{secce} is a known function of these parameters and moments of observed variables. Were $f_0$ and $g_0$ known, the parameters could be recovered by GMM after eliminating the fixed effects. In applications the nuisance functions are unknown and possibly nonlinear and high-dimensional, so directly substituting machine-learning estimates into the moments can introduce regularization and overfitting biases of first-order relevance. The next section removes these biases using DML. To the best of our knowledge, this paper is the first one to extend the DML framework \citep{chernozhukov2018double} originally developed for cross-sectional settings without mediation to causal mediation analysis in panel data with two-way fixed effects.

\section{Estimation}
\label{secest}
\subsection{Estimation algorithm for structural parameters}
\label{secea}
Following the DML principle of \citet{chernozhukov2018double}, our objective is to estimate the low-dimensional parameter vector $\beta_0\coloneq(\theta_0',\gamma_0')'$ by residualizing the treatment through a two-way fixed effects model and residualizing the mediator through the structural mediator equation in \eqref{eq:observed-mediation-model}.

For a user-specified nonnegative integer $r$, consider the treatment model
\begin{equation}
\label{eq:assignment-models}
D_{it}=h_0(\overline{\bX}_{i,t-r:t})+\alpha_i^D+\lambda_t^D+\varepsilon_{it}^D,
\qquad
\E(\varepsilon_{it}^D\mid\overline{\bX}_{iT},\alpha_i^D,\overline{\lambda}_T^D)=0,
\end{equation}
where $h_0$ is an unknown nuisance function, while $\alpha_i^D$ and $\lambda_t^D$ are the individual and period fixed effects in the treatment process. Variables dated before period 1 are set to 0. The conditional mean restriction in \eqref{eq:assignment-models} implies that the treatment innovation satisfies $\E[\varepsilon_{it}^D\varphi(\overline{\bX}_{iT})]=0$ for every square-integrable measurable function $\varphi$. That is, the treatment innovation is orthogonal to the entire covariate trajectory. 

Using the structural mediator equation itself to residualize $M_{it}$ requires a slightly stronger condition than the sequential covariate restriction in \eqref{eq:observed-mediation-model}. Specifically, we further impose
\begin{equation}
\label{eq:mediator-residualization-exogeneity}
\E(\varepsilon_{it}^M\mid\overline{\bX}_{iT},\overline{D}_{it},\alpha_i^M,\overline{\lambda}_t^M)=0.
\end{equation}
Condition \eqref{eq:mediator-residualization-exogeneity} strengthens the mediator equation only with respect to the covariate path and does not require strict exogeneity of the treatment process. This strengthening is required to ensure the Neyman orthogonality of the proposed moment conditions. 

We first take first differences to eliminate the individual fixed effects. For any scalar- or vector-valued period-indexed quantity $Q_{it}$, including an observed variable, a dictionary evaluated at the observed histories, or a nuisance function evaluated at its arguments, let $\Delta Q_{it}\coloneq Q_{it}-Q_{i,t-1}$. Subscripts $0$ and $0k$ indicate that the true or fold-specific estimated function is used. The same convention applies to the period effects and innovations. For $t=2,\ldots,T$, the differenced equations are
\begin{equation}
\label{eq:first-difference-mediation}
\begin{gathered}
\Delta Y_{it}=\Delta\tau_{Y,it}'\theta_0+\Delta f_{0,it}+\Delta\lambda_t^Y+\Delta\varepsilon_{it}^Y, \quad \Delta M_{it}=\Delta\tau_{M,it}'\gamma_0+\Delta g_{0,it}+\Delta\lambda_t^M+\Delta\varepsilon_{it}^M,\\
\Delta D_{it}=\Delta h_{0,it}+\Delta\lambda_t^D+\Delta\varepsilon_{it}^D.
\end{gathered}
\end{equation}

We then randomly split the $N$ individuals into $K\geqslant3$ folds $\{\mathcal{F}_1,\ldots,\mathcal{F}_K\}$, with each individual assigned to exactly one fold. Define $k'=k+1$ if $k<K$ and $k'=1$ if $k=K$. For the score-evaluation fold $\mathcal{F}_k$, the fold $\mathcal{F}_{k'}$ is used only to form period-specific cross-sectional means and is kept separate from both the score-evaluation observations and the nuisance-training observations. For any $Q_{it}$ and $i\in\mathcal{F}_k$, define the combined first-difference and auxiliary-fold demeaning operator
\begin{equation}
\label{eq:auxiliary-fold-demeaning}
\mathcal{D}_kQ_{it}\coloneq(Q_{it}-Q_{i,t-1})-\frac{1}{|\mathcal{F}_{k'}|}\sum_{j\in\mathcal{F}_{k'}}(Q_{jt}-Q_{j,t-1}),
\qquad \Delta Q_{it}^*\coloneq\mathcal{D}_kQ_{it}.
\end{equation}
The operator is applied componentwise to every relevant quantity. Applying it to all terms in \eqref{eq:first-difference-mediation} eliminates the period fixed effects and yields
\begin{equation}
\label{eq:double-difference-mediation}
\begin{gathered}
\Delta Y_{it}^*=\Delta\tau_{Y,it}^{*'}\theta_0+\Delta f_{0,it}^*+\Delta\varepsilon_{it}^{Y*}, \quad \Delta M_{it}^*=\Delta\tau_{M,it}^{*'}\gamma_0+\Delta g_{0,it}^*+\Delta\varepsilon_{it}^{M*},\\
\Delta D_{it}^*=\Delta h_{0,it}^*+\Delta\varepsilon_{it}^{D*}.
\end{gathered}
\end{equation}

For generic $f$, $g$, and $h$, define the transformed residual functions
\begin{gather*}
U_{it}^Y(f,\theta) \coloneq \Delta Y_{it}^*-\Delta\tau_{Y,it}^{*'}\theta-\Delta f_{it}^*, \quad 
U_{it}^M(g,\gamma) \coloneq \Delta M_{it}^*-\Delta\tau_{M,it}^{*'}\gamma-\Delta g_{it}^*,\\
U_{it}^D(h) \coloneq \Delta D_{it}^*-\Delta h_{it}^*.
\end{gather*}
The first available treatment and mediator residuals are dated period 2. Since contemporaneous transformed residuals are not used as instruments under sequential exogeneity, the first equation period for which such lagged residual instruments are available is period 3. Hence, letting $\mathcal{T}\coloneq\{3,\ldots,T\}$, define for $t\in\mathcal{T}$
\begin{align*}
A_{it}^M(h)&\coloneq\big(U_{i2}^D(h),\ldots,U_{i,t-1}^D(h)\big)',\\
A_{it}^Y(h,g,\bar{\gamma})&\coloneq\big(U_{i2}^D(h),\ldots,U_{i,t-1}^D(h),U_{i2}^M(g,\bar{\gamma}),\ldots,U_{i,t-1}^M(g,\bar{\gamma})\big)',
\end{align*}
where $\bar{\gamma}$ is a nuisance copy of $\gamma_0$ used only in the outcome instrument. It is kept distinct from the candidate value $\gamma$ entering the mediator equation moment. Consequently, once the nuisance quantities have been estimated outside the score-evaluation fold, the outcome instrument does not vary with the final GMM candidate $\gamma$. The resulting score is affine in $\beta$, so the final GMM criterion remains quadratic even though the structural mediator residual enters the outcome instrument. Let $W_i$ denote the full observed trajectory of individual $i$, and let $\eta\coloneq(f,g,h,\bar{\gamma})$ and $\eta_0\coloneq(f_0,g_0,h_0,\gamma_0)$. The score functions are
\[
\psi_{it}^Y(W_i;\eta,\theta) \coloneq A_{it}^Y(h,g,\bar{\gamma})U_{it}^Y(f,\theta), \quad \psi_{it}^M(W_i;\eta,\gamma) \coloneq A_{it}^M(h)U_{it}^M(g,\gamma),
\]
and $\psi_{it}(W_i;\eta,\beta)\coloneq(\psi_{it}^Y(W_i;\eta,\theta)',\psi_{it}^M(W_i;\eta,\gamma)')'$. Because every starred quantity also depends on the independent auxiliary fold $\mathcal{F}_{k'}$, expectations involving starred quantities are taken jointly over the target individual and the observations used to form the auxiliary-fold means. At the true parameter and nuisance values, the equation residuals reduce to the transformed outcome and mediator innovations, while the instrument vectors consist of lagged transformed treatment and mediator innovations. Under \eqref{eq:observed-mediation-model}, \eqref{eq:assignment-models}, and \eqref{eq:mediator-residualization-exogeneity},
\begin{equation}
\label{eq:orthogonal-moments-mediation}
\E[\psi_{it}(W_i;\eta_0,\beta_0)]=0,
\qquad t\in\mathcal{T},
\end{equation}
and, for every $\eta$ in a neighborhood of $\eta_0$,
\begin{equation}
\label{eq:neyman-orthogonality-mediation}
\frac{\partial}{\partial a}\E\big[\psi_{it}\{W_i;\eta_0+a(\eta-\eta_0),\beta_0\}\big]\bigg|_{a=0}=0.
\end{equation}
In particular, for every $p<t$, the derivative with respect to the nuisance copy satisfies
\[
-\E\big[\Delta\tau_{M,ip}^*\Delta\varepsilon_{it}^{Y*}\big]=\boldsymbol{0}.
\]
This follows from the sequential outcome restriction because $\Delta\tau_{M,ip}$ is measurable with respect to histories dated no later than $p$, while auxiliary-fold demeaning preserves the corresponding zero covariance. The remaining derivatives vanish by the same sequential restrictions together with the full-covariate-history conditions in \eqref{eq:assignment-models} and \eqref{eq:mediator-residualization-exogeneity}.

Replacing level instruments by lagged differenced residuals removes the period-2 moment and may reduce the number of available moments. Identification therefore requires the population Jacobian of the stacked moments to have full column rank. When necessary, the instrument vectors can be augmented by additional valid residualized transformations or interactions of lagged treatment and mediator variables.

In light of \eqref{eq:orthogonal-moments-mediation} and the sample-splitting construction above, we implement the moments as follows. For notational convenience, suppose that $N$ is divisible by $K$.
\begin{algorithm}[Double/debiased machine learning for the panel mediation model]
\label{ag:dml-mediation}\ 
\begin{enumerate}[leftmargin=0.5cm]
\item Randomly split the $N$ individuals into $K$ folds $\{\mathcal{F}_1,\ldots,\mathcal{F}_K\}$ of equal size.
\item For each $k=1,\ldots,K$:
\begin{enumerate}[leftmargin=0.5cm]
\item Set $k'=k+1$ if $k<K$ and $k'=1$ otherwise, and let $\mathcal{F}_{k,k'}^c\coloneq\{1,\ldots,N\}\backslash(\mathcal{F}_k\cup\mathcal{F}_{k'})$.
\item Using only individuals in $\mathcal{F}_{k,k'}^c$, construct $\widehat{f}_{0k}$, $\widehat{g}_{0k}$, $\widehat{h}_{0k}$, and a preliminary estimator $\widetilde{\gamma}_k$ of $\gamma_0$. 
\item For every $i\in\mathcal{F}_k$ and $t=2,\ldots,T$, compute all starred quantities entering the residuals below using $\mathcal{F}_{k'}$ according to \eqref{eq:auxiliary-fold-demeaning}. This includes the transformed outcomes, treatment and mediator dictionaries, and the fitted values of all nuisance functions.
\end{enumerate}
\item For every $k$, $i\in\mathcal{F}_k$, and $t=2,\ldots,T$, define $\widehat{U}_{it,k}^Y(\theta)\coloneq\Delta Y_{it}^*-\Delta\tau_{Y,it}^{*'}\theta-\Delta\widehat{f}_{0k,it}^*$, $\widehat{U}_{it,k}^M(\gamma) \coloneq \Delta M_{it}^*-\Delta\tau_{M,it}^{*'}\gamma-\Delta\widehat{g}_{0k,it}^*$, $\widehat{U}_{it,k}^D\coloneq\Delta D_{it}^*-\Delta\widehat{h}_{0k,it}^*$. For $t\in\mathcal{T}$, let
\[
\widehat{A}_{it,k}^M \coloneq\big(\widehat{U}_{i2,k}^D,\ldots,\widehat{U}_{i,t-1,k}^D\big)', 
\widehat{A}_{it,k}^Y \coloneq\big(\widehat{U}_{i2,k}^D,\ldots,\widehat{U}_{i,t-1,k}^D,\widehat{U}_{i2,k}^M(\widetilde{\gamma}_k),\ldots,\widehat{U}_{i,t-1,k}^M(\widetilde{\gamma}_k)\big)'.
\]
Define $\widehat{\boldsymbol{A}}_{i,k}^Y\coloneq\mathrm{Diag}(\widehat{A}_{i3,k}^Y,\ldots,\widehat{A}_{iT,k}^Y)$, $\widehat{\boldsymbol{U}}_{i,k}^Y(\theta)\coloneq\big(\widehat{U}_{i3,k}^Y(\theta),\ldots,\widehat{U}_{iT,k}^Y(\theta)\big)'$,\\ $\widehat{\boldsymbol{A}}_{i,k}^M\coloneq\mathrm{Diag}(\widehat{A}_{i3,k}^M,\ldots,\widehat{A}_{iT,k}^M)$, $\widehat{\boldsymbol{U}}_{i,k}^M(\gamma)\coloneq\big(\widehat{U}_{i3,k}^M(\gamma),\ldots,\widehat{U}_{iT,k}^M(\gamma)\big)'$. Here and below, $\mathrm{Diag}(a_3,\ldots,a_T)$ denotes the block-diagonal matrix whose period-$t$ block is the column vector $a_t$.
\item For $\beta=(\theta',\gamma')'$, define
\[
\widehat{\psi}_i^k(\beta)\coloneq
\begin{pmatrix}
\widehat{\boldsymbol{A}}_{i,k}^Y\widehat{\boldsymbol{U}}_{i,k}^Y(\theta)\\
\widehat{\boldsymbol{A}}_{i,k}^M\widehat{\boldsymbol{U}}_{i,k}^M(\gamma)
\end{pmatrix},
\]
and obtain the DML--GMM estimator from
\begin{equation}
\label{eq:dml-gmm-mediation-estimator}
\widehat{\beta}_0^D\coloneq\mathop{\arg\min}_{\beta\in\mathbb{R}^{s_Y+s_M}}\bigg\{\frac{1}{N}\sum_{k=1}^K\sum_{i\in\mathcal{F}_k}\widehat{\psi}_i^k(\beta)\bigg\}'\widehat{\boldsymbol{\Omega}}\bigg\{\frac{1}{N}\sum_{k=1}^K\sum_{i\in\mathcal{F}_k}\widehat{\psi}_i^k(\beta)\bigg\},
\end{equation}
where $\widehat{\boldsymbol{\Omega}}$ is positive definite and held fixed during minimization.
\end{enumerate}
\end{algorithm}

The preliminary coefficient $\widetilde{\gamma}_k$ appears only in the outcome instrument; the reported estimator of $\gamma_0$ is the $\gamma$ component of $\widehat{\beta}_0^D$. Under merely sequential exogeneity, the nuisance functions and $\widetilde{\gamma}_k$ may be learned by regularized IV or GMM methods using valid lagged instruments. 

Taking $\widehat{\boldsymbol{\Omega}}$ to be the identity matrix gives a one-step estimator. A two-step estimator instead uses the inverse of an estimator of the covariance matrix of the first-step moments. Although the outcome moments do not contain the final GMM candidate $\gamma$, jointly stacking the outcome and mediator equations and estimating their full moment covariance retains the cross-equation dependence needed for inference on natural direct and indirect effects. Identification requires the population Jacobian of the stacked moments to have full column rank. 

\subsection{Estimators of causal estimands}
\label{secest-causal}
The structural form of the mediation model allows us to construct counterfactual mediators and outcomes without estimating the individual or period fixed effects. For any treatment history $\overline{d}_t$, define the estimated counterfactual mediator for individual $i$ in period $t$ by
\begin{equation}
\label{eq:estimated-counterfactual-mediator}
\widehat{M}_{it}(\overline{d}_t)\coloneq M_{it}+\big(\tau_M(\overline{d}_{t-L:t},\overline{\bX}_{i,t-L:t})-\tau_M(\overline{D}_{i,t-L:t},\overline{\bX}_{i,t-L:t})\big)'\widehat{\gamma}_0^D,
\end{equation}
and let $\widehat{\overline{M}}_{it}(\overline{d}_t)\coloneq(\widehat{M}_{i1}(d_1),\ldots,\widehat{M}_{it}(\overline{d}_t))'$. Likewise, for a prespecified mediator history $\overline{m}_t$, define
\begin{equation}
\label{eq:estimated-counterfactual-outcome}
\widehat{Y}_{it}(\overline{d}_t,\overline{m}_t)\coloneq Y_{it}+\big(\tau_Y(\overline{d}_{t-L:t},\overline{m}_{t-L:t},\overline{\bX}_{i,t-L:t})-\tau_Y(\overline{D}_{i,t-L:t},\overline{M}_{i,t-L:t},\overline{\bX}_{i,t-L:t})\big)'\widehat{\theta}_0^D.
\end{equation}
The nested counterfactual outcome $\widehat{Y}_{it}\{\overline{a}_t,\widehat{\overline{M}}_{it}(\overline{b}_t)\}$ is obtained by replacing $\overline{d}_t$ in \eqref{eq:estimated-counterfactual-outcome} with $\overline{a}_t$ and replacing $\overline{m}_t$ with $\widehat{\overline{M}}_{it}(\overline{b}_t)$. The nuisance functions and the fixed effects cancel from \eqref{eq:estimated-counterfactual-mediator} and \eqref{eq:estimated-counterfactual-outcome}; therefore, only the DML--GMM estimators of $\theta_0$ and $\gamma_0$ are required for these causal contrasts.

The plug-in estimators of the history-specific causal effects are
\begin{align*}
\widehat{\operatorname{CDE}}_t(\overline{d}_t,\overline{d}_t^*;\overline{m}_t)&\coloneq\frac{1}{N}\sum_{i=1}^N\big\{\widehat{Y}_{it}(\overline{d}_t,\overline{m}_t)-\widehat{Y}_{it}(\overline{d}_t^*,\overline{m}_t)\big\},\\
\widehat{\operatorname{NDE}}_t(\overline{d}_t,\overline{d}_t^*)&\coloneq\frac{1}{N}\sum_{i=1}^N\big[\widehat{Y}_{it}\{\overline{d}_t,\widehat{\overline{M}}_{it}(\overline{d}_t^*)\}-\widehat{Y}_{it}\{\overline{d}_t^*,\widehat{\overline{M}}_{it}(\overline{d}_t^*)\}\big],\\
\widehat{\operatorname{NIE}}_t(\overline{d}_t,\overline{d}_t^*)&\coloneq\frac{1}{N}\sum_{i=1}^N\big[\widehat{Y}_{it}\{\overline{d}_t,\widehat{\overline{M}}_{it}(\overline{d}_t)\}-\widehat{Y}_{it}\{\overline{d}_t,\widehat{\overline{M}}_{it}(\overline{d}_t^*)\}\big],\\
\widehat{\operatorname{TE}}_t(\overline{d}_t,\overline{d}_t^*)&\coloneq \widehat{\operatorname{NDE}}_t(\overline{d}_t,\overline{d}_t^*)+\widehat{\operatorname{NIE}}_t(\overline{d}_t,\overline{d}_t^*).
\end{align*}
When the treatment and mediator are continuous and the dictionaries are differentiable, the derivative effects can be estimated analogously. Specifically,
\begin{gather*}
\widehat{\operatorname{CDD}}_{t,j}(\overline{d}_t,\overline{m}_t)\coloneq\frac{1}{N}\sum_{i=1}^N\frac{\partial}{\partial d_j}\widehat{Y}_{it}(\overline{d}_t,\overline{m}_t),\quad 
\widehat{\operatorname{NDD}}_{t,j}(\overline{d}_t)\coloneq\frac{1}{N}\sum_{i=1}^N\frac{\partial}{\partial a_j}\widehat{Y}_{it}\{\overline{a}_t,\widehat{\overline{M}}_{it}(\overline{b}_t)\}\bigg|_{\overline{a}_t=\overline{b}_t=\overline{d}_t},\\
\widehat{\operatorname{NID}}_{t,j}(\overline{d}_t)\coloneq\frac{1}{N}\sum_{i=1}^N\frac{\partial}{\partial b_j}\widehat{Y}_{it}\{\overline{a}_t,\widehat{\overline{M}}_{it}(\overline{b}_t)\}\bigg|_{\overline{a}_t=\overline{b}_t=\overline{d}_t},\quad \widehat{\operatorname{TD}}_{t,j}(\overline{d}_t)\coloneq\widehat{\operatorname{NDD}}_{t,j}(\overline{d}_t)+\widehat{\operatorname{NID}}_{t,j}(\overline{d}_t).
\end{gather*}
The derivative in $\widehat{\operatorname{NID}}_{t,j}$ is evaluated by the chain rule and therefore incorporates the derivative of \eqref{eq:estimated-counterfactual-mediator} with respect to $b_j$. 

For a smooth causal estimand, inference can be conducted by combining the influence function of $\widehat{\beta}_0^D$ with the empirical variation of the individual plug-in contributions and then applying the multivariate delta method. A convenient alternative, particularly for nonlinear dictionaries and natural indirect effects, is a cross-sectional bootstrap that resamples individuals and repeats the complete nuisance-estimation and DML--GMM procedure in each bootstrap sample.

\section{Asymptotic theory}
\label{secas}
The asymptotic argument proceeds in two stages. We first establish an asymptotic linear representation and root-$N$ asymptotic normality for the joint DML--GMM estimator $\widehat{\beta}_0^D=(\widehat{\theta}_0^{D\prime},\widehat{\gamma}_0^{D\prime})'$ of the structural coefficients. We then translate this coefficient-level result into asymptotic theory for the plug-in estimators of the causal effects. 

Because the coefficient-level analysis is technically involved, its complete conditions, auxiliary results, asymptotic normality theorem, and sandwich variance theory are deferred to Appendix C. We retain in the main text only the sampling framework and the structural-parameter linear representation needed to carry out the second stage. We assume that
\begin{equation}
\label{iid}
\left\{\overline{D}_{iT},\overline{\bX}_{iT},\alpha_i^Y,\alpha_i^M,\overline{\varepsilon}_{iT}^Y,\overline{\varepsilon}_{iT}^M\right\}_{i=1}^N
\text{ are independent and identically distributed (i.i.d.)}.
\end{equation}
This assumption is imposed only across units. Within a given unit, the treatment, mediator, covariate, and innovation sequences may be arbitrarily dependent over time, and the unit fixed effects may be correlated with each other and with the observed histories. Throughout this section, $T\geqslant3$, $K$, $L$, $r$, $s_Y$, and $s_M$ are fixed, $N\to\infty$, and $|\mathcal{F}_k|/N\to1/K$ for every $k$.

The structural-parameter theory in Appendix C is developed under the observed-data moment restrictions in \eqref{eq:observed-mediation-model}--\eqref{eq:mediator-residualization-exogeneity}. At a high level, it requires finite moments and foldwise mean-square consistency of the nuisance estimators, a second-order product-rate restriction on the equation-regression and instrument-construction errors, and stability and full column rank of the population Jacobian. The relevant products of nuisance errors must be $o_p(N^{-1/2})$; $o_p(N^{-1/4})$ rates for all nuisance components are sufficient but not necessary. The appendix also shows that the sample means introduced by auxiliary-fold demeaning are negligible at the first-order scale.

Let $F_0$ denote the distribution of the individual trajectory $W_i$, let $\widetilde{\psi}_i$ denote the oracle stacked score formed from the lagged differenced treatment and mediator innovations and the corresponding outcome and mediator innovations, and let $\boldsymbol{G}$ be the population Jacobian of the stacked moments. If $\widehat{\boldsymbol{\Omega}}\xrightarrow{p}\boldsymbol{\Omega}$, define
\[
\mathcal{A}_0\coloneq(\boldsymbol{G}'\boldsymbol{\Omega}\boldsymbol{G})^{-1}\boldsymbol{G}'\boldsymbol{\Omega},
\qquad
\varphi_{\beta,i}\coloneq-\mathcal{A}_0\widetilde{\psi}_i.
\]
Under the conditions stated formally in Appendix C,
\begin{equation}
\sqrt{N}(\widehat{\beta}_0^D-\beta_0)
=\frac{1}{\sqrt{N}}\sum_{i=1}^N\varphi_{\beta,i}+o_p(1)
\xrightarrow{d}N(\boldsymbol{0},\boldsymbol{V}_\beta),
\qquad
\boldsymbol{V}_\beta\coloneq\E(\varphi_{\beta,i}\varphi_{\beta,i}').
\label{eq:structural-parameter-input}
\end{equation}
The same appendix establishes consistency of the stacked sandwich estimator of $\boldsymbol{V}_\beta$. Equation \eqref{eq:structural-parameter-input} is the structural-parameter input used throughout the remainder of this section.

We now carry out the second stage. The following unit-level contributions connect the plug-in estimators to the causal targets, while Assumption \ref{ass:causal-effect-regularity} supplies the smoothness and moment conditions needed to propagate the structural-parameter representation in \eqref{eq:structural-parameter-input} through these maps.

For a generic value $\beta=(\theta',\gamma')'$, define
\begin{align}
M_{it}^{\beta}(\overline{d}_t)&\coloneq M_{it}+\Big\{\tau_M(\overline{d}_{t-L:t},\overline{\bX}_{i,t-L:t})-\tau_M(\overline{D}_{i,t-L:t},\overline{\bX}_{i,t-L:t})\Big\}'\gamma,\label{eq:generic-counterfactual-mediator}\\
Y_{it}^{\beta}(\overline{d}_t,\overline{m}_t)&\coloneq Y_{it}+\Big\{\tau_Y(\overline{d}_{t-L:t},\overline{m}_{t-L:t},\overline{\bX}_{i,t-L:t})-\tau_Y(\overline{D}_{i,t-L:t},\overline{M}_{i,t-L:t},\overline{\bX}_{i,t-L:t})\Big\}'\theta.\notag
\end{align}
Let $\overline{M}_{it}^{\beta}(\overline{d}_t)\coloneq\{M_{i1}^{\beta}(d_1),\ldots,M_{it}^{\beta}(\overline{d}_t)\}'$. Under the potential-outcome mediation model and consistency, $M_{it}^{\beta_0}(\overline{d}_t)=M_{it}(\overline{d}_t)$, $Y_{it}^{\beta_0}(\overline{d}_t,\overline{m}_t)=Y_{it}(\overline{d}_t,\overline{m}_t)$. Accordingly, define the unit-level contributions to the level effects by
\begin{gather*}
\ell_{i,t}^{\mathrm{CDE}}(\beta)\coloneq Y_{it}^{\beta}(\overline{d}_t,\overline{m}_t)-Y_{it}^{\beta}(\overline{d}_t^*,\overline{m}_t),\quad \ell_{i,t}^{\mathrm{NDE}}(\beta)\coloneq Y_{it}^{\beta}\{\overline{d}_t,\overline{M}_{it}^{\beta}(\overline{d}_t^*)\}-Y_{it}^{\beta}\{\overline{d}_t^*,\overline{M}_{it}^{\beta}(\overline{d}_t^*)\},\\
\ell_{i,t}^{\mathrm{NIE}}(\beta)\coloneq Y_{it}^{\beta}\{\overline{d}_t,\overline{M}_{it}^{\beta}(\overline{d}_t)\}-Y_{it}^{\beta}\{\overline{d}_t,\overline{M}_{it}^{\beta}(\overline{d}_t^*)\},\quad \ell_{i,t}^{\mathrm{TE}}(\beta)\coloneq Y_{it}^{\beta}\{\overline{d}_t,\overline{M}_{it}^{\beta}(\overline{d}_t)\}-Y_{it}^{\beta}\{\overline{d}_t^*,\overline{M}_{it}^{\beta}(\overline{d}_t^*)\}.
\end{gather*}
Their expectations at $\beta_0$ equal the corresponding causal effects in Section \ref{secce}, and $\ell_{i,t}^{\mathrm{TE}}(\beta)=\ell_{i,t}^{\mathrm{NDE}}(\beta)+\ell_{i,t}^{\mathrm{NIE}}(\beta)$ for every $i$ and $\beta$.

For continuous treatments and mediators, define analogously
\begin{gather*}
\ell_{i,t,j}^{\mathrm{CDD}}(\beta)\coloneq\frac{\partial}{\partial d_j}Y_{it}^{\beta}(\overline{d}_t,\overline{m}_t),\quad \ell_{i,t,j}^{\mathrm{NDD}}(\beta)\coloneq\frac{\partial}{\partial a_j}Y_{it}^{\beta}\{\overline{a}_t,\overline{M}_{it}^{\beta}(\overline{b}_t)\}\bigg|_{\overline{a}_t=\overline{b}_t=\overline{d}_t},\\
\ell_{i,t,j}^{\mathrm{NID}}(\beta)\coloneq\frac{\partial}{\partial b_j}Y_{it}^{\beta}\{\overline{a}_t,\overline{M}_{it}^{\beta}(\overline{b}_t)\}\bigg|_{\overline{a}_t=\overline{b}_t=\overline{d}_t},\quad \ell_{i,t,j}^{\mathrm{TD}}(\beta)\coloneq\ell_{i,t,j}^{\mathrm{NDD}}(\beta)+\ell_{i,t,j}^{\mathrm{NID}}(\beta).
\end{gather*}
The derivative in $\ell_{i,t,j}^{\mathrm{NID}}(\beta)$ is a total derivative and therefore differentiates through the counterfactual mediator map in \eqref{eq:generic-counterfactual-mediator}.

Let $\ell(W_i;\beta)\in\mathbb{R}^p$ stack any fixed finite collection of the preceding contributions and define
\begin{equation}
\vartheta_0\coloneq \E\{\ell(W_i;\beta_0)\},\qquad \widehat{\vartheta}\coloneq \mathbb{P}_N\ell(W_i;\widehat{\beta}_0^D),\qquad \mathbb{P}_N a_i\coloneq\frac{1}{N}\sum_{i=1}^N a_i.
\label{eq:generic-causal-estimator}
\end{equation}
Thus, each component of $\vartheta_0$ is a causal estimand in Section \ref{secce}, and each component of $\widehat{\vartheta}$ is its estimator in Section \ref{secest-causal}. Write $\dot{\ell}_\beta(w;\beta)\coloneq\frac{\partial\ell(w;\beta)}{\partial\beta'}$, so that $\dot{\ell}_\beta(w;\beta)$ is a $p\times(s_Y+s_M)$ Jacobian.

\begin{assumption}[Regularity of the causal-effect maps]
\label{ass:causal-effect-regularity}
There exists a compact and convex set $\mathcal{B}$ such that $\beta_0\in\operatorname{int}(\mathcal{B})$ and the following conditions hold:
\begin{enumerate}
\item[(i)] For every $\beta\in\mathcal{B}$, the maps $w\mapsto\ell(w;\beta)$ and $w\mapsto\dot{\ell}_\beta(w;\beta)$ are measurable. For $F_0$-almost every $w$, the map $\beta\mapsto\ell(w;\beta)$ is continuously differentiable on $\mathcal{B}$.
\item[(ii)] $\E\|\ell(W_i;\beta_0)\|^2+\E\sup_{\beta\in\mathcal{B}}\|\dot{\ell}_\beta(W_i;\beta)\|_F^2<\infty$.
\item[(iii)] For every derivative effect included in $\vartheta_0$, the treatment- and mediator-history derivatives entering both $\ell(W_i;\beta)$ and $\dot{\ell}_\beta(W_i;\beta)$ exist in a neighborhood of the histories at which they are evaluated and admit integrable envelopes, uniformly over $\beta\in\mathcal{B}$. In particular, differentiation with respect to the treatment and mediator histories may be interchanged with expectation.
\end{enumerate}
\end{assumption}

Assumption \ref{ass:causal-effect-regularity} controls the fixed-dimensional map that transforms a unit trajectory and the structural parameter into a causal contrast. Parts (i)--(ii) provide the uniform first-order expansion needed when $\beta_0$ is replaced by $\widehat{\beta}_0^D$ and ensure that the resulting influence function has a finite variance; no second derivative of the causal-effect map is required. Part (iii) is needed only for derivative estimands and justifies representing a derivative of a population response surface by the expectation of the corresponding unit-level derivative. For commonly used smooth dictionaries, these requirements reduce to moment conditions on the relevant dictionary terms and their derivatives.

Define $\boldsymbol{J}_\vartheta\coloneq\E\{\dot{\ell}_\beta(W_i;\beta_0)\}$. By \eqref{eq:structural-parameter-input}, $\varphi_{\beta,i}=-\mathcal{A}_0\widetilde{\psi}_i$ is the influence function of $\widehat{\beta}_0^D$. For natural effects, $\boldsymbol{J}_\vartheta$ is a total derivative with respect to both parameter blocks and therefore differentiates through $\overline{M}_{it}^{\beta}(\cdot)$.

\begin{theorem}
\label{th:causal-effect-normality}
Suppose that the potential-outcome mediation model and Assumption \ref{asscon} hold. Suppose also that the structural-parameter conditions stated in Appendix C are satisfied, so that \eqref{eq:structural-parameter-input} holds, and that Assumption \ref{ass:causal-effect-regularity} is satisfied. Then
\begin{equation}
\sqrt{N}(\widehat{\vartheta}-\vartheta_0)=\frac{1}{\sqrt{N}}\sum_{i=1}^N\varphi_{\vartheta,i}+o_p(1)\xrightarrow{d}N(\boldsymbol{0},\boldsymbol{\Sigma}_\vartheta),
\label{eq:causal-effect-linearization}
\end{equation}
where
\begin{equation}
\varphi_{\vartheta,i}\coloneq\ell(W_i;\beta_0)-\vartheta_0+\boldsymbol{J}_\vartheta\varphi_{\beta,i}
=\ell(W_i;\beta_0)-\vartheta_0-\boldsymbol{J}_\vartheta\mathcal{A}_0\widetilde{\psi}_i,\quad
\boldsymbol{\Sigma}_\vartheta\coloneq\E(\varphi_{\vartheta,i}\varphi_{\vartheta,i}').\label{eq:causal-effect-if}
\end{equation}
The conclusion holds jointly for any fixed finite collection of causal effects. The covariance matrix $\boldsymbol{\Sigma}_\vartheta$ is allowed to be singular, as occurs when effects linked by an exact decomposition are stacked simultaneously.
\end{theorem}

The first term in \eqref{eq:causal-effect-if} captures sampling variation in the empirical average of the individual counterfactual contrasts, while $\boldsymbol{J}_\vartheta\varphi_{\beta,i}$ propagates estimation error in $\widehat{\theta}_0^D$ and $\widehat{\gamma}_0^D$. Their covariance is retained in $\boldsymbol{\Sigma}_\vartheta$, which is particularly important for natural effects because the generated counterfactual mediator depends on the estimated mediator parameter.

For feasible analytical inference, let $\widehat{\psi}_i\coloneq\widehat{\psi}_i^k(\widehat{\beta}_0^D)$ for $i\in\mathcal{F}_k$ and define the sample Jacobian
\[
\widehat{\boldsymbol{G}}\coloneq\frac{1}{N}\sum_{k=1}^K\sum_{i\in\mathcal{F}_k}
\frac{\partial\widehat{\psi}_i^k(\widehat{\beta}_0^D)}{\partial\beta'}.
\]
Then define
\begin{align}
\widehat{\mathcal{A}}&\coloneq(\widehat{\boldsymbol{G}}'\widehat{\boldsymbol{\Omega}}\widehat{\boldsymbol{G}})^{-1}\widehat{\boldsymbol{G}}'\widehat{\boldsymbol{\Omega}},\qquad \widehat{\boldsymbol{J}}_\vartheta\coloneq\mathbb{P}_N\dot{\ell}_\beta(W_i;\widehat{\beta}_0^D),\notag\\
\overline{\widehat{\psi}}&\coloneq\mathbb{P}_N\widehat{\psi}_i,\qquad \widehat{\varphi}_{\vartheta,i}\coloneq\ell(W_i;\widehat{\beta}_0^D)-\widehat{\vartheta}-\widehat{\boldsymbol{J}}_\vartheta\widehat{\mathcal{A}}(\widehat{\psi}_i-\overline{\widehat{\psi}}),\notag\\
\widehat{\boldsymbol{\Sigma}}_\vartheta&\coloneq\mathbb{P}_N\widehat{\varphi}_{\vartheta,i}\widehat{\varphi}_{\vartheta,i}'.\label{eq:estimated-causal-if}
\end{align}
Centering the estimated score ensures that $\mathbb{P}_N\widehat{\varphi}_{\vartheta,i}=\boldsymbol{0}$ exactly.

\begin{theorem}[Consistency of the causal-effect variance estimator]
\label{prop:causal-effect-variance}
Suppose that the conditions of Theorem \ref{th:causal-effect-normality} hold and that the additional score-stability conditions in Appendix C for consistency of the structural sandwich estimator are satisfied. In addition, assume that
\begin{equation}
\frac{1}{N}\sum_{k=1}^K\sum_{i\in\mathcal{F}_k}\|\widehat{\psi}_i-\widetilde{\psi}_i\|^2\xrightarrow{p}0.
\label{eq:score-l2-for-causal-variance}
\end{equation}
Then $\widehat{\boldsymbol{J}}_\vartheta\xrightarrow{p}\boldsymbol{J}_\vartheta$, $\widehat{\boldsymbol{\Sigma}}_\vartheta\xrightarrow{p}\boldsymbol{\Sigma}_\vartheta$. Consequently, $\widehat{\boldsymbol{\Sigma}}_\vartheta/N$ consistently estimates the asymptotic covariance matrix of $\widehat{\vartheta}$.
\end{theorem}

For a scalar component $\vartheta_{0q}$ with $\Sigma_{\vartheta,qq}>0$, an asymptotic $(1-\alpha)$ confidence interval is $\widehat{\vartheta}_q\ \pm\ z_{1-\alpha/2}\sqrt{\widehat{\Sigma}_{\vartheta,qq}/N}$. More generally, for a full-row-rank matrix $R$ such that $R\boldsymbol{\Sigma}_\vartheta R'$ is nonsingular, the Wald statistic for $H_0:R\vartheta_0=c$ is $N(R\widehat{\vartheta}-c)'(R\widehat{\boldsymbol{\Sigma}}_\vartheta R')^{-1}(R\widehat{\vartheta}-c)$, which converges to a chi-squared distribution with $\operatorname{rank}(R)$ degrees of freedom.

Although Theorem \ref{prop:causal-effect-variance} provides analytical standard errors, in the empirical analysis we use a cross-sectional bootstrap \citep{kapetanios2008bootstrap}. For each bootstrap replication $b=1,\ldots,B$, we draw $N$ individuals with replacement from the original sample and retain the complete time trajectory of every selected individual. We then repeat the entire estimation procedure to obtain $\widehat{\vartheta}^{*(b)}$. Then, $\{\widehat{\vartheta}^{*(b)}\}_{b=1}^B$ are used to calculate the bootstrap standard error. Because resampling is performed at the individual level, the procedure preserves arbitrary serial dependence within each individual trajectory. Repeating all estimation steps allows the bootstrap variation to reflect nuisance-function estimation, auxiliary-fold transformations, DML--GMM estimation, and the nonlinear counterfactual plug-in calculation.

\section{Simulation study}
\label{sec:simulation}
The Monte Carlo study evaluates the DML--GMM estimator of $\beta_0=(\theta_0',\gamma_0')'$, the plug-in causal-effect estimators, and the influence-function standard errors of Section \ref{secas}. The design is deliberately adverse: it combines nonlinear time-varying confounding, correlated unit effects, equation-specific period effects, misspecified nuisance-function dictionaries, and sequentially endogenous differenced regressors.

\subsection{Simulation design}
\label{sec:simulation-design}
We set $T=10$, $N=1600$, $L_X=5$, and $L=1$. Variables dated before period 1 are zero, and the structural dictionaries and parameters are
\[
\tau_Y=(d_t,d_t^2,d_{t-1},d_tX_{it3},m_t,m_{t-1})',\qquad
\tau_M=(d_t,d_t^2,d_{t-1})',
\]
\[
\theta_0=(2,0.5,2,2,1,0.5)',\qquad \gamma_0=(1,0.5,0.5)'.
\]
Writing $q\in\{D,M,Y\}$, the individual effects are $\alpha_i^q=A_i+\zeta_i^q$ with a common standard-normal factor $A_i$ and independent equation-specific components. The covariates satisfy
\[
X_{i1j}=\nu_{i1j},\qquad X_{itj}=0.5X_{i,t-1,j}+\sqrt{1-0.5^2}\,\nu_{itj},\qquad j=1,\ldots,5,
\]
and the treatment innovation is $\varepsilon_{it}^D=\xi_{it}^D+\varepsilon_{i,t-1}^Y+\varepsilon_{i,t-1}^M$ for $t\geqslant2$. The observed variables are generated recursively from
\begin{align*}
D_{it}&=h_0(X_{it})+\alpha_i^D+\lambda_t^D+\varepsilon_{it}^D,\\
M_{it}&=\tau_M(D_{i,t-1:t},X_{i,t-1:t})'\gamma_0+g_0(X_{it})+\alpha_i^M+\lambda_t^M+\varepsilon_{it}^M,\\
Y_{it}&=\tau_Y(D_{i,t-1:t},M_{i,t-1:t},X_{i,t-1:t})'\theta_0+f_0(X_{it})+\alpha_i^Y+\lambda_t^Y+\varepsilon_{it}^Y.
\end{align*}
The five covariates therefore follow stationary Gaussian AR(1) processes. The unit effects in the treatment, mediator, and outcome equations share a common latent factor, while the period effects differ across equations. The treatment disturbance contains the preceding outcome and mediator innovations. Hence treatment can respond to recent shocks, as a policy maker or firm might do, while current primitive innovations remain independent of the relevant past histories. The level equations satisfy the conditional moment restrictions used for identification, but first differencing induces correlation between the structural regressors and the differenced innovations, so contemporaneous instruments are invalid.

The nuisance functions $f_0$ and $g_0$ contain hyperbolic, trigonometric, interaction, and exponential terms, whereas estimation uses all $55$ monomials of total degree at most three in the five current covariates. Thus the first-stage library does not span either nuisance function. Both methods estimate $f_0$ and $g_0$ from first-differenced and period-demeaned equations by the same $\ell_1$-penalized GMM learner. DML--GMM uses ten folds, a cyclic auxiliary fold, and lagged residualized treatment and mediator innovations; its histories are truncated at two lags and augmented by the treatment-residual square and treatment--covariate interaction, yielding $60$ outcome and $23$ mediator moments. Naive--GMM instead trains once on the full sample, uses full-sample period means, and replaces residualized histories by lagged levels. Both procedures use identity weighting. The comparison therefore isolates cross-fitting, orthogonal residualization, and auxiliary-fold demeaning rather than the quality of the structural nuisance learner. Appendix D.1 gives the complete data-generating process and implementation details of the first-stage, instrument, GMM, and variance calculations.

Level effects are evaluated at $t=T$, comparing $\overline d_T=\boldsymbol1_T$ with $\overline d_T^*=\boldsymbol0_T$ and fixing the mediator history at zero for the controlled direct effect; derivative effects are evaluated at $\overline d_T=\boldsymbol1_T$ with respect to the period-$T$ treatment. The design yields
\begin{equation}
   \label{eq:simulation-truths}
   \operatorname{CDE}_T=\operatorname{NDE}_T=4.5, \operatorname{NIE}_T=3, \operatorname{TE}_T=7.5, \operatorname{NDD}_{T,T}=3, \operatorname{NID}_{T,T}=2, \operatorname{TD}_{T,T}=5.
\end{equation}

We draw $1{,}000$ independent panels. For each structural parameter and causal effect, we report the Monte Carlo mean, signed bias, standard deviation, root mean squared error, average estimated standard error, and coverage of the nominal $95\%$ Wald interval. Structural standard errors use the stacked sandwich estimator developed in Appendix C; causal-effect standard errors use the estimated influence function in \eqref{eq:estimated-causal-if}.

\subsection{Simulation results}
\begin{table}[!ht]
\renewcommand{\arraystretch}{0.5}
\centering
\caption{Monte Carlo results for structural parameters and causal effects}
\label{tab:simulation-results}
\small
\setlength{\tabcolsep}{10pt}
\begin{threeparttable}
\begin{tabular}{lccccccc}
\toprule
Estimand & Method & Mean & Bias & SD & RMSE & ASE & CP \\
\midrule
\multicolumn{8}{l}{\emph{Panel A: Structural parameters}} \\
\multirow{2}{*}{$\theta_{D_t}$}
 & DML--GMM   & 1.9981 & -0.0019 & 0.0274 & 0.0274 & 0.0273 & 0.951 \\
 & Naive--GMM & 1.9895 & -0.0105 & 0.0600 & 0.0609 & 0.0576 & 0.966 \\

\multirow{2}{*}{$\theta_{D_t^2}$}
 & DML--GMM   & 0.5050 &  0.0050 & 0.0150 & 0.0158 & 0.0152 & 0.953 \\
 & Naive--GMM & 0.5228 &  0.0228 & 0.0461 & 0.0514 & 0.0400 & 0.905 \\

\multirow{2}{*}{$\theta_{D_{t-1}}$}
 & DML--GMM   & 1.9989 & -0.0011 & 0.0169 & 0.0169 & 0.0174 & 0.958 \\
 & Naive--GMM & 2.0052 &  0.0052 & 0.0359 & 0.0363 & 0.0334 & 0.955 \\

\multirow{2}{*}{$\theta_{DX}$}
 & DML--GMM   & 1.9950 & -0.0050 & 0.0215 & 0.0221 & 0.0213 & 0.944 \\
 & Naive--GMM & 1.9705 & -0.0295 & 0.0310 & 0.0428 & 0.0247 & 0.747 \\

\multirow{2}{*}{$\theta_{M_t}$}
 & DML--GMM   & 0.9956 & -0.0044 & 0.0204 & 0.0208 & 0.0204 & 0.958 \\
 & Naive--GMM & 0.9781 & -0.0219 & 0.0678 & 0.0713 & 0.0605 & 0.938 \\

\multirow{2}{*}{$\theta_{M_{t-1}}$}
 & DML--GMM   & 0.5007 &  0.0007 & 0.0065 & 0.0065 & 0.0063 & 0.946 \\
 & Naive--GMM & 0.5014 &  0.0014 & 0.0036 & 0.0039 & 0.0033 & 0.915 \\

\multirow{2}{*}{$\gamma_{D_t}$}
 & DML--GMM   & 0.9907 & -0.0093 & 0.0342 & 0.0355 & 0.0350 & 0.955 \\
 & Naive--GMM & 0.8230 & -0.1770 & 0.0271 & 0.1791 & 0.0104 & 0.000 \\

\multirow{2}{*}{$\gamma_{D_t^2}$}
 & DML--GMM   & 0.4996 & -0.0004 & 0.0141 & 0.0141 & 0.0133 & 0.952 \\
 & Naive--GMM & 0.5074 &  0.0074 & 0.0028 & 0.0079 & 0.0026 & 0.201 \\

\multirow{2}{*}{$\gamma_{D_{t-1}}$}
 & DML--GMM   & 0.4964 & -0.0036 & 0.0231 & 0.0233 & 0.0233 & 0.948 \\
 & Naive--GMM & 0.4827 & -0.0173 & 0.0088 & 0.0194 & 0.0082 & 0.449 \\
\addlinespace[2pt]
\multicolumn{8}{l}{\emph{Panel B: Causal effects}} \\
\multirow{2}{*}{CDE}
 & DML--GMM   & 4.5015 &  0.0015 & 0.0731 & 0.0731 & 0.0726 & 0.950 \\
 & Naive--GMM & 4.5169 &  0.0169 & 0.1448 & 0.1457 & 0.1351 & 0.956 \\

\multirow{2}{*}{NDE}
 & DML--GMM   & 4.5015 &  0.0015 & 0.0731 & 0.0731 & 0.0726 & 0.950 \\
 & Naive--GMM & 4.5169 &  0.0169 & 0.1448 & 0.1457 & 0.1351 & 0.956 \\

\multirow{2}{*}{NIE}
 & DML--GMM   & 2.9726 & -0.0274 & 0.0993 & 0.1029 & 0.1008 & 0.944 \\
 & Naive--GMM & 2.6826 & -0.3174 & 0.1302 & 0.3431 & 0.1099 & 0.204 \\

\multirow{2}{*}{TE}
 & DML--GMM   & 7.4741 & -0.0259 & 0.1053 & 0.1084 & 0.1086 & 0.947 \\
 & Naive--GMM & 7.1995 & -0.3005 & 0.0783 & 0.3106 & 0.0652 & 0.012 \\

\multirow{2}{*}{NDD}
 & DML--GMM   & 3.0076 &  0.0076 & 0.0710 & 0.0714 & 0.0713 & 0.954 \\
 & Naive--GMM & 3.0345 &  0.0345 & 0.1546 & 0.1583 & 0.1417 & 0.948 \\

\multirow{2}{*}{NID}
 & DML--GMM   & 1.9811 & -0.0189 & 0.0611 & 0.0639 & 0.0615 & 0.937 \\
 & Naive--GMM & 1.7976 & -0.2024 & 0.1289 & 0.2399 & 0.1119 & 0.498 \\

\multirow{2}{*}{TD}
 & DML--GMM   & 4.9887 & -0.0113 & 0.0725 & 0.0733 & 0.0737 & 0.952 \\
 & Naive--GMM & 4.8322 & -0.1678 & 0.0645 & 0.1798 & 0.0581 & 0.186 \\
\bottomrule
\end{tabular}
\begin{tablenotes}[flushleft]
\footnotesize
\item Notes: Each method uses 1,000 replications with $T=10$. Bias is signed, SD is the Monte Carlo standard deviation, RMSE is the root mean squared error, ASE is the average estimated standard error, and CP is the coverage probability of the nominal 95\% Wald interval. NDE coincides with CDE by construction.
\end{tablenotes}
\end{threeparttable}
\end{table}
Table \ref{tab:simulation-results} shows that DML--GMM is well centered for all nine structural parameters. Its absolute biases range from $0.0004$ to $0.0093$ and never exceed $1.1\%$ of the corresponding truth. The average estimated standard errors track the Monte Carlo standard deviations to within $6\%$, and coverage stays between $0.944$ and $0.958$. Its RMSE is also substantially smaller for six of the nine coefficients. The three exceptions have only a small absolute RMSE advantage for Naive--GMM, and for the two mediator coefficients that apparent gain reflects a smaller dispersion around a displaced center rather than reliable estimation.

The benchmark fails differently across the two equations. In the outcome equation, it is generally much less precise: the standard deviations of the main treatment and contemporaneous mediator coefficients are roughly two to three times those of DML--GMM. The treatment--covariate interaction also has bias $-0.0295$, which reduces coverage to $0.747$. In the mediator equation, the problem is systematic centering and variance estimation. The estimate of $\gamma_{D_t}$ has bias $-0.1770$, and the coverages of $\gamma_{D_t}$, $\gamma_{D_t^2}$, and $\gamma_{D_{t-1}}$ are $0$, $0.201$, and $0.449$, respectively.

Panel B gives the corresponding causal-effect results. Every DML--GMM absolute bias is below $0.028$, its estimated standard errors reproduce the Monte Carlo dispersion to within $4\%$, and coverage ranges from $0.937$ to $0.954$. It also has the smaller RMSE for all seven effects. Naive--GMM remains reasonably centered for CDE, NDE, and NDD, but its errors in the mediator equation pass through the indirect channel: the biases of NIE, TE, NID, and TD are $-0.3174$, $-0.3005$, $-0.2024$, and $-0.1678$, and their coverages fall to $0.204$, $0.012$, $0.498$, and $0.186$. Its standard errors understate the Monte Carlo dispersion of every causal-effect estimator as well.

Two conclusions are important. First, the improvement cannot be attributed to a better first-stage learner, because the methods use the same covariate dictionary and the same penalized-GMM fits; it is generated by cross-fitting, residualized orthogonal moments, and independent auxiliary-fold demeaning. Second, acceptable direct-effect estimation does not guarantee a credible mediation analysis. Natural and total effects depend jointly on the outcome and mediator equations and on their estimation covariance, so a procedure can appear adequate for the direct channel while being unusable for the indirect channel. Appendix D.2 and D.3 show that these conclusions persist across sample sizes and under an alternative specification of the nuisance functions.

\section{Empirical application}
\label{sec:empirical}
We apply our estimator to the question of how the workplace employment rate of a U.S. county affects personal income per capita and how much of that effect is transmitted by the local average wage. Two findings from the regional labor-market literature motivate this decomposition. Since \citet{blanchard1992regional}, regional labor-demand shocks have been understood to move employment quickly while relative wages adjust more slowly, with participation and migration absorbing much of the adjustment. Decompositions in the tradition of \citet{bartik1991benefits} accordingly attribute the response of local income per capita mainly to the employment margin and a smaller part to average wages.

We use the county panel dataset assembled by \citet{cerqua2026misuse} to illustrate the proposed method. The balanced panel contains all $3{,}058$ U.S. counties annually from 2000 to 2019. Variables for the year 2000 are used to construct covariates and differenced variables, hence $T=19$. The outcome $Y_{it}$ is log personal income per capita, the mediator $M_{it}$ is log average wage per job, and the treatment $D_{it}$ is the workplace employment rate. The covariate vector contains eight county characteristics measured in calendar year $t-1$: log population, the share of personal income received as unemployment insurance, population shares under age 18 and over age 65, births and deaths per thousand residents, and net international and domestic migration per thousand residents. Every variable is winsorized at its pooled $1\%$ and $99\%$ quantiles and then standardized, so the reported effects are in outcome standard deviations.

We set $L=r=0$ and consider two structural specifications. Specification I is homogeneous,
\[
\tau_Y(d_t,m_t)=(d_t,m_t)',\qquad \tau_M(d_t)=(d_t)',
\]
with two outcome parameters and one mediator parameter. Specification II adds treatment--covariate interactions to both equations,
\[
\tau_Y(d_t,m_t,\bX_{it})=(d_t,m_t,d_t\bX_{it}')',\qquad
\tau_M(d_t,\bX_{it})=(d_t,d_t\bX_{it}')',
\]
so the direct effect of employment on income and the effect of employment on wages may vary with county demographics and migration. The two specifications use the same cross-fitted first-stage construction. Appendix E reports the $108$-term covariate library, lagged long-difference instruments, moment dimensions, fold choices, and bootstrap implementation.

For each specification we report averages of the period-$t$ derivative effects over the nineteen sample years,
\begin{equation}
\label{eq:empirical-averaged-estimands}
\overline{\operatorname{NDD}}\coloneq\frac{1}{T}\sum_{t=1}^{T}\operatorname{NDD}_{t,t},\qquad
\overline{\operatorname{NID}}\coloneq\frac{1}{T}\sum_{t=1}^{T}\operatorname{NID}_{t,t},\qquad
\overline{\operatorname{TD}}\coloneq\frac{1}{T}\sum_{t=1}^{T}\operatorname{TD}_{t,t}.
\end{equation}
Under Specification I, $\operatorname{NDD}_{t,t}$, $\operatorname{NID}_{t,t}$, and $\operatorname{TD}_{t,t}$ are constant over time. Under Specification II they vary through $\E(\bX_{it})$, although the structural coefficients themselves remain constant. Because treatment enters both dictionaries linearly, the period-$t$ derivatives also equal the corresponding effects of a one-standard-deviation increase in employment in period $t$, holding the other treatment components at zero. The exact equivalence and the period-specific profiles are given in Appendix E.

\begin{table}[!ht]
\renewcommand{\arraystretch}{0.6}
\centering
\caption{County-level mediation estimates, 2001--2019}
\label{tab:empirical-results}
\small
\setlength{\tabcolsep}{18pt}
\begin{threeparttable}
\begin{tabular}{lcccc}
\toprule
Estimand & Estimate & Boot.\ SE & $95\%$ CI & $p$-value \\
\midrule
\multicolumn{5}{l}{\emph{Panel A: Specification I, homogeneous dictionaries}} \\
$\overline{\operatorname{NDD}}$ & $0.2145$ & $0.1063$ & $[0.0062,\,0.4227]$ & $0.0435$ \\
$\overline{\operatorname{NID}}$ & $0.4017$ & $0.0760$ & $[0.2526,\,0.5507]$ & $<0.0001$ \\
$\overline{\operatorname{TD}}$  & $0.6162$ & $0.0951$ & $[0.4298,\,0.8025]$ & $<0.0001$ \\
$\overline{\operatorname{NID}}/\overline{\operatorname{TD}}$ & $0.6519$ & $0.1423$ & $[0.3729,\,0.9308]$ & --- \\
\addlinespace[2pt]
\multicolumn{5}{l}{\emph{Panel B: Specification II, treatment--covariate interactions}} \\
$\overline{\operatorname{NDD}}$ & $0.3597$ & $0.1261$ & $[0.1125,\,0.6068]$ & $0.0043$ \\
$\overline{\operatorname{NID}}$ & $0.2502$ & $0.0608$ & $[0.1312,\,0.3693]$ & $<0.0001$ \\
$\overline{\operatorname{TD}}$  & $0.6099$ & $0.1126$ & $[0.3892,\,0.8306]$ & $<0.0001$ \\
$\overline{\operatorname{NID}}/\overline{\operatorname{TD}}$ & $0.4103$ & $0.1742$ & $[0.0688,\,0.7518]$ & --- \\
\addlinespace[2pt]
\multicolumn{5}{l}{\emph{Panel C: Paired difference, Specification II minus Specification I}} \\
$\overline{\operatorname{NDD}}$ & $0.1452$ & $0.0943$ & $[-0.0397,\,0.3300]$ & $0.1238$ \\
$\overline{\operatorname{NID}}$ & $-0.1514$ & $0.0616$ & $[-0.2722,\,-0.0306]$ & $0.0140$ \\
$\overline{\operatorname{TD}}$  & $-0.0063$ & $0.0774$ & $[-0.1579,\,0.1454]$ & $0.9356$ \\
$\overline{\operatorname{NID}}/\overline{\operatorname{TD}}$ & $-0.2416$ & $0.1460$ & $[-0.5278,\,0.0446]$ & $0.0980$ \\
\bottomrule
\end{tabular}
\begin{tablenotes}[flushleft]
\footnotesize
\item Notes: Standard errors are the standard deviations of $500$ county-trajectory bootstrap replications; intervals are Wald intervals and $p$-values are two-sided. The two specifications are bootstrapped on the same county resamples, so Panel C is the distribution of the within-replication difference and accounts for the correlation between the two sets of estimates; no $p$-value is reported for the mediated share in Panels A and B because a test against zero is not of interest there.
\end{tablenotes}
\end{threeparttable}
\end{table}

Table \ref{tab:empirical-results} shows that the total effect is large and precisely estimated under both specifications. A one-standard-deviation increase in the workplace employment rate raises log personal income per capita by $0.6162$ standard deviations under Specification I and $0.6099$ under Specification II. On the original scale, one treatment standard deviation equals $0.39$ jobs per working-age resident and one outcome standard deviation equals $0.28$ log units. The estimate therefore corresponds to income per capita about $4.5\%$ higher for ten additional jobs per hundred working-age residents. Both the direct and wage channels are positive at the $5\%$ level in either specification, so wages are quantitatively relevant but not the sole pathway.

The decomposition is much less stable than the total. Because the two specifications are estimated on the same county bootstrap resamples, Panel C reports their within-replication difference and incorporates the strong dependence between the estimates. The total effect changes by only $-0.0063$, with a bootstrap standard error of $0.0774$, whereas the indirect effect changes by $-0.1514$, with a standard error of $0.0616$. The mediated share falls from $65.19\%$ to $41.03\%$. Imposing homogeneous coefficients therefore leaves the total effect statistically indistinguishable while reallocating about one quarter of it between the two channels and reversing which channel is larger. Specification I assigns roughly two thirds of the effect to wages, contrary to the ordering suggested by the regional literature, whereas Specification II assigns about $41\%$ to that channel.

The period-specific estimates under Specification II in Appendix E show what the averages conceal. All three profiles are mildly U-shaped. The total derivative falls from $0.635$ in 2001 to $0.564$ in 2010 and rises to $0.692$ in 2019; the direct and indirect channels reach their minima in 2008 and 2010, respectively. Since the coefficients do not vary by year, this pattern reflects changes in the cross-sectional distribution of county demographics and migration rather than time-varying structural parameters.

Although a causal interpretation relies on Assumption \ref{assscia}---whose plausibility could be verified via single-world intervention graphs \citep{intro_wu2026}---we forgo such verification here for conciseness. Even so, the current application offers a useful demonstration of the estimator on a large regional panel and, in particular, of how the direct-indirect decomposition is sensitive to model specification.

\section{Conclusion}
\label{sec:conclusion}
This paper proposes a framework for causal mediation analysis within a partially linear panel data model with two-way fixed effects. We define history-specific controlled direct, natural direct, natural indirect, and total effects, together with their derivative counterparts for continuous treatments, and identify them under a sequential conditional independence assumption that conditions on the unit and period effects and therefore permits arbitrary time-invariant confounding among treatment, mediator, and outcome. Estimation combines first differencing, a three-group sample splitting scheme, and cross-fitted machine learning of the confounding functions in a GMM problem whose score is Neyman orthogonal in every nuisance component, including the mediator coefficient in the outcome instrument. Asymptotic normality is established for the structural parameters and for all causal estimands, with consistent variance estimators. In simulations, our estimator is nearly unbiased with accurate coverage, whereas a benchmark without cross-fitting, auxiliary-fold demeaning, or residualized instruments performs poorly for the mediator equation and hence for the indirect and total effects. In the U.S. county application, allowing treatment--covariate interactions leaves the total effect essentially unchanged but lowers the estimated mediated share through wages, yielding a decomposition consistent with the regional labor-market literature.

Several promising avenues remain for future research. First, since verifying the identifying assumption via causal graphs is often nontrivial in practice, a sensitivity analysis that examines the potential variation of the reported decomposition under assumption violations would be useful—particularly because, as both our simulations and application show, the decomposition is sensitive to specification choices. Second, replacing the dictionaries by a fully nonparametric specification, with the debiasing term estimated rather than derived in closed form, would remove the functional-form restriction that remains. Third, when the stable unit treatment value assumption is violated because treatments or mediators spill over across units, adapting our approach to account for such interference represents an important direction. Finally, developing an asymptotic theory that allows the number of periods to grow with the number of units would substantially broaden the scope of the framework. 

\setcounter{equation}{0}
\setcounter{theorem}{0}
\setcounter{proposition}{0}
\setcounter{assumption}{0}
\setcounter{corollary}{0}
\setcounter{figure}{0}

\renewcommand{\theequation}{A.\arabic{equation}}
\renewcommand{\thetheorem}{A\arabic{theorem}}
\renewcommand{\theproposition}{A\arabic{proposition}}
\renewcommand{\theassumption}{A\arabic{assumption}}
\renewcommand{\thecorollary}{A\arabic{corollary}}
\renewcommand{\thefigure}{A.\arabic{figure}}

\renewcommand{\theHequation}{appA.\arabic{equation}}
\renewcommand{\theHtheorem}{appA.\arabic{theorem}}
\renewcommand{\theHproposition}{appA.\arabic{proposition}}
\renewcommand{\theHassumption}{appA.\arabic{assumption}}
\renewcommand{\theHcorollary}{appA.\arabic{corollary}}
\renewcommand{\theHfigure}{appA.\arabic{figure}}

\begin{appendices}
\section{Vector and High-Dimensional Mediators}
\label{sec:vector-highdim-mediators}
This section first extends the framework to a fixed-dimensional vector mediator. We then discuss what changes when the number of mediator coordinates increases with the cross-sectional sample size. The latter discussion is not a formal extension of the theory in Section \ref{secas}; its purpose is to clarify the additional identification, regularization, and inferential issues that arise in panel mediation analysis with high-dimensional mediators. 

\subsection{A fixed-dimensional vector mediator}
\label{sec:fixed-vector-mediator}

Let
\[
\boldsymbol{M}_{it}=(M_{it,1},\ldots,M_{it,q_M})'\in\mathbb{R}^{q_M}, \qquad q_M<\infty,
\]
where $q_M$ is fixed as $N\to\infty$, and define
\[
\overline{\boldsymbol{M}}_{it}\coloneq(\boldsymbol{M}_{i1}',\ldots,\boldsymbol{M}_{it}')', \qquad \overline{\boldsymbol{m}}_t\coloneq(\boldsymbol{m}_1',\ldots,\boldsymbol{m}_t')'.
\]
The potential outcome is written as $Y_{it}(\overline{d}_t,\overline{\boldsymbol{m}}_t)$, while $\boldsymbol{M}_{it}(\overline{d}_t)$ denotes the vector of potential mediators under treatment history $\overline{d}_t$. Its natural history is
\[
\overline{\boldsymbol{M}}_{it}(\overline{d}_t)\coloneq\big\{\boldsymbol{M}_{i1}(d_1)', \boldsymbol{M}_{i2}(\overline{d}_2)',\ldots, \boldsymbol{M}_{it}(\overline{d}_t)'\big\}'.
\]
SUTVA, no anticipation, and consistency are understood jointly for the entire mediator vector. We treat its coordinates as parallel mediators: they may be statistically dependent, but no contemporaneous causal ordering among them is imposed.

\paragraph{Joint level effects.}
The mediator vector is treated as one jointly intervenable block. For two treatment histories $\overline{d}_t$ and $\overline{d}_t^*$, define
\begin{align}
\operatorname{CDE}^{(v)}_t(\overline{d}_t,\overline{d}_t^*;\overline{\boldsymbol{m}}_t)&\coloneq \E\{Y_{it}(\overline{d}_t,\overline{\boldsymbol{m}}_t)-Y_{it}(\overline{d}_t^*,\overline{\boldsymbol{m}}_t)\}, \label{eq:vector-cde}\\
\operatorname{NDE}^{(v)}_t(\overline{d}_t,\overline{d}_t^*)&\coloneq \E\big[Y_{it}\{\overline{d}_t,\overline{\boldsymbol{M}}_{it}(\overline{d}_t^*)\}-Y_{it}\{\overline{d}_t^*,\overline{\boldsymbol{M}}_{it}(\overline{d}_t^*)\}\big], \notag\\
\operatorname{NIE}^{(v)}_t(\overline{d}_t,\overline{d}_t^*)&\coloneq \E\big[Y_{it}\{\overline{d}_t,\overline{\boldsymbol{M}}_{it}(\overline{d}_t)\}-Y_{it}\{\overline{d}_t,\overline{\boldsymbol{M}}_{it}(\overline{d}_t^*)\}\big], \label{eq:vector-nie}\\
\operatorname{TE}^{(v)}_t(\overline{d}_t,\overline{d}_t^*)&\coloneq \E\big[Y_{it}\{\overline{d}_t,\overline{\boldsymbol{M}}_{it}(\overline{d}_t)\}-Y_{it}\{\overline{d}_t^*,\overline{\boldsymbol{M}}_{it}(\overline{d}_t^*)\}\big]. \label{eq:vector-te}
\end{align}
As in the scalar case,
\[
\operatorname{TE}^{(v)}_t(\overline{d}_t,\overline{d}_t^*)=\operatorname{NDE}^{(v)}_t(\overline{d}_t,\overline{d}_t^*)+\operatorname{NIE}^{(v)}_t(\overline{d}_t,\overline{d}_t^*).
\]

\paragraph{Joint derivative effects.}
Suppose that the treatment and every mediator coordinate are continuous, the relevant maps are continuously differentiable, and differentiation may be interchanged with expectation. Define
\[
\Psi_t^{(v)}(\overline{a}_t,\overline{b}_t)\coloneq\E\big[Y_{it}\{\overline{a}_t,\overline{\boldsymbol{M}}_{it}(\overline{b}_t)\}\big].
\]
For $j\leqslant t$, define
\begin{align*}
\operatorname{CDD}^{(v)}_{t,j}(\overline{d}_t,\overline{\boldsymbol{m}}_t)&\coloneq \frac{\partial}{\partial d_j}\E\{Y_{it}(\overline{d}_t,\overline{\boldsymbol{m}}_t)\}, \\
\operatorname{NDD}^{(v)}_{t,j}(\overline{d}_t)&\coloneq \frac{\partial}{\partial a_j}\Psi_t^{(v)}(\overline{a}_t,\overline{b}_t)\big|_{\overline{a}_t=\overline{b}_t=\overline{d}_t}, \\
\operatorname{NID}^{(v)}_{t,j}(\overline{d}_t)&\coloneq \frac{\partial}{\partial b_j}\Psi_t^{(v)}(\overline{a}_t,\overline{b}_t)\big|_{\overline{a}_t=\overline{b}_t=\overline{d}_t}, \\
\operatorname{TD}^{(v)}_{t,j}(\overline{d}_t)&\coloneq \frac{\partial}{\partial d_j}\E\big[Y_{it}\{\overline{d}_t,\overline{\boldsymbol{M}}_{it}(\overline{d}_t)\}\big].
\end{align*}
The chain rule gives
\[
\operatorname{TD}^{(v)}_{t,j}(\overline{d}_t)=\operatorname{NDD}^{(v)}_{t,j}(\overline{d}_t)+\operatorname{NID}^{(v)}_{t,j}(\overline{d}_t),
\]
and
\begin{align}
\operatorname{NID}^{(v)}_{t,j}(\overline{d}_t)&=\sum_{\ell=1}^{q_M}\operatorname{NID}^{(v)}_{t,j,\ell}(\overline{d}_t), \label{eq:vector-component-nid}\\
\operatorname{NID}^{(v)}_{t,j,\ell}(\overline{d}_t)&\coloneq \E\left[\sum_{r=j}^t\left.\frac{\partial Y_{it}(\overline{d}_t,\overline{\boldsymbol{m}}_t)}{\partial m_{r\ell}}\right|_{\overline{\boldsymbol{m}}_t=\overline{\boldsymbol{M}}_{it}(\overline{d}_t)}\frac{\partial M_{ir,\ell}(\overline{d}_r)}{\partial d_j}\right]. \notag
\end{align}
Under the finite-lag model below, terms outside the relevant lag window are zero.

\begin{remark}[Mediator-specific decompositions]
\label{rem:joint-component-mediation}
The level effects above treat the mediator vector jointly. The joint natural indirect effect in \eqref{eq:vector-nie} is therefore well defined without assigning portions of the effect to individual mediator coordinates. Such an assignment is generally order dependent if the coordinates are changed one at a time. For illustration, suppress the unit, time, and treatment arguments and suppose that
\[
Y(m_1,m_2)=m_1+m_2+m_1m_2,
\]
and that the natural mediator vector changes from $(0,0)'$ under $d^*$ to $(1,1)'$ under $d$. The joint natural indirect effect is
\[
Y(1,1)-Y(0,0)=3.
\]
If $M_1$ is changed first and $M_2$ second, the two increments are
\[
Y(1,0)-Y(0,0)=1, \qquad Y(1,1)-Y(1,0)=2,
\]
so the decomposition assigns $1$ to $M_1$ and $2$ to $M_2$. If $M_2$ is changed first, the increments are
\[
Y(0,1)-Y(0,0)=1, \qquad Y(1,1)-Y(0,1)=2,
\]
and the assignment is reversed: $M_2$ receives $1$ and $M_1$ receives $2$. Thus, the joint level effect remains $3$, but its mediator-specific allocation depends on the chosen replacement order. Without additional structure fixing such an allocation, the joint natural indirect effect is the primary level estimand.

The derivative decomposition in \eqref{eq:vector-component-nid} is different. All coordinate contributions are evaluated at the same mediator history, and the chain rule adds their infinitesimal contributions without changing the mediator coordinates sequentially. Consequently, the coordinatewise derivative decomposition is invariant to the order in which the mediator coordinates are listed.
\end{remark}

\paragraph{Vector partially linear mediation model.}
A convenient stacked representation is
\begin{equation*}
\begin{gathered}
Y_{it}(\overline{d}_t,\overline{\boldsymbol{m}}_t)=\tau_Y^{(v)}(\overline{d}_{t-L:t},\overline{\boldsymbol{m}}_{t-L:t},\overline{\bX}_{i,t-L:t})'\theta_0+f_0(\overline{\bX}_{i,t-L:t})+\alpha_i^Y+\lambda_t^Y+\varepsilon_{it}^Y,\\
\boldsymbol{M}_{it}(\overline{d}_t)=\mathcal{T}_M(\overline{d}_{t-L:t},\overline{\bX}_{i,t-L:t})\gamma_0+\boldsymbol{g}_0(\overline{\bX}_{i,t-L:t})+\boldsymbol{\alpha}_i^M+\boldsymbol{\lambda}_t^M+\boldsymbol{\varepsilon}_{it}^M.
\end{gathered}
\end{equation*}
Here $\tau_Y^{(v)}\in\mathbb{R}^{s_Y}$, $\mathcal{T}_M\in\mathbb{R}^{q_M\times s_M}$, $\theta_0\in\mathbb{R}^{s_Y}$, and $\gamma_0\in\mathbb{R}^{s_M}$, with $s_Y$ and $s_M$ fixed, and $\beta_0\coloneq(\theta_0',\gamma_0')'$.
The quantities $\boldsymbol{g}_0$, $\boldsymbol{\alpha}_i^M$, $\boldsymbol{\lambda}_t^M$, and $\boldsymbol{\varepsilon}_{it}^M$ are $q_M$-vectors. Separate dictionaries and coefficient blocks for the mediator equations can be represented by assigning disjoint subvectors of $\gamma_0$ to the rows of $\mathcal{T}_M$. The outcome dictionary may contain coordinate-specific mediator terms, mediator--mediator interactions, treatment--mediator interactions, and lags, provided its dimension remains fixed. The mediator innovations may be arbitrarily correlated across coordinates.

The potential-outcome disturbances satisfy
\[
\E(\varepsilon_{it}^Y\mid\overline{\bX}_{it},\alpha_i^Y,\overline{\lambda}_t^Y)=0, \qquad \E(\boldsymbol{\varepsilon}_{it}^M\mid\overline{\bX}_{it},\boldsymbol{\alpha}_i^M,\overline{\boldsymbol{\lambda}}_t^M)=\boldsymbol{0},
\]
where $\overline{\boldsymbol{\lambda}}_t^M\coloneq(\boldsymbol{\lambda}_1^{M\prime},\ldots,\boldsymbol{\lambda}_t^{M\prime})'$.
The vector analogue of Assumption \ref{assscia} replaces every scalar mediator by the complete vector and interprets the conditional independence statements jointly. Coordinatewise independence conditions do not generally imply the required joint restriction, and the support condition in Theorem \ref{th1} must hold for the joint mediator history. Under these conditions and consistency, the observed system is
\begin{equation}
\label{eq:vector-observed-mediation-model}
\begin{gathered}
Y_{it}=\tau_Y^{(v)}(\overline{D}_{i,t-L:t},\overline{\boldsymbol{M}}_{i,t-L:t},\overline{\bX}_{i,t-L:t})'\theta_0+f_0(\overline{\bX}_{i,t-L:t})+\alpha_i^Y+\lambda_t^Y+\varepsilon_{it}^Y,\\
\E(\varepsilon_{it}^Y\mid\overline{\bX}_{it},\overline{D}_{it},\overline{\boldsymbol{M}}_{it},\alpha_i^Y,\overline{\lambda}_t^Y)=0,\\
\boldsymbol{M}_{it}=\mathcal{T}_M(\overline{D}_{i,t-L:t},\overline{\bX}_{i,t-L:t})\gamma_0+\boldsymbol{g}_0(\overline{\bX}_{i,t-L:t})+\boldsymbol{\alpha}_i^M+\boldsymbol{\lambda}_t^M+\boldsymbol{\varepsilon}_{it}^M,\\
\E(\boldsymbol{\varepsilon}_{it}^M\mid\overline{\bX}_{it},\overline{D}_{it},\boldsymbol{\alpha}_i^M,\overline{\boldsymbol{\lambda}}_t^M)=\boldsymbol{0}.
\end{gathered}
\end{equation}
For DML--GMM estimation, \eqref{eq:mediator-residualization-exogeneity} is replaced by
\begin{equation}
\E(\boldsymbol{\varepsilon}_{it}^M\mid\overline{\bX}_{iT},\overline{D}_{it},\boldsymbol{\alpha}_i^M,\overline{\boldsymbol{\lambda}}_t^M)=\boldsymbol{0}.
\label{eq:vector-mediator-residualization}
\end{equation}
The treatment model \eqref{eq:assignment-models} is unchanged. The cross-sectional sampling condition is imposed on the complete vector trajectory, with arbitrary temporal and cross-coordinate dependence within an individual. If one mediator coordinate causally precedes another within period $t$, a recursive mediator system should instead be specified, with sequential restrictions and nested counterfactuals that follow that ordering.

\paragraph{DML--GMM moments.}
First differencing and auxiliary-fold demeaning are applied componentwise. With $\Delta\tau_{Y,it}^{(v)}$, $\Delta\mathcal{T}_{M,it}$, and $\Delta\boldsymbol{g}_{it}$ denoting the corresponding time differences and the superscript $*$ denoting the componentwise transformation in \eqref{eq:auxiliary-fold-demeaning}, define
\begin{align*}
U_{it}^{Y,v}(f,\theta)&\coloneq \Delta Y_{it}^*-\Delta\tau_{Y,it}^{(v)*\prime}\theta-\Delta f_{it}^*,\\
\boldsymbol{U}_{it}^{M,v}(\boldsymbol{g},\gamma)&\coloneq \Delta\boldsymbol{M}_{it}^*-\Delta\mathcal{T}_{M,it}^*\gamma-\Delta\boldsymbol{g}_{it}^*,\\
U_{it}^{D}(h)&\coloneq \Delta D_{it}^*-\Delta h_{it}^*.
\end{align*}
For $t=3,\ldots,T$, let
\begin{align*}
A_{it}^{M,v}(h)&\coloneq \big(U_{i2}^D(h),\ldots,U_{i,t-1}^D(h)\big)',\\
A_{it}^{Y,v}(h,\boldsymbol{g},\overline{\gamma})&\coloneq \big(U_{i2}^D(h),\ldots,U_{i,t-1}^D(h),\boldsymbol{U}_{i2}^{M,v}(\boldsymbol{g},\overline{\gamma})',\ldots,\boldsymbol{U}_{i,t-1}^{M,v}(\boldsymbol{g},\overline{\gamma})'\big)',
\end{align*}
where $\overline{\gamma}$ is the nuisance copy used only to construct the outcome instrument. Let $W_i$ denote the complete observed trajectory containing the vector mediator. For $\eta^{(v)}\coloneq(f,\boldsymbol{g},h,\overline{\gamma})$ and $\eta_0^{(v)}\coloneq(f_0,\boldsymbol{g}_0,h_0,\gamma_0)$, define
\begin{align}
\psi_{it}^{Y,v}(W_i;\eta^{(v)},\theta)&\coloneq A_{it}^{Y,v}(h,\boldsymbol{g},\overline{\gamma})U_{it}^{Y,v}(f,\theta), \label{eq:vector-outcome-score}\\
\psi_{it}^{M,v}(W_i;\eta^{(v)},\gamma)&\coloneq \operatorname{vec}\{A_{it}^{M,v}(h)\boldsymbol{U}_{it}^{M,v}(\boldsymbol{g},\gamma)'\}=\boldsymbol{U}_{it}^{M,v}(\boldsymbol{g},\gamma)\otimes A_{it}^{M,v}(h). \label{eq:vector-mediator-score}
\end{align}
For $\beta=(\theta',\gamma')'$, let $\psi_i^{(v)}(W_i;\eta^{(v)},\beta)$ denote the vector obtained by stacking \eqref{eq:vector-outcome-score} and \eqref{eq:vector-mediator-score} over $t=3,\ldots,T$. Under \eqref{eq:vector-observed-mediation-model}, \eqref{eq:assignment-models}, and \eqref{eq:vector-mediator-residualization},
\[
\E\{\psi_i^{(v)}(W_i;\eta_0^{(v)},\beta_0)\}=\boldsymbol{0},
\]
and the Gateaux derivative of this expectation with respect to every component of $f$, $\boldsymbol{g}$, $h$, and $\overline{\gamma}$ is zero at the truth.

Let $\widehat{\psi}_i^{k,(v)}(\beta)$ be the cross-fitted version of this individual score, constructed as in Algorithm \ref{ag:dml-mediation} with a vector-valued estimate of $\boldsymbol{g}_0$ and a preliminary estimate $\widetilde{\gamma}_k$. Define
\begin{equation*}
\widehat{\beta}_0^{D,v}\coloneq\arg\min_{\beta\in\mathbb{R}^{s_Y+s_M}}\widehat{m}_N^{(v)}(\beta)'\widehat{\boldsymbol{\Omega}}^{(v)}\widehat{m}_N^{(v)}(\beta), \qquad \widehat{m}_N^{(v)}(\beta)\coloneq\frac{1}{N}\sum_{k=1}^K\sum_{i\in\mathcal{F}_k}\widehat{\psi}_i^{k,(v)}(\beta),
\end{equation*}
where $\widehat{\boldsymbol{\Omega}}^{(v)}$ is positive definite and fixed during minimization. The nuisance copy keeps the score affine in the final candidate $\beta$.

For $t=3,\ldots,T$, define the uncentered innovation instruments
\begin{align*}
A_{it,0}^{M,v}&\coloneq (\Delta\varepsilon_{i2}^D,\ldots,\Delta\varepsilon_{i,t-1}^D)',\\
A_{it,0}^{Y,v}&\coloneq (\Delta\varepsilon_{i2}^D,\ldots,\Delta\varepsilon_{i,t-1}^D,\Delta\boldsymbol{\varepsilon}_{i2}^{M\prime},\ldots,\Delta\boldsymbol{\varepsilon}_{i,t-1}^{M\prime})'.
\end{align*}
The oracle individual score that enters the limiting representation is
\begin{equation*}
\widetilde{\psi}_i^{(v)}\coloneq\begin{pmatrix}
\{A_{it,0}^{Y,v}\Delta\varepsilon_{it}^Y\}_{t=3}^T\\[2mm]
\{\operatorname{vec}(A_{it,0}^{M,v}\Delta\boldsymbol{\varepsilon}_{it}^{M\prime})\}_{t=3}^T
\end{pmatrix},
\end{equation*}
where braces indexed by $t$ mean vertical stacking. This is the vector counterpart of $\widetilde{\psi}_i$ in \eqref{eq:oracle-score-mediation}; it is written in terms of uncentered innovations because the auxiliary-fold mean-subtraction terms are asymptotically negligible. In the display below, $I_{q_M}$ denotes the $q_M\times q_M$ identity matrix. Let
\begin{equation*}
\boldsymbol{G}^{(v)}\coloneq-\E\begin{pmatrix}
\{A_{it,0}^{Y,v}\Delta\tau_{Y,it}^{(v)\prime}\}_{t=3}^T&\boldsymbol{0}\\
\boldsymbol{0}&\{(I_{q_M}\otimes A_{it,0}^{M,v})\Delta\mathcal{T}_{M,it}\}_{t=3}^T
\end{pmatrix},
\boldsymbol{\Psi}^{(v)}\coloneq\E(\widetilde{\psi}_i^{(v)}\widetilde{\psi}_i^{(v)\prime}),
\end{equation*}
For a positive-definite probability limit $\boldsymbol{\Omega}^{(v)}$ of $\widehat{\boldsymbol{\Omega}}^{(v)}$, define
\begin{equation*}
\mathcal{A}_0^{(v)}\coloneq\{\boldsymbol{G}^{(v)\prime}\boldsymbol{\Omega}^{(v)}\boldsymbol{G}^{(v)}\}^{-1}\boldsymbol{G}^{(v)\prime}\boldsymbol{\Omega}^{(v)}, \boldsymbol{V}_\beta^{(v)}\coloneq\mathcal{A}_0^{(v)}\boldsymbol{\Psi}^{(v)}\mathcal{A}_0^{(v)\prime}.
\end{equation*}

The counterfactual maps are
\begin{align}
\boldsymbol{M}_{it}^{\beta,v}(\overline{d}_t)&\coloneq \boldsymbol{M}_{it}+\{\mathcal{T}_M(\overline{d}_{t-L:t},\overline{\bX}_{i,t-L:t})-\mathcal{T}_M(\overline{D}_{i,t-L:t},\overline{\bX}_{i,t-L:t})\}\gamma, \label{eq:vector-counterfactual-mediator}\\
Y_{it}^{\beta,v}(\overline{d}_t,\overline{\boldsymbol{m}}_t)&\coloneq Y_{it}+\{\tau_Y^{(v)}(\overline{d}_{t-L:t},\overline{\boldsymbol{m}}_{t-L:t},\overline{\bX}_{i,t-L:t})-\tau_Y^{(v)}(\overline{D}_{i,t-L:t},\overline{\boldsymbol{M}}_{i,t-L:t},\overline{\bX}_{i,t-L:t})\}'\theta. \label{eq:vector-counterfactual-outcome}
\end{align}
For any fixed finite collection of the joint level or derivative effects, let $\ell^{(v)}(W_i;\beta)$ denote the corresponding individual plug-in contribution constructed from these maps, with natural mediator histories formed by stacking $\boldsymbol{M}_{is}^{\beta,v}(\overline{d}_s)$ over $s\leqslant t$. Define
\begin{align*}
\vartheta_0^{(v)}&\coloneq \E\{\ell^{(v)}(W_i;\beta_0)\}, &\boldsymbol{J}_\vartheta^{(v)}&\coloneq \E\left\{\left.\frac{\partial\ell^{(v)}(W_i;\beta)}{\partial\beta'}\right|_{\beta=\beta_0}\right\}, \\
\varphi_{\vartheta,i}^{(v)}&\coloneq \ell^{(v)}(W_i;\beta_0)-\vartheta_0^{(v)}-\boldsymbol{J}_\vartheta^{(v)}\mathcal{A}_0^{(v)}\widetilde{\psi}_i^{(v)}, &\boldsymbol{\Sigma}_\vartheta^{(v)}&\coloneq \E\{\varphi_{\vartheta,i}^{(v)}\varphi_{\vartheta,i}^{(v)\prime}\}.
\end{align*}

\begin{proposition}[Fixed-dimensional vector extension]
\label{prop:fixed-vector-extension}
Suppose $q_M$, $s_Y$, $s_M$, $T$, $K$, $L$, and $r$ are fixed. Assume the vector models, joint identification and support conditions, cross-sectional sampling condition, and conditional moment restrictions above, together with the treatment model \eqref{eq:assignment-models} and the vector analogues of Assumptions \ref{ass:mediation-l2}--\ref{ass:mediation-jacobian}. If
$\widehat{\boldsymbol{\Omega}}^{(v)}\xrightarrow{p}\boldsymbol{\Omega}^{(v)}$ for a positive-definite $\boldsymbol{\Omega}^{(v)}$ and $\boldsymbol{G}^{(v)}$ has full column rank, then
\begin{equation}
\sqrt{N}(\widehat{\beta}_0^{D,v}-\beta_0)=-\mathcal{A}_0^{(v)}\frac{1}{\sqrt{N}}\sum_{i=1}^N\widetilde{\psi}_i^{(v)}+o_p(1)\xrightarrow{d}N(\boldsymbol{0},\boldsymbol{V}_\beta^{(v)}).
\label{eq:vector-beta-clt}
\end{equation}
If the vector analogue of Assumption \ref{ass:causal-effect-regularity} also holds, then
$\widehat{\vartheta}^{(v)}\coloneq\mathbb{P}_N\ell^{(v)}(W_i;\widehat{\beta}_0^{D,v})$ satisfies
\begin{equation}
\sqrt{N}(\widehat{\vartheta}^{(v)}-\vartheta_0^{(v)})=\frac{1}{\sqrt{N}}\sum_{i=1}^N\varphi_{\vartheta,i}^{(v)}+o_p(1)\xrightarrow{d}N(\boldsymbol{0},\boldsymbol{\Sigma}_\vartheta^{(v)}).
\label{eq:vector-causal-clt}
\end{equation}
The covariance matrix $\boldsymbol{\Sigma}_\vartheta^{(v)}$ may be singular when effects linked by exact decompositions are stacked together.
\end{proposition}

Under the vector analogues of the additional conditions in Theorems \ref{prop:mediation-sandwich} and \ref{prop:causal-effect-variance}, the corresponding analytical covariance estimators are consistent. A cross-sectional bootstrap can be implemented by resampling complete vector trajectories and repeating the full procedure, but no separate bootstrap-consistency result is asserted here. In conclusion, for fixed $q_M$, the extension requires only fixed-dimensional stacking and the associated rank condition.

\subsection{Discussion of a growing mediator dimension}
\label{sec:high-dimensional-mediator}

The preceding subsection keeps the mediator dimension $q_M$ fixed. To formulate a high-dimensional extension precisely, consider a sequence of panel models indexed by the cross-sectional sample size $N$. In the $N$th model, let
\[
\boldsymbol{M}_{it,N}=\big(M_{it,1,N},\ldots,M_{it,q_{M,N},N}\big)'\in\mathbb{R}^{q_{M,N}}, \qquad q_{M,N}\longrightarrow\infty,
\]
where $q_{M,N}$ is the number of mediator coordinates observed at sample size $N$. The mediator histories and their counterfactual versions are defined by stacking $\boldsymbol{M}_{it,N}$ over time as in the fixed-dimensional case. For each $N$, the joint effects in \eqref{eq:vector-cde}--\eqref{eq:vector-te} remain well defined, and the decomposition of the total effect into natural direct and indirect effects remains exact. These effects nevertheless form a sequence, such as $\operatorname{NIE}^{(v)}_{t,N}$, because the mediator intervention changes with $N$. If the growing vector is regarded as an approximation to an underlying infinite-dimensional mediator, a fixed limiting effect must be defined separately, and the approximation error must be negligible at the relevant inferential scale.

Identification becomes more demanding as the mediator dimension grows. The vector version of SCIA must concern the current mediator vector jointly, conditional on the relevant treatment and mediator histories; coordinatewise conditional-independence statements are not sufficient in general. The relevant-support condition must likewise cover the joint counterfactual mediator histories entering the natural effects. For continuous mediators, such joint overlap can deteriorate rapidly with $q_{M,N}$, while strong dependence or near-collinearity among mediator coordinates may make some mediated directions weakly identified. A high-dimensional extension would therefore require joint identification and support conditions that remain sufficiently informative along the sequence of models.

The growth of $q_{M,N}$ does not by itself imply a high-dimensional structural problem. If the mediator vector enters the model only through a fixed number of known summaries or a common fixed-dimensional parameterization, the preceding analysis can essentially be applied after reparameterization. The difficult case is the one in which the mediator-related dictionaries and coefficient vectors also grow. A schematic extension of the partially linear model is
\begin{align*}
Y_{it}(\overline{d}_t,\overline{\boldsymbol{m}}_{t,N})=&\tau_{Y,N}(\overline{d}_{t-L:t},\overline{\boldsymbol{m}}_{t-L:t,N},\overline{\bX}_{i,t-L:t})'\theta_{0,N}+f_{0,N}(\overline{\bX}_{i,t-L:t})+\alpha_i^Y+\lambda_t^Y+\varepsilon_{it}^Y,\\
\boldsymbol{M}_{it,N}(\overline{d}_t)=&\mathcal{T}_{M,N}(\overline{d}_{t-L:t},\overline{\bX}_{i,t-L:t})\gamma_{0,N}+\boldsymbol{g}_{0,N}(\overline{\bX}_{i,t-L:t})+\boldsymbol{\alpha}_{i,N}^M+\boldsymbol{\lambda}_{t,N}^M+\boldsymbol{\varepsilon}_{it,N}^M,
\end{align*}
where $\tau_{Y,N}\in\mathbb{R}^{s_{Y,N}}$, $\theta_{0,N}\in\mathbb{R}^{s_{Y,N}}$, $\mathcal{T}_{M,N}\in\mathbb{R}^{q_{M,N}\times s_{M,N}}$, and $\gamma_{0,N}\in\mathbb{R}^{s_{M,N}}$. A genuinely high-dimensional specification allows $s_{Y,N}$, $s_{M,N}$, or both to increase with $N$. Since unrestricted inclusion of all mediator main effects, lags, and interactions would lead to explosive dictionary growth, some substantively defensible low-complexity structure is essential. Possible examples include approximate sparsity or group sparsity, shared structure across mediator equations, a low-rank representation, or dependence on a small number of mediator summaries.

Even a simple linear model illustrates the inferential difficulty. Suppose, only for illustration, that the outcome depends on the mediator only through the contemporaneous linear term $b_{0,N}'\boldsymbol{m}_{t,N}$, with no lagged mediator terms or treatment--mediator interactions, where $b_{0,N}\in\mathbb{R}^{q_{M,N}}$. Define
\[
\boldsymbol{\Delta}_{M,t,N}(\overline{d}_t,\overline{d}_t^*)\coloneq\E\big\{\boldsymbol{M}_{it,N}(\overline{d}_t)-\boldsymbol{M}_{it,N}(\overline{d}_t^*)\big\}.
\]
The joint natural indirect effect is then
\[
b_{0,N}'\boldsymbol{\Delta}_{M,t,N}(\overline{d}_t,\overline{d}_t^*).
\]
Thus, the target is a bilinear functional of two high-dimensional objects. Screening only by outcome prediction or only by treatment--mediator association is not aligned with this target, and shrinkage bias in either component can produce nonnegligible bias in their plug-in inner product. Nonlinear mediator terms, lags, and treatment--mediator interactions make the relevant low-complexity structure still more target dependent.

The Neyman orthogonality used in the main text controls estimation errors in the nuisance functions when the structural parameter is fixed dimensional; it does not automatically remove regularization bias from the growing vectors $\theta_{0,N}$ and $\gamma_{0,N}$. A plausible route would combine regularized moment estimation with a second, target-specific debiasing step. Writing $\beta_{0,N}=(\theta_{0,N}',\gamma_{0,N}')'$ and $\vartheta_{0,N}=\E\{\ell_N(W_i;\beta_{0,N})\}$ for a scalar causal effect, one would seek a loading vector $a_{0,N}$ satisfying the schematic orthogonality condition
\[
\frac{\partial}{\partial\beta'}\E\big\{\ell_N(W_i;\beta)-a_{0,N}'\psi_N(W_i;\eta_{0,N},\beta)\big\}\bigg|_{\beta=\beta_{0,N}}=0,
\]
where $\psi_N$ is the moment vector and $\eta_{0,N}$ collects the nuisance functions. Such a construction would require the target derivative to be represented by the moment Jacobian and the loading $a_{0,N}$ to admit an estimable low-complexity representation. Cross-fitting alone does not provide this additional orthogonality.

The panel structure adds further complications. The mediator equations, nuisance functions, and auxiliary-fold averages used to remove the vector of time effects must be controlled uniformly over $q_{M,N}$ coordinates. Coordinatewise consistency is insufficient when the final score aggregates many mediator components, particularly under strong cross-coordinate dependence. The instrument and moment dimensions may also increase, making an unrestricted inverse of the sample moment covariance matrix unstable or undefined. A feasible procedure would therefore require a restricted moment set or a structured and regularized weighting scheme, together with appropriate identification conditions for the penalized moment problem. The structural complexity, nuisance-estimation errors, and approximation errors would all have to grow slowly enough for their combined contribution to be negligible relative to the sampling fluctuation of the target.

Finally, the causal target must remain well scaled. An aggregate indirect effect that contains nonnegligible contributions from an increasing number of mediators may diverge, vanish, or require a normalization different from the usual $\sqrt{N}$ scale. If a growing collection of mediator-specific effects is reported, simultaneous inference and multiplicity control are also necessary. Selecting a small subset of mediators and then applying the fixed-dimensional procedure is not sufficient in general: sample splitting removes dependence created by selection, but it does not eliminate bias from omitted mediated pathways. Cross-sectional high-dimensional mediation methods, such as \citet{wei2025debiased}, illustrate the need for structured regularization and debiasing, while the dynamic histories, two-way fixed effects, and auxiliary-fold construction considered here create additional panel-specific issues.

These considerations show that an unrestricted growing mediator vector is not a mechanical extension of the present framework. A complete treatment would need to combine uniform identification of the joint mediator intervention, substantively justified low-complexity structure, regularized estimation of the growing panel system, target-specific debiasing, and appropriately scaled inference. We leave this development to future research.

\section{Calculation of Causal Effects}
Suppose
\[
\tau_Y(\overline{d}_{t-L:t},\overline{m}_{t-L:t},\overline{\bX}_{i,t-L:t})=(d_t,d_{t-1},d_t^2,d_tX_{it1},m_t,d_tm_t,m_tX_{it2})'
\]
and
\[
\tau_M(\overline{d}_{t-L:t},\overline{\bX}_{t-L:t})=(d_t,d_{t-1},d_t^2,d_tX_{it3})'.
\]
Write $\theta_0=(\theta_{01},\ldots,\theta_{07})'$ and $\gamma_0=(\gamma_{01},\ldots,\gamma_{04})'$, and consider two treatment histories that differ only in period $t$: $\overline{d}_t=(\overline{d}_{t-1}',d_t)'$ and $\overline{d}_t^*=(\overline{d}_{t-1}',d_t^*)'$. Since
\[
\E\{M_{it}(\overline{d}_t^*)\}-\E(M_{it})=\E\big[\tau_M(\overline{d}^*_{t-L:t},\overline{\bX}_{i,t-L:t})'\gamma_0-\tau_M(\overline{D}_{i,t-L:t},\overline{\bX}_{i,t-L:t})'\gamma_0\big],
\]
define
\[
\begin{aligned}
\mu_{M,t}(\overline{d}_t^*)\coloneq{}&\E\{M_{it}(\overline{d}_t^*)\}=\E(M_{it})+\gamma_{01}\{d_t^*-\E(D_{it})\}+\gamma_{02}\{d_{t-1}-\E(D_{i,t-1})\}\\
&+\gamma_{03}\{(d_t^*)^2-\E(D_{it}^2)\}+\gamma_{04}\{d_t^*\E(X_{it3})-\E(D_{it}X_{it3})\}.
\end{aligned}
\]
Then, for any prespecified mediator history $\overline{m}_t$,
\begin{align*}
\operatorname{CDE}_t(\overline{d}_t,\overline{d}_t^*;\overline{m}_t)&=(d_t-d_t^*)\{\theta_{01}+(d_t+d_t^*)\theta_{03}+\E(X_{it1})\theta_{04}+m_t\theta_{06}\},\\
\operatorname{NDE}_t(\overline{d}_t,\overline{d}_t^*)&=(d_t-d_t^*)\{\theta_{01}+(d_t+d_t^*)\theta_{03}+\E(X_{it1})\theta_{04}+\mu_{M,t}(\overline{d}_t^*)\theta_{06}\}.
\end{align*}
For derivatives with respect to the current treatment,
\begin{align*}
\operatorname{CDD}_{t,t}(\overline{d}_t,\overline{m}_t)&=\theta_{01}+2d_t\theta_{03}+\E(X_{it1})\theta_{04}+m_t\theta_{06},\\
\operatorname{NDD}_{t,t}(\overline{d}_t)&=\theta_{01}+2d_t\theta_{03}+\E(X_{it1})\theta_{04}+\mu_{M,t}(\overline{d}_t^*)\theta_{06}.
\end{align*}

\section{Asymptotic Theory for the DML--GMM Estimator}
\label{sec:structural-asymptotic-supp}
This appendix gives the complete regularity conditions and limit theory for the structural estimator $\widehat{\beta}_0^D$ used in Section \ref{secas}. We maintain the cross-sectional sampling condition \eqref{iid}, the fixed-$T$ asymptotic framework described there, and the observed-data moment restrictions in \eqref{eq:observed-mediation-model}--\eqref{eq:mediator-residualization-exogeneity}. The analysis below also makes explicit why auxiliary-fold demeaning does not affect the first-order distribution.

For notational convenience, define $\Delta\rho_{0,it}=\Delta\tau_{M,it}'\gamma_0+\Delta g_{0,it}$, $\Delta\widehat{\rho}_{0k,it}=\Delta\tau_{M,it}'\widetilde{\gamma}_k+\Delta\widehat{g}_{0k,it}$. Define their starred versions by \eqref{eq:auxiliary-fold-demeaning}; hence, $\displaystyle \Delta\widehat{\rho}_{0k,it}^*
\coloneq\Delta\widehat{\rho}_{0k,it}-\frac{1}{|\mathcal{F}_{k'}|}\sum_{j\in\mathcal{F}_{k'}}\Delta\widehat{\rho}_{0k,jt}$, with $\Delta\rho_{0,it}^*$ defined analogously. Thus, $\Delta M_{ip}^*-\Delta\widehat{\rho}_{0k,ip}^*=\widehat{U}_{ip,k}^M(\widetilde{\gamma}_k)$. 

For $t=3,\ldots,T$, write
\begin{align*}
B_{it}^Y&\coloneq(\Delta D_{i2},\ldots,\Delta D_{i,t-1},\Delta M_{i2},\ldots,\Delta M_{i,t-1})',\\
B_{it}^M&\coloneq(\Delta D_{i2},\ldots,\Delta D_{i,t-1})',\\
C_{it}^Y(h,\rho)&\coloneq(-\Delta h_{i2},\ldots,-\Delta h_{i,t-1},-\Delta\rho_{i2},\ldots,-\Delta\rho_{i,t-1})',\\
C_{it}^M(h)&\coloneq(-\Delta h_{i2},\ldots,-\Delta h_{i,t-1})'.
\end{align*}
A superscript $*$ on these vectors means that every component is transformed at its own period according to \eqref{eq:auxiliary-fold-demeaning}; for example,
\begin{align*}
B_{it}^{Y*}&\coloneq(\Delta D_{i2}^*,\ldots,\Delta D_{i,t-1}^*,\Delta M_{i2}^*,\ldots,\Delta M_{i,t-1}^*)',\\
C_{it}^{Y*}(h,\rho)&\coloneq(-\Delta h_{i2}^*,\ldots,-\Delta h_{i,t-1}^*,-\Delta\rho_{i2}^*,\ldots,-\Delta\rho_{i,t-1}^*)',
\end{align*}
with $B_{it}^{M*}$ and $C_{it}^{M*}(h)$ defined analogously. Let $\boldsymbol{B}_i^q\coloneq\mathrm{Diag}(B_{i3}^q,\ldots,B_{iT}^q)$ for $q\in\{Y,M\}$, and define $\boldsymbol{C}_i^Y(h,\rho)$ and $\boldsymbol{C}_i^M(h)$ analogously; their starred versions are formed from the corresponding starred blocks. The residual vectors from periods $3,\ldots,T$ are
\[
\varepsilon^Y(W_i,f,\theta)\coloneq\big(U_{i3}^Y(f,\theta),\ldots,U_{iT}^Y(f,\theta)\big)',
\qquad
\varepsilon^M(W_i,g,\gamma)\coloneq\big(U_{i3}^M(g,\gamma),\ldots,U_{iT}^M(g,\gamma)\big)'.
\]
For $\eta=(f,g,h,\rho)$ and $\beta=(\theta',\gamma')'$, define
\begin{gather*}
m(W_i;f,g,\beta) \coloneq \begin{pmatrix}\boldsymbol{B}_i^{Y*}\varepsilon^Y(W_i,f,\theta)\\ \boldsymbol{B}_i^{M*}\varepsilon^M(W_i,g,\gamma)\end{pmatrix}, \phi(W_i;f,g,h,\rho,\beta) \coloneq \begin{pmatrix}\boldsymbol{C}_i^{Y*}(h,\rho)\varepsilon^Y(W_i,f,\theta)\\ \boldsymbol{C}_i^{M*}(h)\varepsilon^M(W_i,g,\gamma)\end{pmatrix},\\
\psi(W_i;\eta,\beta) \coloneq m(W_i;f,g,\beta)+\phi(W_i;f,g,h,\rho,\beta).
\end{gather*}
Then $\widehat{\psi}_i^k(\beta)$ in \eqref{eq:dml-gmm-mediation-estimator} satisfies $\widehat{\psi}_i^k(\beta)=\psi(W_i;\widehat{\eta}_k,\beta)$ for $i\in\mathcal{F}_k$, where $\eta_0\coloneq(f_0,g_0,h_0,\rho_0)$, $\widehat{\eta}_k\coloneq(\widehat{f}_{0k},\widehat{g}_{0k},\widehat{h}_{0k},\widehat{\rho}_{0k})$.

For $t=3,\ldots,T$, define the uncentered innovation instruments
\[
A_{it}^Y \coloneq(\Delta\varepsilon_{i2}^D,\ldots,\Delta\varepsilon_{i,t-1}^D,\Delta\varepsilon_{i2}^M,\ldots,\Delta\varepsilon_{i,t-1}^M)', \quad A_{it}^M \coloneq(\Delta\varepsilon_{i2}^D,\ldots,\Delta\varepsilon_{i,t-1}^D)',
\]
and let $\boldsymbol{A}_i^q\coloneq\mathrm{Diag}(A_{i3}^q,\ldots,A_{iT}^q)$ for $q\in\{Y,M\}$. Also set $e_i^Y\coloneq(\Delta\varepsilon_{i3}^Y,\ldots,\Delta\varepsilon_{iT}^Y)'$, $e_i^M\coloneq(\Delta\varepsilon_{i3}^M,\ldots,\Delta\varepsilon_{iT}^M)'$, $\boldsymbol{\mathcal{T}}_i^Y\coloneq(\Delta\tau_{Y,i3},\ldots,\Delta\tau_{Y,iT})'$, $\boldsymbol{\mathcal{T}}_i^M\coloneq(\Delta\tau_{M,i3},\ldots,\Delta\tau_{M,iT})'$.
\begin{equation}
\widetilde{\psi}_i\coloneq\begin{pmatrix}\boldsymbol{A}_i^Ye_i^Y\\ \boldsymbol{A}_i^Me_i^M\end{pmatrix},\qquad \boldsymbol{\Psi}\coloneq\E(\widetilde{\psi}_i\widetilde{\psi}_i'),\qquad \boldsymbol{G}\coloneq-\E\begin{pmatrix}\boldsymbol{A}_i^Y\boldsymbol{\mathcal{T}}_i^Y&\boldsymbol{0}\\ \boldsymbol{0}&\boldsymbol{A}_i^M\boldsymbol{\mathcal{T}}_i^M\end{pmatrix}.
\label{eq:oracle-score-mediation}
\end{equation}

Let $F_0$ denote the distribution of $W_i$, and write $\|a\|_{F_0,2}\coloneq\{\int\|a(w)\|^2F_0(\mathrm{d}w)\}^{1/2}$ for a vector- or matrix-valued function $a$, where $\|\cdot\|$ is the Euclidean norm for vectors and $\|\cdot\|_F$ is the Frobenius norm for matrices. Norms involving estimated nuisance functions are understood conditionally on the nuisance-training observations and, for starred quantities, the auxiliary fold.
\begin{assumption}[Moment bounds and first-stage consistency]
\label{ass:mediation-l2}
For every fold $k$, the following conditions hold:\\
(i) $\displaystyle \E\bigg\{\|\widetilde{\psi}_i\|^2+\sum_{q\in\{Y,M\}}\big(\|\boldsymbol{A}_i^q\|_F^2+\|e_i^q\|^2\big)\bigg\}<\infty,\qquad \E\big(\|\boldsymbol{\mathcal{T}}_i^Y\|_F+\|\boldsymbol{\mathcal{T}}_i^M\|_F\big)<\infty$. \\
(ii) $\displaystyle \begin{aligned}[t]
&\|m(\cdot;\widehat{f}_{0k},\widehat{g}_{0k},\beta_0)-m(\cdot;f_0,g_0,\beta_0)\|_{F_0,2}\\
&\quad+\|\phi(\cdot;\widehat{f}_{0k},\widehat{g}_{0k},h_0,\rho_0,\beta_0)-\phi(\cdot;f_0,g_0,h_0,\rho_0,\beta_0)\|_{F_0,2}=o_p(1).
\end{aligned}$\\
(iii) $\|\phi(\cdot;f_0,g_0,\widehat{h}_{0k},\widehat{\rho}_{0k},\beta_0)-\phi(\cdot;f_0,g_0,h_0,\rho_0,\beta_0)\|_{F_0,2}=o_p(1)$. \\
(iv) $\displaystyle \begin{aligned}[t]
&\|\varepsilon^Y(\cdot,\widehat{f}_{0k},\theta_0)-\varepsilon^Y(\cdot,f_0,\theta_0)\|_{F_0,2}+\|\varepsilon^M(\cdot,\widehat{g}_{0k},\gamma_0)-\varepsilon^M(\cdot,g_0,\gamma_0)\|_{F_0,2}\\
&\quad+\|\boldsymbol{C}_i^Y(\widehat{h}_{0k},\widehat{\rho}_{0k})-\boldsymbol{C}_i^Y(h_0,\rho_0)\|_{F_0,2}+\|\boldsymbol{C}_i^M(\widehat{h}_{0k})-\boldsymbol{C}_i^M(h_0)\|_{F_0,2}=o_p(1).
\end{aligned}$
\end{assumption}
Assumption \ref{ass:mediation-l2} collects the finite-moment and mean-square consistency conditions needed for the score and its instrument and residual components. Since
\[
\Delta\widehat{\rho}_{0k,it}-\Delta\rho_{0,it}=\Delta\widehat{g}_{0k,it}-\Delta g_{0,it}+\Delta\tau_{M,it}'(\widetilde{\gamma}_k-\gamma_0),
\]
its conditions on $\widehat{\rho}_{0k}$ concern only the existing estimates $\widehat{g}_{0k}$ and $\widetilde{\gamma}_k$.

The remaining second-order term is the product of errors from the equation regressions and from instrument construction. For $i\in\mathcal{F}_k$, define
\begin{align*}
\widehat{\delta}_{ik}(W_i)\coloneq&\phi(W_i;\widehat{f}_{0k},\widehat{g}_{0k},\widehat{h}_{0k},\widehat{\rho}_{0k},\beta_0)-\phi(W_i;f_0,g_0,\widehat{h}_{0k},\widehat{\rho}_{0k},\beta_0)\\
&-\phi(W_i;\widehat{f}_{0k},\widehat{g}_{0k},h_0,\rho_0,\beta_0)+\phi(W_i;f_0,g_0,h_0,\rho_0,\beta_0).
\end{align*}
\begin{assumption}[Second-order nuisance remainder]
\label{ass:mediation-product}
For every $k$, at least one of the following conditions is satisfied: (i) $\displaystyle \sqrt{N}\int\widehat{\delta}_{ik}(w)F_0(\mathrm{d}w)\xrightarrow{p}0$ and $\displaystyle \|\widehat{\delta}_{ik}\|_{F_0,2}=o_p(1)$; (ii) $\displaystyle N^{-1/2}\sum_{i\in\mathcal{F}_k}\|\widehat{\delta}_{ik}(W_i)\|\xrightarrow{p}0$; (iii) $\displaystyle N^{-1/2}\sum_{i\in\mathcal{F}_k}\widehat{\delta}_{ik}(W_i)\xrightarrow{p}0$.
\end{assumption}
Define 
\begin{align*}
\|\widehat{f}_{0k}-f_0\|_{F_0,2}&\coloneq\left\|\big(\Delta\widehat{f}_{0k,i3}-\Delta f_{0,i3},\ldots,\Delta\widehat{f}_{0k,iT}-\Delta f_{0,iT}\big)'\right\|_{F_0,2},\\
\|\widehat{g}_{0k}-g_0\|_{F_0,2}&\coloneq\left\|\big(\Delta\widehat{g}_{0k,i2}-\Delta g_{0,i2},\ldots,\Delta\widehat{g}_{0k,iT}-\Delta g_{0,iT}\big)'\right\|_{F_0,2},\\
\|\widehat{h}_{0k}-h_0\|_{F_0,2}&\coloneq\left\|\big(\Delta\widehat{h}_{0k,i2}-\Delta h_{0,i2},\ldots,\Delta\widehat{h}_{0k,i,T-1}-\Delta h_{0,i,T-1}\big)'\right\|_{F_0,2},\\
\|\Delta\tau_M'(\widetilde{\gamma}_k-\gamma_0)\|_{F_0,2}&\coloneq\left\|\big(\Delta\tau_{M,i2}'(\widetilde{\gamma}_k-\gamma_0),\ldots,\Delta\tau_{M,i,T-1}'(\widetilde{\gamma}_k-\gamma_0)\big)'\right\|_{F_0,2}.
\end{align*}
Under standard moment bounds, a primitive sufficient condition for Assumption \ref{ass:mediation-product} is
\begin{align*}
&\|\widehat{f}_{0k}-f_0\|_{F_0,2}\Big\{\|\widehat{h}_{0k}-h_0\|_{F_0,2}+\|\widehat{g}_{0k}-g_0\|_{F_0,2}+\|\Delta\tau_M'(\widetilde{\gamma}_k-\gamma_0)\|_{F_0,2}\Big\}\\
&\qquad+\|\widehat{g}_{0k}-g_0\|_{F_0,2}\|\widehat{h}_{0k}-h_0\|_{F_0,2}=o_p(N^{-1/2}).
\end{align*}
Thus, $o_p(N^{-1/4})$ rates for all nuisance components are sufficient but not necessary. If $\widetilde{\gamma}_k$ is root-$N$ consistent, its contribution is negligible whenever $\widehat{f}_{0k}$ is consistent.

The preceding assumptions yield the oracle reduction of the estimated score.
\begin{lemma}
\label{lem:mediation-score-reduction}
Suppose \eqref{iid}, \eqref{eq:observed-mediation-model}, \eqref{eq:assignment-models}, and \eqref{eq:mediator-residualization-exogeneity} hold. Under Assumptions \ref{ass:mediation-l2} and \ref{ass:mediation-product},
\[
\frac{1}{\sqrt{N}}\sum_{k=1}^K\sum_{i\in\mathcal{F}_k}\widehat{\psi}_i^k(\beta_0)=\frac{1}{\sqrt{N}}\sum_{k=1}^K\sum_{i\in\mathcal{F}_k}\psi(W_i;\eta_0,\beta_0)+o_p(1).
\]
\end{lemma}

\begin{assumption}[Jacobian stability]
\label{ass:mediation-jacobian}
The following conditions hold:
\begin{enumerate}
\item[(i)] $\displaystyle \E\left\|\partial\psi(W_i;\eta_0,\beta_0)/\partial\beta'\right\|_F<\infty$.
\item[(ii)] For every $k$,
\[
\int\left\|\frac{\partial\psi(w;\widehat{\eta}_k,\beta_0)}{\partial\beta'}-\frac{\partial\psi(w;\eta_0,\beta_0)}{\partial\beta'}\right\|_F F_0(\mathrm{d}w)\xrightarrow{p}0.
\]
\end{enumerate}
\end{assumption}
\begin{lemma}
\label{lem:mediation-jacobian}
Under \eqref{iid}, \eqref{eq:observed-mediation-model}, \eqref{eq:assignment-models}, \eqref{eq:mediator-residualization-exogeneity}, Assumption \ref{ass:mediation-l2}(i), and Assumption \ref{ass:mediation-jacobian}, for any possibly data-dependent vector $\overline{\beta}\in\mathbb{R}^{s_Y+s_M}$,
\[
\frac{1}{N}\sum_{k=1}^K\sum_{i\in\mathcal{F}_k}\frac{\partial\widehat{\psi}_i^k(\overline{\beta})}{\partial\beta'}\xrightarrow{p}\boldsymbol{G}.
\]
\end{lemma}

\begin{theorem}
\label{th:mediation-beta-normality}
Suppose \eqref{iid}, \eqref{eq:observed-mediation-model}, \eqref{eq:assignment-models}, and \eqref{eq:mediator-residualization-exogeneity} hold, and Assumptions \ref{ass:mediation-l2}--\ref{ass:mediation-jacobian} are satisfied. If $\widehat{\boldsymbol{\Omega}}\xrightarrow{p}\boldsymbol{\Omega}$, where $\boldsymbol{\Omega}$ is positive definite, and $\boldsymbol{G}'\boldsymbol{\Omega}\boldsymbol{G}$ is nonsingular, then
\begin{equation}
\sqrt{N}(\widehat{\beta}_0^D-\beta_0)=-(\boldsymbol{G}'\boldsymbol{\Omega}\boldsymbol{G})^{-1}\boldsymbol{G}'\boldsymbol{\Omega}\frac{1}{\sqrt{N}}\sum_{i=1}^N\widetilde{\psi}_i+o_p(1)\xrightarrow{d}N(\boldsymbol{0},\boldsymbol{V}_\beta),
\label{eq:mediation-beta-clt}
\end{equation}
where $\boldsymbol{V}_\beta\coloneq(\boldsymbol{G}'\boldsymbol{\Omega}\boldsymbol{G})^{-1}\boldsymbol{G}'\boldsymbol{\Omega}\boldsymbol{\Psi}\boldsymbol{\Omega}\boldsymbol{G}(\boldsymbol{G}'\boldsymbol{\Omega}\boldsymbol{G})^{-1}$.
\end{theorem}
Note that the preliminary coefficients $\widetilde{\gamma}_k$ enter only through $\widehat{\rho}_{0k}$ and disappear from the first-order representation by Neyman orthogonality. In addition, joint estimation of $\theta_0$ and $\gamma_0$ is important because their limiting covariance enters inference for natural direct and indirect effects.

Let $\widehat{\psi}_i\coloneq\widehat{\psi}_i^k(\widehat{\beta}_0^D)$ for $i\in\mathcal{F}_k$ and define $\displaystyle \widehat{\boldsymbol{G}}\coloneq\frac{1}{N}\sum_{k=1}^K\sum_{i\in\mathcal{F}_k}\frac{\partial\widehat{\psi}_i^k(\widehat{\beta}_0^D)}{\partial\beta'},
\widehat{\boldsymbol{\Psi}}\coloneq\frac{1}{N}\sum_{k=1}^K\sum_{i\in\mathcal{F}_k}\widehat{\psi}_i\widehat{\psi}_i'$. A sandwich estimator of the asymptotic variance of $\sqrt{N}(\widehat{\beta}_0^D-\beta_0)$ is
\begin{equation}
\label{eq:sandwich-mediation}
\widehat{\boldsymbol{V}}_\beta\coloneq(\widehat{\boldsymbol{G}}'\widehat{\boldsymbol{\Omega}}\widehat{\boldsymbol{G}})^{-1}\widehat{\boldsymbol{G}}'\widehat{\boldsymbol{\Omega}}\widehat{\boldsymbol{\Psi}}\widehat{\boldsymbol{\Omega}}\widehat{\boldsymbol{G}}(\widehat{\boldsymbol{G}}'\widehat{\boldsymbol{\Omega}}\widehat{\boldsymbol{G}})^{-1},
\end{equation}
so the estimated covariance matrix of $\widehat{\beta}_0^D$ is $\widehat{\boldsymbol{V}}_\beta/N$.

The following proposition establishes the consistency of the sandwich variance estimator.
\begin{theorem}
\label{prop:mediation-sandwich}
Suppose the conditions of Theorem \ref{th:mediation-beta-normality} hold. In addition, assume that, for every $k$, $\|\psi(\cdot;\widehat{\eta}_k,\widehat{\beta}_0^D)-\psi(\cdot;\widehat{\eta}_k,\beta_0)\|_{F_0,2}=o_p(1), \|\widehat{\delta}_{ik}\|_{F_0,2}=o_p(1)$. Then $\widehat{\boldsymbol{\Psi}}\xrightarrow{p}\boldsymbol{\Psi}$, $\widehat{\boldsymbol{G}}\xrightarrow{p}\boldsymbol{G}$, and $\widehat{\boldsymbol{V}}_\beta\xrightarrow{p}\boldsymbol{V}_\beta$.
\end{theorem}

\section{Supplementary Simulation Results}
\label{sec:supplementary-simulation}
\subsection{Baseline data-generating process and implementation details}
\label{sec:baseline-simulation-details}
The baseline design uses $N=1600$ individuals, $T=10$ periods, five time-varying covariates, and one lag in each structural dictionary; variables dated before period 1 are set to zero. The outcome and mediator dictionaries are
\begin{align*}
\tau_Y(\overline d_{t-1:t},\overline m_{t-1:t},\overline\bX_{i,t-1:t})
&=(d_t,d_t^2,d_{t-1},d_tX_{it3},m_t,m_{t-1})',\\
\tau_M(\overline d_{t-1:t},\overline\bX_{i,t-1:t})
&=(d_t,d_t^2,d_{t-1})'.
\end{align*}
with $\theta_0=(2,0.5,2,2,1,0.5)'$ and $\gamma_0=(1,0.5,0.5)'$. Thus the stacked structural parameter has nine components.

For each individual, draw mutually independent standard normal variables $A_i$, $\zeta_i^D$, $\zeta_i^M$, $\zeta_i^Y$, $\xi_{it}^D$, $\varepsilon_{it}^M$, $\varepsilon_{it}^Y$, and $\nu_{itj}$, $j=1,\ldots,5$, independently across individuals and periods. Set $\alpha_i^q=A_i+\zeta_i^q$ for $q\in\{D,M,Y\}$, so the three equations share a time-invariant latent factor while retaining equation-specific heterogeneity. The covariates have stationary standard normal marginals and obey
\[
X_{i1j}=\nu_{i1j},\qquad X_{itj}=0.5X_{i,t-1,j}+\sqrt{1-0.5^2}\,\nu_{itj},\qquad t=2,\ldots,T.
\]
The deterministic period effects are $\lambda_t^Y=\sin(2\pi t/T)$ and $\lambda_t^M=\lambda_t^D=0.5\cos(2\pi t/T)$. The nuisance functions are
\begin{equation}
\label{eq:simulation-nuisances}
\begin{gathered}
f_0(x)=(x_1^2-1)+\{\cosh(x_1)-e^{1/2}\}+\sin(2x_2)+\tanh(2x_3)+\sin(x_1x_2),\\
g_0(x)=(x_1^2-1)+2\{\cos(2x_2)-e^{-2}\}+\sin(2x_4)+2\{e^{x_5}-e^{1/2}\},\\
h_0(x)=(x_1^2-1)-x_1+(x_2^2-1)+x_3.
\end{gathered}
\end{equation}
Every displayed component is centered under the standard normal marginal distribution. The functions $f_0$ and $g_0$ deliberately lie outside the finite cubic library used for estimation, whereas $h_0$ lies in its span.

The panel is generated recursively. At $t=1$, let $\varepsilon_{i1}^D=\xi_{i1}^D$; for $t\geqslant2$, let $\varepsilon_{it}^D=\xi_{it}^D+\varepsilon_{i,t-1}^Y+\varepsilon_{i,t-1}^M$. Then
\begin{align*}
D_{it}&=h_0(X_{it})+\alpha_i^D+\lambda_t^D+\varepsilon_{it}^D,\\
M_{it}&=D_{it}+0.5D_{it}^2+0.5D_{i,t-1}+g_0(X_{it})+\alpha_i^M+\lambda_t^M+\varepsilon_{it}^M,\\
Y_{it}&=2D_{it}+0.5D_{it}^2+2D_{i,t-1}+2D_{it}X_{it3}+M_{it}+0.5M_{i,t-1}
+f_0(X_{it})+\alpha_i^Y+\lambda_t^Y+\varepsilon_{it}^Y.
\end{align*}
The feedback in $\varepsilon_{it}^D$ makes treatment respond to the preceding outcome and mediator shocks. Current primitive innovations remain independent of the relevant histories, so the level equations satisfy the conditional moment restrictions used in the paper, but the differenced structural regressors are correlated with differenced innovations. This creates sequential rather than strict exogeneity.

The covariate library contains every monomial of total degree one, two, or three in the five current-period covariates, without an intercept. Its dimension is $\binom{5+3}{3}-1=55$. The code first constructs the structural dictionaries and this covariate basis in levels and then takes first differences. For a nuisance-training sample, every differenced response, structural regressor, basis term, and instrument is demeaned by its training-sample cross-sectional mean separately in each period. Because valid lagged instruments first become available in period 3, the penalized structural first stages use the equations for $t=3,\ldots,T$.

The outcome first stage regresses the period-demeaned $\Delta Y_{it}$ on the six differenced outcome-dictionary terms and the 55 differenced basis terms. The contemporaneous treatment, squared treatment, treatment--$X_3$ interaction, and mediator terms are treated as endogenous. The excluded instrument set contains the period-demeaned long differences $D_{i,t-1}-D_{i1}$, $D_{i,t-1}^2-D_{i1}^2$, $M_{i,t-1}-M_{i1}$, and $D_{i,t-1}X_{i,t-1,3}-D_{i1}X_{i1,3}$, together with the lagged first differences $\Delta D_{i,t-1}^2$ and $\Delta(D_{i,t-1}X_{i,t-1,3})$. The lagged treatment and mediator terms, which are not designated endogenous, and all 55 basis terms are also included in the instrument matrix. The mediator first stage is constructed analogously: it regresses $\Delta M_{it}$ on the three differenced mediator-dictionary terms and the basis, treats the contemporaneous treatment and squared treatment terms as endogenous, and uses $D_{i,t-1}-D_{i1}$, $D_{i,t-1}^2-D_{i1}^2$, and $\Delta D_{i,t-1}^2$ as excluded instruments, along with the lagged treatment term and the basis.

Computationally, the code obtains a QR basis for each instrument matrix, projects the response and designated endogenous columns onto its span, and then fits an $\ell_1$-penalized regression without an intercept. All coefficients, including the structural and basis coefficients, are penalized. For an equation with $p$ regressors and $n_{\mathrm{obs}}$ stacked observations, the penalty is $\sqrt{2\log(2p)/n_{\mathrm{obs}}}$; the implementation uses \texttt{glmnet} with $\alpha=1$, standardized regressors, and a maximum of $100{,}000$ iterations. The first six outcome coefficients and first three mediator coefficients are not used as final estimates. Instead, the mediator coefficients supply the fold-specific preliminary value $\widetilde\gamma_k$, while the estimated basis coefficients generate $\widehat f_{0k}$ and $\widehat g_{0k}$. The treatment nuisance $h_0$ is estimated separately by ordinary least squares of the training-sample period-demeaned $\Delta D_{it}$ on the 55 period-demeaned differenced basis terms over $t=2,\ldots,T$, again without an intercept.

For DML--GMM, the individuals are randomly partitioned into ten equal folds. For target fold $\mathcal F_k$, the cyclic successor $\mathcal F_{k'}$ is used only for period-specific auxiliary means, and the remaining eight folds train the nuisance functions. For every target-fold first difference and fitted nuisance value, the code subtracts the corresponding auxiliary-fold mean. This produces the starred outcomes, dictionaries, and fitted values in Algorithm \ref{ag:dml-mediation}. It then forms the treatment residual $\widehat U_{ip,k}^D=\Delta D_{ip}^*-\Delta\widehat h_{0k,ip}^*$ and the preliminary mediator residual $\widehat U_{ip,k}^M(\widetilde\gamma_k)=\Delta M_{ip}^*-\Delta\tau_{M,ip}^{*'}\widetilde\gamma_k-\Delta\widehat g_{0k,ip}^*$.

For the period-$t$ outcome moment, $t=3,\ldots,T$, the instrument history retains at most the two most recent dates $p<t$. For each retained $p$, the outcome instrument includes $\widehat U_{ip,k}^D$, $\widehat U_{ip,k}^M(\widetilde\gamma_k)$, $\widehat U_{ip,k}^DX_{ip3}$, and a squared treatment-residual term. Only the squared-residual block is centered and standardized across the combined evaluation sample. Thus the outcome equation contributes four moments at $t=3$ and eight in each subsequent period, for a total of 60. The mediator instrument contains the retained treatment residuals and one centered and standardized square of the most recent treatment residual. It contributes two moments at $t=3$ and three thereafter, for a total of 23. The 83 moments are stacked jointly for the nine parameters and minimized with the identity weighting matrix. Structural standard errors use the full empirical covariance of the stacked individual score, so cross-equation and cross-period dependence is retained.

Naive--GMM keeps the same differenced equations, 55-term library, penalized structural nuisance learner, history truncation, instrument augmentations, and identity weighting. It differs in three code paths: $\widehat f_0$ and $\widehat g_0$ are fitted once on all individuals; period demeaning uses the full sample; and the treatment and mediator residual histories in the final instruments are replaced by the corresponding period-demeaned treatment and mediator levels. No separate treatment nuisance is therefore needed for this benchmark.

The level effects are evaluated at $t=T$ for $\overline d_T=\boldsymbol1_T$ versus $\overline d_T^*=\boldsymbol0_T$. For each individual, the code computes
\begin{align*}
\ell_i^{\mathrm{CDE}}(\beta)=\ell_i^{\mathrm{NDE}}(\beta)&=\theta_1+\theta_2+\theta_3+\theta_4X_{iT3},\\
\ell_i^{\mathrm{NIE}}(\beta)&=(\gamma_1+\gamma_2+\gamma_3)(\theta_5+\theta_6),\\
\ell_i^{\mathrm{TE}}(\beta)&=\ell_i^{\mathrm{NDE}}(\beta)+\ell_i^{\mathrm{NIE}}(\beta),\\
\ell_i^{\mathrm{NDD}}(\beta)&=\theta_1+2\theta_2+\theta_4X_{iT3},\\
\ell_i^{\mathrm{NID}}(\beta)&=(\gamma_1+2\gamma_2)\theta_5,\qquad
\ell_i^{\mathrm{TD}}(\beta)=\ell_i^{\mathrm{NDD}}(\beta)+\ell_i^{\mathrm{NID}}(\beta).
\end{align*}
The estimated effect is the cross-sectional average of its individual contribution. Its influence function is computed as the centered individual contribution plus the estimated influence function of $\widehat\beta_0^D$ multiplied by the analytic gradient of the effect map. Standard errors are the square roots of the empirical second moments of these influence functions divided by $N$. Each design cell uses $1{,}000$ independent replications, and coverage records whether the resulting two-sided $95\%$ Wald interval contains the corresponding truth in \eqref{eq:simulation-truths}.

\subsection{Alternative sample sizes}
\label{sec:simulation-alternative-sample-sizes}
We repeat the baseline experiment of Section \ref{sec:baseline-simulation-details} with $N=800$ and $N=3200$, holding $T$, the dictionaries, the data-generating process, the first-stage learner, the number of folds and the number of replications fixed.

\begin{table}[!ht]
\centering
\caption{Monte Carlo results under alternative sample sizes}
\label{tab:simulation-alternative-sample-sizes}
\small
\setlength{\tabcolsep}{8pt}
\renewcommand{\arraystretch}{0.5}
\begin{threeparttable}
\begin{tabular}{lrrrrrrrr}
\toprule
& \multicolumn{4}{c}{$N=800$} & \multicolumn{4}{c}{$N=3200$} \\
\cmidrule(lr){2-5}\cmidrule(lr){6-9}
Estimand & Bias & SD & ASE & CP & Bias & SD & ASE & CP \\
\midrule
\multicolumn{9}{l}{\emph{Panel A: DML--GMM structural parameters}} \\
$\theta_{D_t}$     & -0.0091 & 0.0386 & 0.0381 & 0.942 &  0.0001 & 0.0194 & 0.0193 & 0.943 \\
$\theta_{D_t^2}$   &  0.0060 & 0.0190 & 0.0185 & 0.944 &  0.0039 & 0.0121 & 0.0118 & 0.928 \\
$\theta_{D_{t-1}}$ & -0.0040 & 0.0250 & 0.0242 & 0.936 &  0.0003 & 0.0119 & 0.0123 & 0.957 \\
$\theta_{DX}$      & -0.0072 & 0.0282 & 0.0273 & 0.933 & -0.0035 & 0.0171 & 0.0164 & 0.940 \\
$\theta_{M_t}$     & -0.0034 & 0.0262 & 0.0256 & 0.955 & -0.0038 & 0.0157 & 0.0155 & 0.951 \\
$\theta_{M_{t-1}}$ &  0.0015 & 0.0085 & 0.0080 & 0.930 &  0.0005 & 0.0048 & 0.0048 & 0.956 \\
$\gamma_{D_t}$     & -0.0215 & 0.0492 & 0.0512 & 0.940 & -0.0023 & 0.0230 & 0.0242 & 0.966 \\
$\gamma_{D_t^2}$   & -0.0009 & 0.0187 & 0.0182 & 0.958 & -0.0001 & 0.0091 & 0.0096 & 0.959 \\
$\gamma_{D_{t-1}}$ & -0.0099 & 0.0326 & 0.0338 & 0.953 & -0.0009 & 0.0156 & 0.0161 & 0.952 \\
\addlinespace[2pt]
\multicolumn{9}{l}{\emph{Panel B: Naive--GMM structural parameters}} \\
$\theta_{D_t}$     & -0.0234 & 0.0703 & 0.0665 & 0.948 & -0.0040 & 0.0538 & 0.0485 & 0.952 \\
$\theta_{D_t^2}$   &  0.0180 & 0.0475 & 0.0397 & 0.912 &  0.0227 & 0.0428 & 0.0373 & 0.913 \\
$\theta_{D_{t-1}}$ & -0.0009 & 0.0429 & 0.0388 & 0.952 &  0.0077 & 0.0330 & 0.0280 & 0.957 \\
$\theta_{DX}$      & -0.0299 & 0.0371 & 0.0310 & 0.809 & -0.0255 & 0.0235 & 0.0198 & 0.738 \\
$\theta_{M_t}$     & -0.0138 & 0.0705 & 0.0604 & 0.946 & -0.0230 & 0.0639 & 0.0560 & 0.941 \\
$\theta_{M_{t-1}}$ &  0.0013 & 0.0048 & 0.0044 & 0.935 &  0.0011 & 0.0027 & 0.0025 & 0.925 \\
$\gamma_{D_t}$     & -0.2238 & 0.0408 & 0.0152 & 0.000 & -0.1391 & 0.0180 & 0.0072 & 0.000 \\
$\gamma_{D_t^2}$   &  0.0054 & 0.0045 & 0.0040 & 0.680 &  0.0089 & 0.0018 & 0.0017 & 0.004 \\
$\gamma_{D_{t-1}}$ & -0.0227 & 0.0131 & 0.0117 & 0.515 & -0.0136 & 0.0064 & 0.0058 & 0.356 \\
\addlinespace[2pt]
\multicolumn{9}{l}{\emph{Panel C: DML--GMM causal effects}} \\
CDE & -0.0054 & 0.1031 & 0.1000 & 0.943 &  0.0052 & 0.0528 & 0.0521 & 0.941 \\
NDE & -0.0054 & 0.1031 & 0.1000 & 0.943 &  0.0052 & 0.0528 & 0.0521 & 0.941 \\
NIE & -0.0522 & 0.1375 & 0.1407 & 0.935 & -0.0116 & 0.0715 & 0.0722 & 0.948 \\
TE  & -0.0576 & 0.1546 & 0.1544 & 0.938 & -0.0064 & 0.0739 & 0.0769 & 0.957 \\
NDD &  0.0046 & 0.0980 & 0.0962 & 0.947 &  0.0088 & 0.0532 & 0.0520 & 0.941 \\
NID & -0.0300 & 0.0821 & 0.0825 & 0.928 & -0.0101 & 0.0454 & 0.0451 & 0.943 \\
TD  & -0.0254 & 0.1062 & 0.1028 & 0.940 & -0.0013 & 0.0514 & 0.0530 & 0.951 \\
\addlinespace[2pt]
\multicolumn{9}{l}{\emph{Panel D: Naive--GMM causal effects}} \\
CDE & -0.0047 & 0.1616 & 0.1524 & 0.957 &  0.0273 & 0.1300 & 0.1163 & 0.946 \\
NDE & -0.0047 & 0.1616 & 0.1524 & 0.957 &  0.0273 & 0.1300 & 0.1163 & 0.946 \\
NIE & -0.3837 & 0.1427 & 0.1100 & 0.137 & -0.2565 & 0.1205 & 0.1026 & 0.280 \\
TE  & -0.3883 & 0.1155 & 0.0934 & 0.036 & -0.2292 & 0.0553 & 0.0449 & 0.004 \\
NDD &  0.0142 & 0.1665 & 0.1539 & 0.957 &  0.0422 & 0.1400 & 0.1255 & 0.946 \\
NID & -0.2376 & 0.1328 & 0.1091 & 0.393 & -0.1646 & 0.1216 & 0.1055 & 0.600 \\
TD  & -0.2234 & 0.0914 & 0.0815 & 0.248 & -0.1224 & 0.0461 & 0.0410 & 0.166 \\
\bottomrule
\end{tabular}
\begin{tablenotes}[flushleft]
\footnotesize
\item Notes: Each design cell contains 1,000 replications with $T=10$. Bias is signed, SD is the Monte Carlo standard deviation, ASE is the average estimated standard error, and CP is the coverage probability of the nominal 95\% Wald interval. NDE coincides with CDE by construction.
\end{tablenotes}
\end{threeparttable}
\end{table}

Table \ref{tab:simulation-alternative-sample-sizes} shows that DML--GMM behaves as the asymptotic theory predicts. Quadrupling $N$ divides its Monte Carlo standard deviation by a factor between $0.47$ and $0.64$, against the $0.5$ implied by the $\sqrt{N}$ rate; the largest absolute bias falls from $0.0215$ to $0.0039$ across the nine coefficients and from $0.0576$ to $0.0116$ across the seven causal effects; and the average estimated standard error stays within $6\%$ of the Monte Carlo dispersion in all thirty-two cells. Coverage is already reliable at $N=800$, ranging from $0.928$ to $0.958$.

The one cell that falls slightly short is $\theta_{D_t^2}$ at $N=3200$, where coverage is $0.928$ while every other cell at that sample size stays between $0.940$ and $0.966$. Its bias declines far more slowly than the biases of the mediator equation, and the cause is the deliberately misspecified first-stage library rather than the orthogonalization: $h_0$ in \eqref{eq:simulation-nuisances} lies in the span of the $55$ monomials, whereas $f_0$ and $g_0$ contain hyperbolic, trigonometric and exponential terms that no cubic dictionary can reproduce. The product condition following Assumption \ref{ass:mediation-product} therefore keeps holding for the mediator moment, whose remainder involves $\|\widehat{h}_{0k}-h_0\|_{F_0,2}$ and whose largest bias falls by an order of magnitude, from $0.0215$ at $N=800$ to $0.0023$ at $N=3200$, but fails in the limit for the outcome moment, whose remainder involves $\|\widehat{f}_{0k}-f_0\|_{F_0,2}\|\widehat{g}_{0k}-g_0\|_{F_0,2}$ and whose leading bias falls only from $0.0060$ to $0.0039$ while the corresponding standard deviation falls from $0.0190$ to $0.0121$. Orthogonality removes the first-order effect of first-stage sampling error, not the effect of a first stage that converges to the wrong function.

Naive--GMM does not improve with the sample size and in several respects deteriorates: coverage for $\gamma_{D_t}$ is $0.000$ at all three sample sizes, coverage for $\gamma_{D_t^2}$ falls from $0.680$ to $0.004$, and coverage for TE never exceeds $0.036$. Two mechanisms are visible. The mediator-equation estimates concentrate at essentially the parametric rate but around a wrong center, so a larger panel only buys shorter intervals that miss the truth more often. The outcome-equation estimates instead disperse far more slowly than $N^{-1/2}$, five of their six standard deviations falling by a factor of only $0.63$ to $0.91$ when $N$ quadruples, as expected of an estimation error that is first order in the first-stage error and, without cross-fitting, common to all individuals. Neither defect is a small-sample artifact.

\subsection{Alternative data-generating process}
\label{sec:simulation-alternative-dgp}
The second experiment changes the baseline design in Section \ref{sec:baseline-simulation-details} in one place only: the nuisance functions \eqref{eq:simulation-nuisances} are replaced by
\begin{equation}
\label{eq:simulation-nuisances-alt}
\begin{gathered}
f_0(x)=(x_1^2-1)+(x_2^2-1)+0.2\,x_1^3|x_1|+\frac{2x_1}{1+x_1^2}
 +\operatorname{arcsinh}(x_3)+\arctan(x_1x_2),\\
g_0(x)=(x_2^2-1)+10\{e^{-x_1^2}-3^{-1/2}\}+\frac{2x_4}{1+x_4^2}
 +2[\{\max(x_5,0)\}^2-1/2],\\
h_0(x)=1.5(x_1^2-1)+(x_2^2-1)+(x_3^2-1)+x_3,
\end{gathered}
\end{equation}
each term again having mean zero under the standard normal marginals. All other simulation settings are identical to those in Section \ref{sec:baseline-simulation-details}. The replacement changes the way the nuisances escape the dictionary used to approximate them but keeps the design adverse. Because the causal estimands are functions of $\theta_0$ and $\gamma_0$ alone, the true values in \eqref{eq:simulation-truths} continue to apply.

\begin{table}[!ht]
\centering
\caption{Monte Carlo results under the alternative data-generating process}
\label{tab:simulation-alternative-dgp}
\small
\setlength{\tabcolsep}{8pt}
\renewcommand{\arraystretch}{0.5}
\begin{threeparttable}
\begin{tabular}{lccccccc}
\toprule
Estimand & Method & Mean & Bias & SD & RMSE & ASE & CP \\
\midrule
\multicolumn{8}{l}{\emph{Panel A: Structural parameters}} \\
\multirow{2}{*}{$\theta_{D_t}$}
 & DML--GMM   & 2.0014 &  0.0014 & 0.0327 & 0.0327 & 0.0323 & 0.950 \\
 & Naive--GMM & 2.0451 &  0.0451 & 0.0766 & 0.0889 & 0.0618 & 0.834 \\

\multirow{2}{*}{$\theta_{D_t^2}$}
 & DML--GMM   & 0.5047 &  0.0047 & 0.0172 & 0.0178 & 0.0171 & 0.939 \\
 & Naive--GMM & 0.5075 &  0.0075 & 0.0890 & 0.0893 & 0.0638 & 0.880 \\

\multirow{2}{*}{$\theta_{D_{t-1}}$}
 & DML--GMM   & 2.0003 &  0.0003 & 0.0186 & 0.0186 & 0.0191 & 0.969 \\
 & Naive--GMM & 2.0070 &  0.0070 & 0.0774 & 0.0777 & 0.0571 & 0.903 \\

\multirow{2}{*}{$\theta_{DX}$}
 & DML--GMM   & 1.9917 & -0.0083 & 0.0211 & 0.0227 & 0.0207 & 0.929 \\
 & Naive--GMM & 1.9709 & -0.0291 & 0.0283 & 0.0406 & 0.0208 & 0.610 \\

\multirow{2}{*}{$\theta_{M_t}$}
 & DML--GMM   & 0.9964 & -0.0036 & 0.0230 & 0.0233 & 0.0236 & 0.953 \\
 & Naive--GMM & 0.9884 & -0.0116 & 0.1526 & 0.1530 & 0.1107 & 0.885 \\

\multirow{2}{*}{$\theta_{M_{t-1}}$}
 & DML--GMM   & 0.4999 & -0.0001 & 0.0048 & 0.0048 & 0.0049 & 0.953 \\
 & Naive--GMM & 0.5000 &  0.0000 & 0.0038 & 0.0038 & 0.0032 & 0.935 \\

\multirow{2}{*}{$\gamma_{D_t}$}
 & DML--GMM   & 0.9793 & -0.0207 & 0.0478 & 0.0521 & 0.0491 & 0.928 \\
 & Naive--GMM & 0.6076 & -0.3924 & 0.0528 & 0.3959 & 0.0147 & 0.000 \\

\multirow{2}{*}{$\gamma_{D_t^2}$}
 & DML--GMM   & 0.4985 & -0.0015 & 0.0167 & 0.0168 & 0.0173 & 0.964 \\
 & Naive--GMM & 0.5196 &  0.0196 & 0.0039 & 0.0199 & 0.0034 & 0.000 \\

\multirow{2}{*}{$\gamma_{D_{t-1}}$}
 & DML--GMM   & 0.4916 & -0.0084 & 0.0325 & 0.0335 & 0.0325 & 0.943 \\
 & Naive--GMM & 0.4754 & -0.0246 & 0.0105 & 0.0268 & 0.0089 & 0.231 \\
\addlinespace[2pt]
\multicolumn{8}{l}{\emph{Panel B: Causal effects}} \\
\multirow{2}{*}{CDE}
 & DML--GMM   & 4.5066 &  0.0066 & 0.0822 & 0.0824 & 0.0795 & 0.940 \\
 & Naive--GMM & 4.5597 &  0.0597 & 0.2429 & 0.2500 & 0.1830 & 0.895 \\

\multirow{2}{*}{NDE}
 & DML--GMM   & 4.5066 &  0.0066 & 0.0822 & 0.0824 & 0.0795 & 0.940 \\
 & Naive--GMM & 4.5597 &  0.0597 & 0.2429 & 0.2500 & 0.1830 & 0.895 \\

\multirow{2}{*}{NIE}
 & DML--GMM   & 2.9469 & -0.0531 & 0.1290 & 0.1394 & 0.1323 & 0.935 \\
 & Naive--GMM & 2.3855 & -0.6145 & 0.2621 & 0.6680 & 0.1789 & 0.161 \\

\multirow{2}{*}{TE}
 & DML--GMM   & 7.4535 & -0.0465 & 0.1341 & 0.1419 & 0.1349 & 0.931 \\
 & Naive--GMM & 6.9453 & -0.5547 & 0.1172 & 0.5670 & 0.0713 & 0.002 \\

\multirow{2}{*}{NDD}
 & DML--GMM   & 3.0110 &  0.0110 & 0.0816 & 0.0823 & 0.0780 & 0.938 \\
 & Naive--GMM & 3.0602 &  0.0602 & 0.2535 & 0.2605 & 0.1891 & 0.880 \\

\multirow{2}{*}{NID}
 & DML--GMM   & 1.9692 & -0.0308 & 0.0740 & 0.0802 & 0.0773 & 0.938 \\
 & Naive--GMM & 1.6280 & -0.3720 & 0.2597 & 0.4536 & 0.1829 & 0.388 \\

\multirow{2}{*}{TD}
 & DML--GMM   & 4.9801 & -0.0199 & 0.0829 & 0.0852 & 0.0827 & 0.939 \\
 & Naive--GMM & 4.6882 & -0.3118 & 0.0834 & 0.3228 & 0.0583 & 0.016 \\
\bottomrule
\end{tabular}
\begin{tablenotes}[flushleft]
\footnotesize
\item Notes: The design differs from the baseline design in Section \ref{sec:baseline-simulation-details} only in the nuisance functions, which are given by \eqref{eq:simulation-nuisances-alt} instead of \eqref{eq:simulation-nuisances}. Each method uses 1,000 replications with $N=1600$ and $T=10$. Bias is signed, SD is the Monte Carlo standard deviation, RMSE is the root mean squared error, ASE is the average estimated standard error, and CP is the coverage probability of the nominal 95\% Wald interval. NDE coincides with CDE by construction.
\end{tablenotes}
\end{threeparttable}
\end{table}

Table \ref{tab:simulation-alternative-dgp} reproduces the pattern of the baseline design reported in Table \ref{tab:simulation-results} of the main text and of the sample-size experiment in Table \ref{tab:simulation-alternative-sample-sizes}. DML--GMM remains centered and correctly sized under the new nuisance functions: its largest absolute bias is $0.0207$ across the nine structural parameters and $0.0531$ across the seven causal effects, at most $2.1\%$ of the corresponding true value; its average estimated standard errors track the Monte Carlo dispersion to within $5\%$ in every one of the sixteen rows; and its coverage ranges from $0.928$ to $0.969$, against $0.937$ to $0.958$ in the main design. Relative to Naive--GMM it has the smaller absolute bias for fifteen of the sixteen quantities, the exception being $\theta_{M_{t-1}}$, for which both biases are below $0.0001$; the coverage closer to nominal for all sixteen; and the smaller root mean squared error for seven of the nine structural parameters and for all seven causal effects, where the reduction ranges from $16\%$ to $87\%$.

Naive--GMM fails in the same two ways as in Section \ref{sec:simulation}, and here more severely. Its outcome equation stays roughly centered but is far noisier, the standard deviations of the three treatment coefficients and of the contemporaneous mediator coefficient being $2.3$ to $6.6$ times those of DML--GMM, and its estimated standard errors understate even that dispersion by $19\%$ to $28\%$, so coverage for those four coefficients runs from $0.834$ to $0.903$. The exception is $\theta_{DX}$, whose bias of $-0.0291$ exceeds its own standard deviation and pushes coverage down to $0.610$. Its mediator equation is displaced instead: $\widehat{\gamma}_{D_t}$ carries a bias of $-0.3924$, or $39\%$ of the truth, and coverage for $\gamma_{D_t}$, $\gamma_{D_t^2}$ and $\gamma_{D_{t-1}}$ is $0.000$, $0.000$ and $0.231$. The damage to the causal estimands is again concentrated in the indirect channel: CDE, NDE and NDD keep biases below $0.061$, with coverage of $0.895$, $0.895$ and $0.880$, while NIE, TE, NID and TD carry biases of $-0.6145$, $-0.5547$, $-0.3720$ and $-0.3118$ with coverage of $0.161$, $0.002$, $0.388$ and $0.016$, and the naive standard errors understate the Monte Carlo dispersion of every causal estimand by $25\%$ to $39\%$. The conclusions of Section \ref{sec:simulation} are therefore not an artifact of the particular nuisance functions used there.

\section{Supplementary Information on the Empirical Application}
\subsection{Empirical estimands and implementation}
\label{sec:empirical-implementation-details}
The two empirical specifications use the same first-stage covariate library. It consists of the 108 monomials of total degree at most three in the eight lagged county characteristics that involve at most two distinct covariates. The nuisance functions $f_0$ and $g_0$ are fitted from first-differenced, period-demeaned equations by $\ell_1$-penalized GMM with lagged long-difference instruments; the mediator fit also supplies the preliminary coefficient $\widetilde\gamma_k$. The treatment nuisance $h_0$ is fitted by least squares on the same library. Algorithm \ref{ag:dml-mediation} uses five folds, a cyclic auxiliary fold, and the identity weighting matrix. Instrument histories are truncated at their two most recent lags, producing 99 moments for the three parameters in Specification I and 627 moments for the nineteen parameters in Specification II.

For inference, each bootstrap replication samples $3{,}058$ counties with replacement and retains the complete 2000--2019 trajectory of every selected county. All first-stage fits, fold-specific transformations, GMM estimates, and causal-effect calculations are then repeated. The two specifications use exactly the same county resample within a replication, so the standard errors for their differences account for the resulting dependence. The reported standard errors use 500 replications.

Because treatment enters both empirical dictionaries linearly, the period-$t$ derivative effects equal the corresponding one-period level contrasts. In particular,
\[
\operatorname{NDD}_{t,t}=\operatorname{NDE}_t\bigl((\boldsymbol0_{t-1}',1)',\boldsymbol0_t\bigr),\qquad
\operatorname{NID}_{t,t}=\operatorname{NIE}_t\bigl((\boldsymbol0_{t-1}',1)',\boldsymbol0_t\bigr),
\]
with the analogous identity between $\operatorname{TD}_{t,t}$ and $\operatorname{TE}_t$. Since all variables are standardized, the entries of Table \ref{tab:empirical-results} may therefore be read either as average derivatives or as effects of increasing the period-$t$ employment rate by one standard deviation while setting the other treatment components to zero.

\subsection{Period-specific empirical effects}
\label{sec:empirical-period-specific}
\begin{figure}[!ht]
\centering
\captionsetup[subfigure]{font=small}
\begin{subfigure}[t]{0.48\textwidth}
\centering
\includegraphics[width=\linewidth]{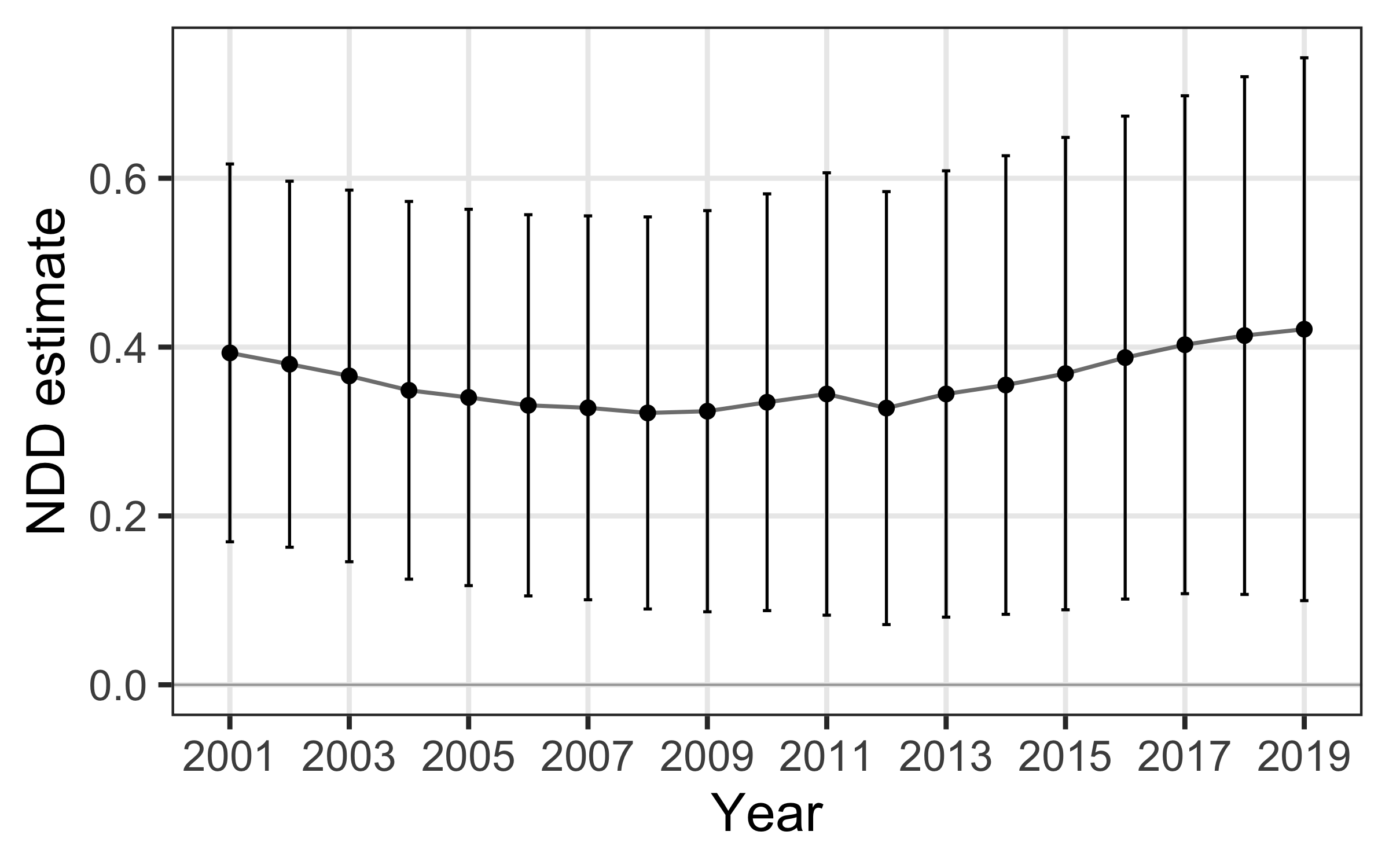}
\caption{$\operatorname{NDD}_{t,t}$: direct channel}
\label{fig:empirical-time-ndd}
\end{subfigure}\hfill
\begin{subfigure}[t]{0.48\textwidth}
\centering
\includegraphics[width=\linewidth]{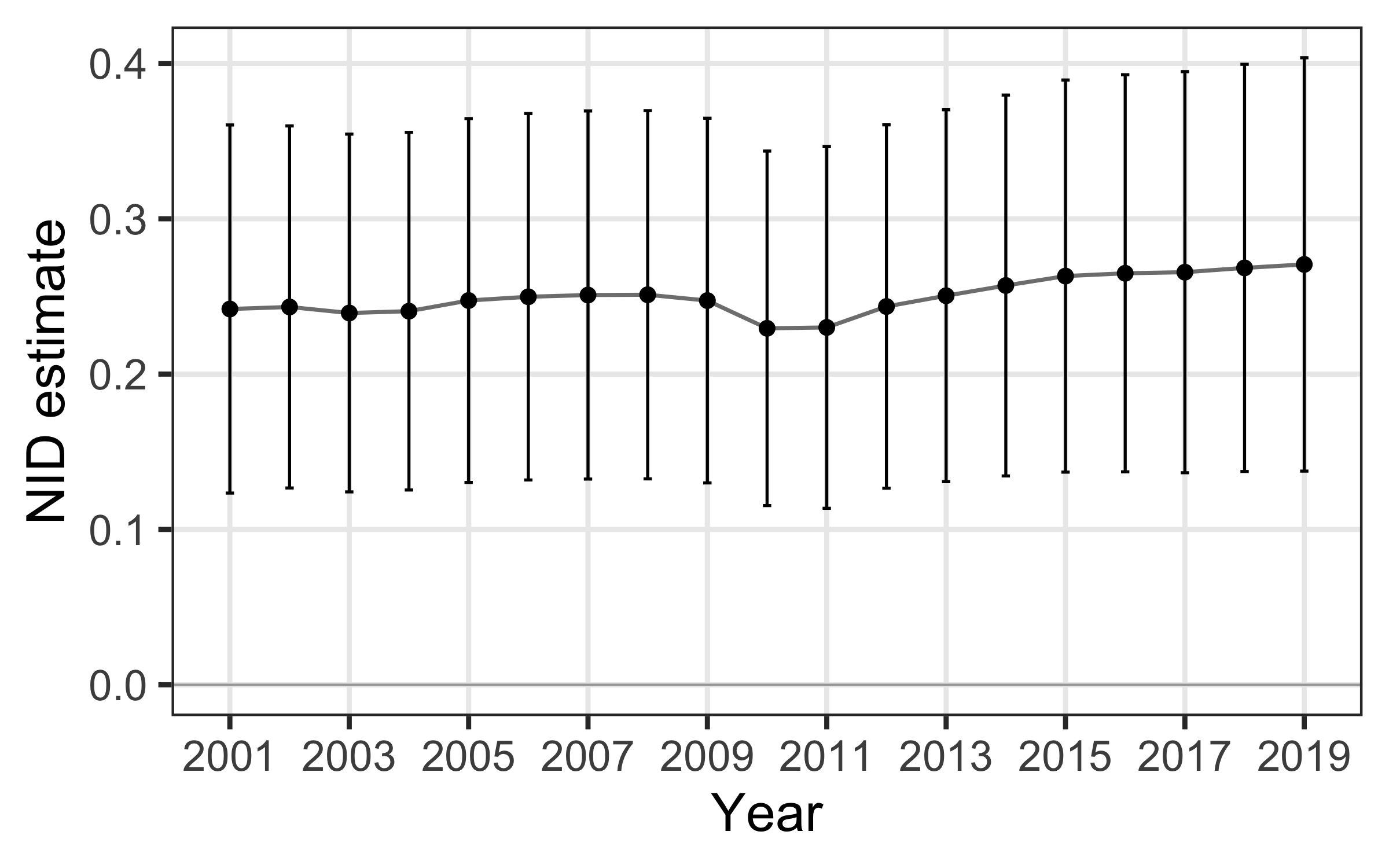}
\caption{$\operatorname{NID}_{t,t}$: wage channel}
\label{fig:empirical-time-nid}
\end{subfigure}

\vspace{1em}

\begin{subfigure}[t]{0.48\textwidth}
\centering
\includegraphics[width=\linewidth]{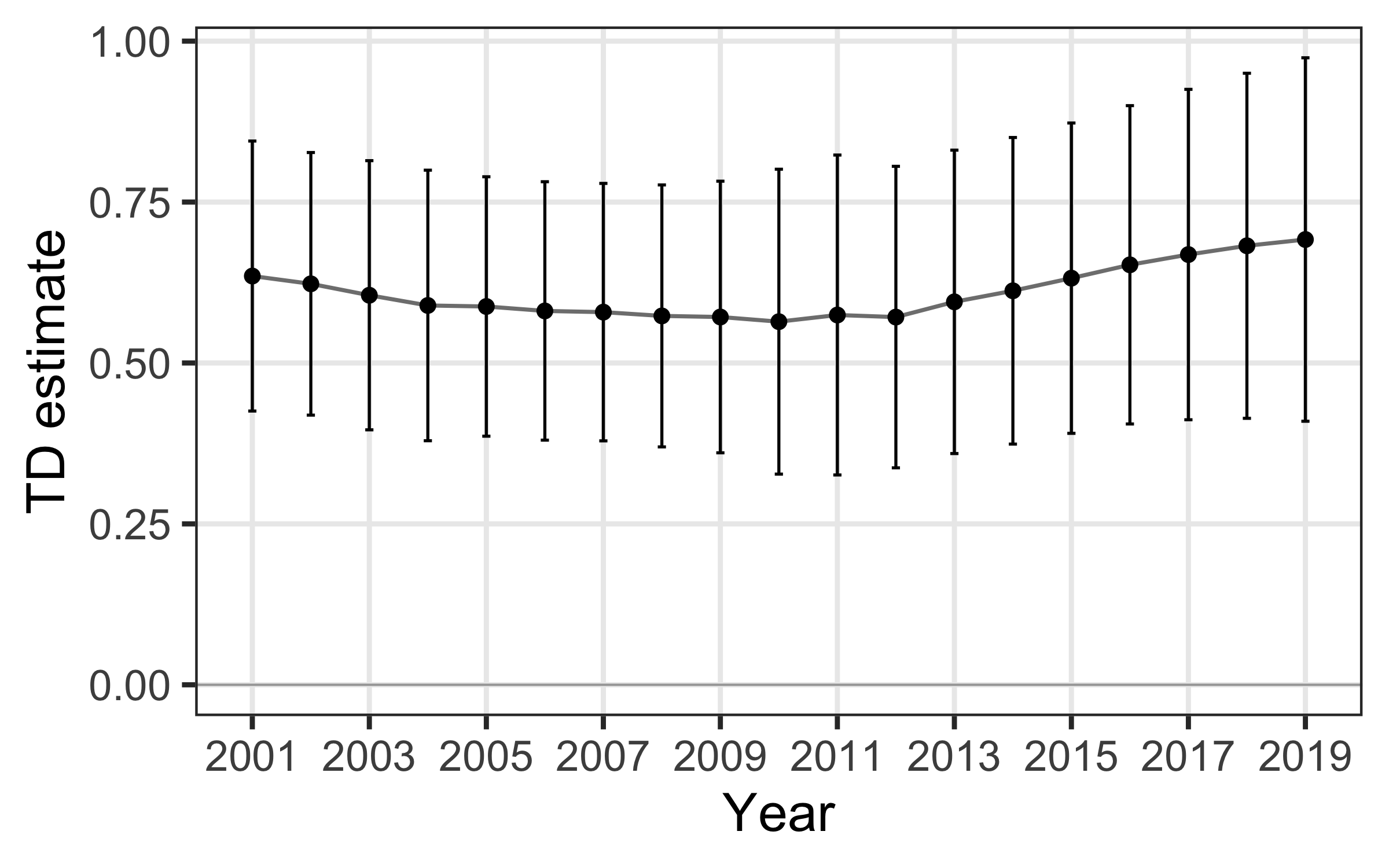}
\caption{$\operatorname{TD}_{t,t}$: total effect}
\label{fig:empirical-time-td}
\end{subfigure}
\caption{Period-$t$ derivative effects under Specification II, with $95\%$ county-trajectory bootstrap intervals.}
\label{fig:empirical-time}
\end{figure}

Figure \ref{fig:empirical-time} traces the path concealed by the averages of Panel B. Since the coefficients do not vary with $t$, the profile measures how the evolving demographic and migration composition of U.S.\ counties shifted the period-$t$ effect, not a change in the structural parameters themselves. All three profiles are U-shaped: $\operatorname{TD}_{t,t}$ falls from $0.635$ in 2001 to a minimum of $0.564$ in 2010 and recovers to $0.692$ by 2019. The two channels reach their minima two years apart, $\operatorname{NDD}_{t,t}$ in 2008 at $0.322$ and $\operatorname{NID}_{t,t}$ in 2010 at $0.229$. The ordering of the troughs matches what slow local wage adjustment would predict, and it is a pattern that the averaged estimands alone would not reveal.

\section{Proofs}
\subsection{Proof of Theorem \ref{th1}}
Note that $X \indep (Y,W) \mid Z$ implies $X \indep Y \mid W,Z$. Then, \eqref{eqsciad} implies
\begin{equation}
Y_{it}(\overline{d}_t,\overline{m}_t) \indep D_{is} \mid \overline{\bX}_{it}, \overline{D}_{i,s-1}=\overline{d}_{s-1}, \overline{M}_{i,s-1}=\overline{m}_{s-1}, \alpha_i^Y,\overline{\lambda}_t^Y, \quad \forall s \leqslant t,
\label{A1}
\end{equation}
and \eqref{eqsciam} implies
\begin{equation}
Y_{it}(\overline{d}_t,\overline{m}_t) \indep M_{is} \mid \overline{\bX}_{it}, \overline{D}_{is}=\overline{d}_s, \overline{M}_{i,s-1}=\overline{m}_{s-1}, \alpha_i^Y,\overline{\lambda}_t^Y, \qquad \forall s\leqslant t.
\label{A2}
\end{equation}
Under Assumption \ref{asscon}, if $\overline{D}_{it}=\overline{d}_t$ and $\overline{M}_{it}=\overline{m}_t$, then $Y_{it}=Y_{it}(\overline{d}_t,\overline{m}_t)$. Therefore,
\[
\E(Y_{it} \mid \overline{\bX}_{it}, \overline{D}_{it}=\overline{d}_t, \overline{M}_{it}=\overline{m}_t,\alpha_i^Y,\overline{\lambda}_t^Y)=\E\{Y_{it}(\overline{d}_t,\overline{m}_t) \mid \overline{\bX}_{it},\overline{D}_{it}=\overline{d}_t,\overline{M}_{it}=\overline{m}_t,\alpha_i^Y,\overline{\lambda}_t^Y\}.
\]
Repeatedly using \eqref{A2} and \eqref{A1}, in this order, from $s=t$ down to $s=1$, eliminates $M_{is}=m_s$ and then $D_{is}=d_s$ in each period, yielding
\[
\E\{Y_{it}(\overline{d}_t,\overline{m}_t) \mid \overline{\bX}_{it},\overline{D}_{it}=\overline{d}_t, \overline{M}_{it}=\overline{m}_t,\alpha_i^Y,\overline{\lambda}_t^Y\}=\E\{Y_{it}(\overline{d}_t,\overline{m}_t) \mid \overline{\bX}_{it}, \alpha_i^Y,\overline{\lambda}_t^Y\}.
\]
Similarly, \eqref{eqsciam2} implies
\begin{equation}
M_{it}(\overline{d}_t) \indep D_{is} \mid \overline{\bX}_{it}, \overline{D}_{i,s-1}=\overline{d}_{s-1}, \alpha_i^M,\overline{\lambda}_t^M, \quad \forall s\leqslant t.
\label{A3}
\end{equation}
Consistency and repeated application of \eqref{A3} give
\[
\E(M_{it} \mid \overline{\bX}_{it},\overline{D}_{it}=\overline{d}_t,\alpha_i^M,\overline{\lambda}_t^M)=\E\{M_{it}(\overline{d}_t) \mid \overline{\bX}_{it},\alpha_i^M,\overline{\lambda}_t^M\}.
\]
This establishes the two conditional-mean identities.

Under the mediation model for the potential outcomes,
\[
\E\{Y_{it}(\overline{d}_t,\overline{m}_t)\mid\overline{\bX}_{it},\alpha_i^Y,\overline{\lambda}_t^Y\}=\tau_Y(\overline{d}_{t-L:t},\overline{m}_{t-L:t},\overline{\bX}_{i,t-L:t})^\prime\theta_0+f_0(\overline{\bX}_{i,t-L:t})+\alpha_i^Y+\lambda_t^Y
\]
and
\[
\E\{M_{it}(\overline{d}_t)\mid\overline{\bX}_{it},\alpha_i^M,\overline{\lambda}_t^M\}=\tau_M(\overline{d}_{t-L:t},\overline{\bX}_{i,t-L:t})^\prime\gamma_0+g_0(\overline{\bX}_{i,t-L:t})+\alpha_i^M+\lambda_t^M.
\]
Combining these equations with the conditional-mean identities gives
\begin{align}
\E(Y_{it}\mid\overline{\bX}_{it},\overline{D}_{it},\overline{M}_{it},\alpha_i^Y,\overline{\lambda}_t^Y)&=\tau_Y(\overline{D}_{i,t-L:t},\overline{M}_{i,t-L:t},\overline{\bX}_{i,t-L:t})^\prime\theta_0+f_0(\overline{\bX}_{i,t-L:t})+\alpha_i^Y+\lambda_t^Y,\label{A4}\\
\E(M_{it}\mid\overline{\bX}_{it},\overline{D}_{it},\alpha_i^M,\overline{\lambda}_t^M)&=\tau_M(\overline{D}_{i,t-L:t},\overline{\bX}_{i,t-L:t})^\prime\gamma_0+g_0(\overline{\bX}_{i,t-L:t})+\alpha_i^M+\lambda_t^M.\label{A5}
\end{align}
Equations \eqref{A4} and \eqref{A5} are equivalent to \eqref{eq:observed-mediation-model}. Q.E.D.

\subsection{Proof of Theorem \ref{th:causal-effect-normality}}
Let $\mathbb{P}$ denote expectation under $F_0$ and write $\Delta_N\coloneq\widehat{\beta}_0^D-\beta_0$. The proof first verifies that the contribution functions target the causal estimands, then establishes the uniform first-order expansion needed for the plug-in estimator.

\paragraph{Step 1: Causal interpretation of the contribution functions.}
Under the potential-outcome mediator model, for any admissible treatment history $\overline{d}_t$,
\begin{align*}
M_{it}(\overline{d}_t)-M_{it}(\overline{D}_{it})=&\Big\{\tau_M(\overline{d}_{t-L:t},\overline{\bX}_{i,t-L:t})-\tau_M(\overline{D}_{i,t-L:t},\overline{\bX}_{i,t-L:t})\Big\}'\gamma_0.
\end{align*}
The nuisance regression, fixed effects, and structural disturbance cancel because they are common to the two potential mediators. By Assumption \ref{asscon}, $M_{it}(\overline{D}_{it})=M_{it}$, and hence
\begin{equation}
M_{it}^{\beta_0}(\overline{d}_t)=M_{it}(\overline{d}_t).
\label{eq:proof-counterfactual-mediator-reconstruction}
\end{equation}
The same argument applied to the potential-outcome equation gives
\begin{align*}
&Y_{it}(\overline{d}_t,\overline{m}_t)-Y_{it}(\overline{D}_{it},\overline{M}_{it})\\
&\quad=\Big\{\tau_Y(\overline{d}_{t-L:t},\overline{m}_{t-L:t},\overline{\bX}_{i,t-L:t})
-\tau_Y(\overline{D}_{i,t-L:t},\overline{M}_{i,t-L:t},\overline{\bX}_{i,t-L:t})\Big\}'\theta_0.
\end{align*}
Consistency therefore implies
\begin{equation}
Y_{it}^{\beta_0}(\overline{d}_t,\overline{m}_t)=Y_{it}(\overline{d}_t,\overline{m}_t).
\label{eq:proof-counterfactual-outcome-reconstruction}
\end{equation}
Applying \eqref{eq:proof-counterfactual-mediator-reconstruction} period by period and substituting the resulting mediator history into \eqref{eq:proof-counterfactual-outcome-reconstruction} yields
\begin{equation*}
Y_{it}^{\beta_0}\{\overline{a}_t,\overline{M}_{it}^{\beta_0}(\overline{b}_t)\}=Y_{it}\{\overline{a}_t,\overline{M}_{it}(\overline{b}_t)\}.
\end{equation*}
Thus, taking expectations of the level-effect contributions at $\beta_0$ gives the CDE, NDE, NIE, and TE in Section \ref{secce}. For example,
\begin{align*}
\E\{\ell_{i,t}^{\mathrm{NDE}}(\beta_0)\}&=\E\Big[Y_{it}\{\overline{d}_t,\overline{M}_{it}(\overline{d}_t^*)\}-Y_{it}\{\overline{d}_t^*,\overline{M}_{it}(\overline{d}_t^*)\}\Big]\\
&=\operatorname{NDE}_t(\overline{d}_t,\overline{d}_t^*).
\end{align*}
For derivative effects, Assumption \ref{ass:causal-effect-regularity}(iii) permits differentiation to pass through expectation. In particular,
\begin{align*}
\E\{\ell_{i,t,j}^{\mathrm{NID}}(\beta_0)\}&=\E\left[\frac{\partial}{\partial b_j}Y_{it}\{\overline{a}_t,\overline{M}_{it}(\overline{b}_t)\}\bigg|_{\overline{a}_t=\overline{b}_t=\overline{d}_t}\right]\\
&=\frac{\partial}{\partial b_j}\E\left[Y_{it}\{\overline{a}_t,\overline{M}_{it}(\overline{b}_t)\}\right]\bigg|_{\overline{a}_t=\overline{b}_t=\overline{d}_t}\\
&=\operatorname{NID}_{t,j}(\overline{d}_t),
\end{align*}
and the CDD, NDD, and TD cases follow analogously. Hence the vector $\vartheta_0$ in \eqref{eq:generic-causal-estimator} is the stated stack of causal estimands.

\paragraph{Step 2: Uniform control of the causal-effect Jacobian.}
Set
\[
G(W_i;\beta)\coloneq\dot{\ell}_\beta(W_i;\beta),\qquad F(W_i)\coloneq\sup_{\beta\in\mathcal{B}}\|G(W_i;\beta)\|_F.
\]
Assumption \ref{ass:causal-effect-regularity}(ii) gives $\E\{F(W_i)^2\}<\infty$, and therefore $\E F(W_i)<\infty$. For $a>0$, define
\[
\omega_a(W_i)\coloneq\sup_{\substack{\beta,\widetilde{\beta}\in\mathcal{B}\\\|\beta-\widetilde{\beta}\|\leqslant a}}\|G(W_i;\beta)-G(W_i;\widetilde{\beta})\|_F.
\]
For almost every $W_i$, continuity of $G(W_i;\cdot)$ on the compact set $\mathcal{B}$ implies $\omega_a(W_i)\to0$ as $a\downarrow0$. Since $\omega_a(W_i)\leqslant2F(W_i)$, dominated convergence gives
\begin{equation*}
\E\{\omega_a(W_i)\}\longrightarrow0\qquad\text{as }a\downarrow0.
\end{equation*}
For fixed $a$, cover $\mathcal{B}$ by finitely many balls of radius $a$ with centers $\beta_1,\ldots,\beta_{J(a)}$. Then
\begin{align}
\sup_{\beta\in\mathcal{B}}\|(\mathbb{P}_N-\mathbb{P})G(W_i;\beta)\|_F\leq&\max_{1\leqslant j\leqslant J(a)}\|(\mathbb{P}_N-\mathbb{P})G(W_i;\beta_j)\|_F\nonumber\\
&+\mathbb{P}_N\omega_a(W_i)+\mathbb{P}\omega_a(W_i).
\label{eq:proof-causal-jacobian-net}
\end{align}
For fixed $a$, the first term converges in probability to zero by the ordinary law of large numbers, and $\mathbb{P}_N\omega_a(W_i)\xrightarrow{p}\mathbb{P}\omega_a(W_i)$. Letting $a\downarrow0$ after taking the probability limit superior in \eqref{eq:proof-causal-jacobian-net} gives
\begin{equation}
\sup_{\beta\in\mathcal{B}}\|(\mathbb{P}_N-\mathbb{P})G(W_i;\beta)\|_F\xrightarrow{p}0.
\label{eq:proof-causal-jacobian-ulln}
\end{equation}
Moreover, pointwise continuity of $G(W_i;\beta)$ and domination by $F(W_i)$ imply that $\beta\mapsto\mathbb{P} G(W_i;\beta)$ is continuous at $\beta_0$. Therefore, for every deterministic sequence $\delta_N\downarrow0$,
\begin{equation}
\sup_{\substack{\beta\in\mathcal{B}\\\|\beta-\beta_0\|\leqslant\delta_N}}\|\mathbb{P}_NG(W_i;\beta)-\boldsymbol{J}_\vartheta\|_F\xrightarrow{p}0.
\label{eq:proof-causal-local-jacobian}
\end{equation}

\paragraph{Step 3: Linearization of the plug-in term.}
Using \eqref{eq:generic-causal-estimator}, add and subtract $\mathbb{P}_N\ell(W_i;\beta_0)$ to obtain
\begin{align}
\sqrt{N}(\widehat{\vartheta}-\vartheta_0)=&\sqrt{N}(\mathbb{P}_N-\mathbb{P})\ell(W_i;\beta_0)+\sqrt{N}\,\mathbb{P}_N\{\ell(W_i;\widehat{\beta}_0^D)-\ell(W_i;\beta_0)\}.
\label{eq:proof-causal-decomposition-full}
\end{align}
Theorem \ref{th:mediation-beta-normality} gives $\sqrt{N}\Delta_N=O_p(1)$. Because $\beta_0$ is an interior point of the compact convex set $\mathcal{B}$, the line segment $\{\beta_0+u\Delta_N:0\leqslant u\leqslant1\}$ lies in $\mathcal{B}$ with probability approaching one. On this event, the fundamental theorem of calculus yields
\begin{equation}
\ell(W_i;\widehat{\beta}_0^D)-\ell(W_i;\beta_0)=\left[\int_0^1\dot{\ell}_\beta(W_i;\beta_0+u\Delta_N)\,\mathrm{d}u\right]\Delta_N.
\label{eq:proof-causal-integral-expansion-full}
\end{equation}
Let
\[
\overline{\boldsymbol{J}}_N\coloneq\int_0^1\mathbb{P}_N\dot{\ell}_\beta(W_i;\beta_0+u\Delta_N)\,\mathrm{d}u.
\]
Choose $\delta_N=N^{-1/4}$. Root-$N$ consistency implies $\P(\|\Delta_N\|\leqslant\delta_N)\to1$, and \eqref{eq:proof-causal-local-jacobian} then gives $\overline{\boldsymbol{J}}_N=\boldsymbol{J}_\vartheta+o_p(1)$. Averaging \eqref{eq:proof-causal-integral-expansion-full} and multiplying by $\sqrt{N}$ therefore yields
\begin{equation}
\sqrt{N}\,\mathbb{P}_N\{\ell(W_i;\widehat{\beta}_0^D)-\ell(W_i;\beta_0)\}=\boldsymbol{J}_\vartheta\sqrt{N}\Delta_N+o_p(1).
\label{eq:proof-causal-plug-in-full}
\end{equation}

\paragraph{Step 4: Influence-function representation and central limit theorem.}
By \eqref{eq:mediation-beta-clt},
\begin{equation}
\sqrt{N}\Delta_N=-\mathcal{A}_0\frac{1}{\sqrt{N}}\sum_{i=1}^N\widetilde{\psi}_i+o_p(1).
\label{eq:proof-beta-if-recalled}
\end{equation}
Substituting \eqref{eq:proof-causal-plug-in-full} and \eqref{eq:proof-beta-if-recalled} into \eqref{eq:proof-causal-decomposition-full} gives
\begin{align*}
\sqrt{N}(\widehat{\vartheta}-\vartheta_0)&=\frac{1}{\sqrt{N}}\sum_{i=1}^N\{\ell(W_i;\beta_0)-\vartheta_0\}-\boldsymbol{J}_\vartheta\mathcal{A}_0\frac{1}{\sqrt{N}}\sum_{i=1}^N\widetilde{\psi}_i+o_p(1)\\
&=\frac{1}{\sqrt{N}}\sum_{i=1}^N\varphi_{\vartheta,i}+o_p(1).
\end{align*}
The oracle moment conditions imply $\E(\widetilde{\psi}_i)=\boldsymbol{0}$, while $\E\{\ell(W_i;\beta_0)-\vartheta_0\}=\boldsymbol{0}$ by definition. Hence $\E(\varphi_{\vartheta,i})=\boldsymbol{0}$. Furthermore,
\begin{align*}
\E\|\varphi_{\vartheta,i}\|^2&\leqslant2\E\|\ell(W_i;\beta_0)-\vartheta_0\|^2+2\|\boldsymbol{J}_\vartheta\mathcal{A}_0\|_F^2\E\|\widetilde{\psi}_i\|^2<\infty
\end{align*}
by Assumption \ref{ass:causal-effect-regularity}(ii) and Assumption \ref{ass:mediation-l2}(i). Under \eqref{iid}, the influence functions are i.i.d.; the multivariate central limit theorem and Slutsky's theorem therefore establish \eqref{eq:causal-effect-linearization}. The multivariate central limit theorem remains valid when $\boldsymbol{\Sigma}_\vartheta$ is singular, in which case the limiting normal law is degenerate on the corresponding linear subspace. Q.E.D.

\subsection{Proof of Theorem \ref{prop:causal-effect-variance}}
We first establish consistency of each ingredient of the estimated influence function and then prove convergence of its empirical second moment.

\paragraph{Step 1: Consistency of $\widehat{\mathcal{A}}$ and $\widehat{\boldsymbol{J}}_\vartheta$.}
Theorem \ref{prop:mediation-sandwich} gives $\widehat{\boldsymbol{G}}\xrightarrow{p}\boldsymbol{G}$ and $\widehat{\boldsymbol{\Omega}}\xrightarrow{p}\boldsymbol{\Omega}$. Since $\boldsymbol{G}'\boldsymbol{\Omega}\boldsymbol{G}$ is nonsingular, its smallest eigenvalue is strictly positive. Weyl's inequality and continuity of matrix multiplication imply that $\widehat{\boldsymbol{G}}'\widehat{\boldsymbol{\Omega}}\widehat{\boldsymbol{G}}$ is nonsingular with probability approaching one. The continuous mapping theorem therefore gives
\begin{equation*}
\widehat{\mathcal{A}}\xrightarrow{p}\mathcal{A}_0.
\end{equation*}
Further, Theorem \ref{th:mediation-beta-normality} implies $\widehat{\beta}_0^D\xrightarrow{p}\beta_0$. The uniform law of large numbers in \eqref{eq:proof-causal-jacobian-ulln}, together with continuity of $\beta\mapsto\E\{\dot{\ell}_\beta(W_i;\beta)\}$ established in the proof of Theorem \ref{th:causal-effect-normality}, gives
\begin{align*}
\|\widehat{\boldsymbol{J}}_\vartheta-\boldsymbol{J}_\vartheta\|_F&\leqslant\sup_{\beta\in\mathcal{B}}\|(\mathbb{P}_N-\E)\dot{\ell}_\beta(W_i;\beta)\|_F\\
&\quad+\left\|\E\{\dot{\ell}_\beta(W_i;\widehat{\beta}_0^D)\}-\E\{\dot{\ell}_\beta(W_i;\beta_0)\}\right\|_F=o_p(1),
\end{align*}
on the event $\widehat{\beta}_0^D\in\mathcal{B}$, whose probability tends to one. Hence
\begin{equation}
\widehat{\boldsymbol{J}}_\vartheta\xrightarrow{p}\boldsymbol{J}_\vartheta.
\label{eq:proof-causal-Jhat-full}
\end{equation}
In particular, $\widehat{\boldsymbol{J}}_\vartheta\widehat{\mathcal{A}}=O_p(1)$ and
\begin{equation}
\widehat{\boldsymbol{J}}_\vartheta\widehat{\mathcal{A}}-\boldsymbol{J}_\vartheta\mathcal{A}_0=o_p(1).
\label{eq:proof-causal-product-matrix}
\end{equation}

\paragraph{Step 2: $L^2$ control of the plug-in causal contributions.}
On the event that the line segment between $\widehat{\beta}_0^D$ and $\beta_0$ is contained in $\mathcal{B}$, the integral expansion \eqref{eq:proof-causal-integral-expansion-full} and the Cauchy--Schwarz inequality imply
\begin{align}
&\mathbb{P}_N\|\ell(W_i;\widehat{\beta}_0^D)-\ell(W_i;\beta_0)\|^2\nonumber\\
&\qquad\leqslant\|\widehat{\beta}_0^D-\beta_0\|^2\mathbb{P}_N\sup_{\beta\in\mathcal{B}}\|\dot{\ell}_\beta(W_i;\beta)\|_F^2=o_p(1).
\label{eq:proof-causal-contribution-l2-full}
\end{align}
Indeed, the first factor is $O_p(N^{-1})$, while the second is $O_p(1)$ by Assumption \ref{ass:causal-effect-regularity}(ii) and the law of large numbers. It follows that
\begin{align*}
\|\widehat{\vartheta}-\vartheta_0\|&\leqslant\left[\mathbb{P}_N\|\ell(W_i;\widehat{\beta}_0^D)-\ell(W_i;\beta_0)\|^2\right]^{1/2}+\|\mathbb{P}_N\ell(W_i;\beta_0)-\vartheta_0\|\\
&=o_p(1),
\end{align*}
where the second term is $o_p(1)$ by the law of large numbers. Thus
\begin{equation}
\widehat{\vartheta}\xrightarrow{p}\vartheta_0.
\label{eq:proof-causal-theta-consistency-full}
\end{equation}

\paragraph{Step 3: $L^2$ control of the centered estimated score.}
Let $r_i\coloneq\widehat{\psi}_i-\widetilde{\psi}_i$ and $\overline{r}\coloneq\mathbb{P}_Nr_i$. Then
\[
\overline{\widehat{\psi}}=\mathbb{P}_N\widetilde{\psi}_i+\overline{r}.
\]
Condition \eqref{eq:score-l2-for-causal-variance} and Jensen's inequality give $\|\overline{r}\|^2\leqslant\mathbb{P}_N\|r_i\|^2=o_p(1)$. Moreover, the i.i.d. central limit theorem and $\E\|\widetilde{\psi}_i\|^2<\infty$ give $\mathbb{P}_N\widetilde{\psi}_i=O_p(N^{-1/2})$. Therefore,
\begin{align}
&\mathbb{P}_N\|\widehat{\psi}_i-\overline{\widehat{\psi}}-\widetilde{\psi}_i\|^2\nonumber\\
&\quad=\mathbb{P}_N\|r_i-\overline{r}-\mathbb{P}_N\widetilde{\psi}_i\|^2\nonumber\\
&\quad\leqslant3\mathbb{P}_N\|r_i\|^2+3\|\overline{r}\|^2+3\|\mathbb{P}_N\widetilde{\psi}_i\|^2=o_p(1).
\label{eq:proof-centered-score-l2-full}
\end{align}

\paragraph{Step 4: $L^2$ convergence of the estimated influence functions.}
Set
\[
L_i\coloneq\ell(W_i;\widehat{\beta}_0^D)-\ell(W_i;\beta_0),\qquad S_i\coloneq(\widehat{\psi}_i-\overline{\widehat{\psi}})-\widetilde{\psi}_i.
\]
Subtracting \eqref{eq:causal-effect-if} from \eqref{eq:estimated-causal-if} gives the exact decomposition
\begin{align*}
\widehat{\varphi}_{\vartheta,i}-\varphi_{\vartheta,i}=&L_i-(\widehat{\vartheta}-\vartheta_0)-\left(\widehat{\boldsymbol{J}}_\vartheta\widehat{\mathcal{A}}-\boldsymbol{J}_\vartheta\mathcal{A}_0\right)\widetilde{\psi}_i-\widehat{\boldsymbol{J}}_\vartheta\widehat{\mathcal{A}}S_i.
\end{align*}
Using $\|a_1+\cdots+a_4\|^2\leqslant4\sum_{j=1}^4\|a_j\|^2$, we obtain
\begin{align*}
\mathbb{P}_N\|\widehat{\varphi}_{\vartheta,i}-\varphi_{\vartheta,i}\|^2\leqslant&4\mathbb{P}_N\|L_i\|^2+4\|\widehat{\vartheta}-\vartheta_0\|^2\\
&+4\left\|\widehat{\boldsymbol{J}}_\vartheta\widehat{\mathcal{A}}-\boldsymbol{J}_\vartheta\mathcal{A}_0\right\|_F^2\mathbb{P}_N\|\widetilde{\psi}_i\|^2\\
&+4\|\widehat{\boldsymbol{J}}_\vartheta\widehat{\mathcal{A}}\|_F^2\mathbb{P}_N\|S_i\|^2.
\end{align*}
The first two terms are $o_p(1)$ by \eqref{eq:proof-causal-contribution-l2-full} and \eqref{eq:proof-causal-theta-consistency-full}. The third term is $o_p(1)$ by \eqref{eq:proof-causal-product-matrix} and $\mathbb{P}_N\|\widetilde{\psi}_i\|^2=O_p(1)$. The fourth term is $o_p(1)$ by \eqref{eq:proof-centered-score-l2-full} and $\widehat{\boldsymbol{J}}_\vartheta\widehat{\mathcal{A}}=O_p(1)$. Hence
\begin{equation}
\mathbb{P}_N\|\widehat{\varphi}_{\vartheta,i}-\varphi_{\vartheta,i}\|^2=o_p(1).
\label{eq:proof-causal-if-l2-full}
\end{equation}

\paragraph{Step 5: Convergence of the covariance estimator.}
Let $\Delta\varphi_i\coloneq\widehat{\varphi}_{\vartheta,i}-\varphi_{\vartheta,i}$. The identity $aa'-bb'=(a-b)a'+b(a-b)'$ and the Cauchy--Schwarz inequality imply
\begin{align*}
&\left\|\widehat{\boldsymbol{\Sigma}}_\vartheta-\mathbb{P}_N\varphi_{\vartheta,i}\varphi_{\vartheta,i}'\right\|_F\\
&\quad\leqslant\left\{\mathbb{P}_N\|\Delta\varphi_i\|^2\right\}^{1/2}\left[\left\{\mathbb{P}_N\|\widehat{\varphi}_{\vartheta,i}\|^2\right\}^{1/2}+\left\{\mathbb{P}_N\|\varphi_{\vartheta,i}\|^2\right\}^{1/2}\right].
\end{align*}
The first factor is $o_p(1)$ by \eqref{eq:proof-causal-if-l2-full}. The law of large numbers gives $\mathbb{P}_N\|\varphi_{\vartheta,i}\|^2=O_p(1)$, and
\[
\mathbb{P}_N\|\widehat{\varphi}_{\vartheta,i}\|^2\leqslant2\mathbb{P}_N\|\varphi_{\vartheta,i}\|^2+2\mathbb{P}_N\|\Delta\varphi_i\|^2=O_p(1).
\]
Thus the displayed covariance difference is $o_p(1)$. Finally,
\[
\mathbb{P}_N\varphi_{\vartheta,i}\varphi_{\vartheta,i}'\xrightarrow{p}\E(\varphi_{\vartheta,i}\varphi_{\vartheta,i}')=\boldsymbol{\Sigma}_\vartheta
\]
by the law of large numbers. Together with \eqref{eq:proof-causal-Jhat-full}, this proves the proposition. Q.E.D.

\subsection{Proof of Proposition \ref{prop:fixed-vector-extension}}
Because $q_M$, $s_Y$, $s_M$, $T$, and $K$ are fixed, the dimensions of the stacked individual score and its Jacobian do not vary with $N$. Under the vector counterparts of Assumptions \ref{ass:mediation-l2} and \ref{ass:mediation-product}, the argument of Lemma \ref{lem:mediation-score-reduction} applies componentwise and gives
\[
\frac{1}{\sqrt{N}}\sum_{k=1}^K\sum_{i\in\mathcal{F}_k}\widehat{\psi}_i^{k,(v)}(\beta_0)=\frac{1}{\sqrt{N}}\sum_{i=1}^N\widetilde{\psi}_i^{(v)}+o_p(1).
\]
The vector counterpart of Assumption \ref{ass:mediation-jacobian} and the argument of Lemma \ref{lem:mediation-jacobian} likewise imply that the sample Jacobian converges in probability to $\boldsymbol{G}^{(v)}$. Expanding the GMM first-order condition around $\beta_0$ and using
$\widehat{\boldsymbol{\Omega}}^{(v)}\xrightarrow{p}\boldsymbol{\Omega}^{(v)}$ therefore yields
\[
\sqrt{N}(\widehat{\beta}_0^{D,v}-\beta_0)=-\mathcal{A}_0^{(v)}\frac{1}{\sqrt{N}}\sum_{i=1}^N\widetilde{\psi}_i^{(v)}+o_p(1).
\]
The oracle scores are i.i.d. across $i$, have mean zero, and have a finite second moment under the vector moment conditions. The multivariate central limit theorem and Slutsky's theorem then give \eqref{eq:vector-beta-clt}.

For the causal estimands, the vector counterfactual maps in \eqref{eq:vector-counterfactual-mediator}--\eqref{eq:vector-counterfactual-outcome} equal the corresponding potential variables at $\beta_0$. The vector counterpart of Assumption \ref{ass:causal-effect-regularity} permits the same empirical-process and mean-value expansion as in the proof of Theorem \ref{th:causal-effect-normality}:
\[
\sqrt{N}(\widehat{\vartheta}^{(v)}-\vartheta_0^{(v)})=\frac{1}{\sqrt{N}}\sum_{i=1}^N\{\ell^{(v)}(W_i;\beta_0)-\vartheta_0^{(v)}\}+\boldsymbol{J}_\vartheta^{(v)}\sqrt{N}(\widehat{\beta}_0^{D,v}-\beta_0)+o_p(1).
\]
Substituting the preceding expansion for $\widehat{\beta}_0^{D,v}$ gives \eqref{eq:vector-causal-clt}, and the multivariate central limit theorem gives the stated covariance matrix. Q.E.D.

\subsection{Proof of Lemma \ref{lem:mediation-score-reduction}}
Condition on the realized fold partition and fix a fold $k$. Write $n=|\mathcal{F}_k|=|\mathcal{F}_{k'}|$, $\widehat{\eta}_k=(\widehat{f}_{0k},\widehat{g}_{0k},\widehat{h}_{0k},\widehat{\rho}_{0k})$, $\mathcal{G}_k=\sigma(W_j:j\in\mathcal{F}_{k,k'}^c)$, and $\E_k(\cdot)=\E(\cdot\mid\mathcal{G}_k)$. Conditional on $\mathcal{G}_k$, the nuisance estimators are fixed, and the observations in $\mathcal{F}_k$ and $\mathcal{F}_{k'}$ are independent i.i.d. draws from $F_0$. For $i\in\mathcal{F}_k$, define
\begin{align*}
\widehat{R}_{1ki}&\coloneq m(W_i;\widehat{f}_{0k},\widehat{g}_{0k},\beta_0)-m(W_i;f_0,g_0,\beta_0),\\
\widehat{R}_{2ki}&\coloneq\phi(W_i;\widehat{f}_{0k},\widehat{g}_{0k},h_0,\rho_0,\beta_0)-\phi(W_i;f_0,g_0,h_0,\rho_0,\beta_0),\\
\widehat{R}_{3ki}&\coloneq\phi(W_i;f_0,g_0,\widehat{h}_{0k},\widehat{\rho}_{0k},\beta_0)-\phi(W_i;f_0,g_0,h_0,\rho_0,\beta_0).
\end{align*}
By the definition of $\widehat{\delta}_{ik}(W_i)$,
\begin{equation}
\widehat{\psi}_i^k(\beta_0)-\psi(W_i;\eta_0,\beta_0)=\widehat{R}_{1ki}+\widehat{R}_{2ki}+\widehat{R}_{3ki}+\widehat{\delta}_{ik}(W_i).
\label{eq:lem1-decomposition-short}
\end{equation}

Let $d_\psi$ denote the fixed dimension of the stacked score, and fix a component $\ell\in\{1,\ldots,d_\psi\}$. Direct substitution of \eqref{eq:first-difference-mediation} and \eqref{eq:double-difference-mediation} shows that this component has one of the following three forms for some $t\in\{3,\ldots,T\}$ and $p<t$:

\noindent outcome--treatment:
\begin{align*}
\widehat{R}_{1ki,\ell}+\widehat{R}_{2ki,\ell}&=\Delta\varepsilon_{ip}^{D*}\{\Delta f_{0,it}^*-\Delta\widehat{f}_{0k,it}^*\},\\
\widehat{R}_{3ki,\ell}&=\{\Delta h_{0,ip}^*-\Delta\widehat{h}_{0k,ip}^*\}\Delta\varepsilon_{it}^{Y*},\\
\widehat{\delta}_{ik,\ell}(W_i)&=\{\Delta h_{0,ip}^*-\Delta\widehat{h}_{0k,ip}^*\}\{\Delta f_{0,it}^*-\Delta\widehat{f}_{0k,it}^*\};
\end{align*}

\noindent outcome--mediator:
\begin{align*}
\widehat{R}_{1ki,\ell}+\widehat{R}_{2ki,\ell}&=\Delta\varepsilon_{ip}^{M*}\{\Delta f_{0,it}^*-\Delta\widehat{f}_{0k,it}^*\},\\
\widehat{R}_{3ki,\ell}&=\{\Delta\rho_{0,ip}^*-\Delta\widehat{\rho}_{0k,ip}^*\}\Delta\varepsilon_{it}^{Y*},\\
\widehat{\delta}_{ik,\ell}(W_i)&=\{\Delta\rho_{0,ip}^*-\Delta\widehat{\rho}_{0k,ip}^*\}\{\Delta f_{0,it}^*-\Delta\widehat{f}_{0k,it}^*\};
\end{align*}

\noindent mediator--treatment:
\begin{align*}
\widehat{R}_{1ki,\ell}+\widehat{R}_{2ki,\ell}&=\Delta\varepsilon_{ip}^{D*}\{\Delta g_{0,it}^*-\Delta\widehat{g}_{0k,it}^*\},\\
\widehat{R}_{3ki,\ell}&=\{\Delta h_{0,ip}^*-\Delta\widehat{h}_{0k,ip}^*\}\Delta\varepsilon_{it}^{M*},\\
\widehat{\delta}_{ik,\ell}(W_i)&=\{\Delta h_{0,ip}^*-\Delta\widehat{h}_{0k,ip}^*\}\{\Delta g_{0,it}^*-\Delta\widehat{g}_{0k,it}^*\}.
\end{align*}
These three cases exhaust all coordinates of the stacked score.

We first record a generic consequence of auxiliary-fold demeaning. Let $Q_i$ and $R_i$ be two scalar quantities, let $\overline{Q}_l=n^{-1}\sum_{i\in\mathcal{F}_l}Q_i$ and $\overline{R}_l=n^{-1}\sum_{i\in\mathcal{F}_l}R_i$, and write $Q_i^*=Q_i-\overline{Q}_{k'}$ and $R_i^*=R_i-\overline{R}_{k'}$ for $i\in\mathcal{F}_k$. Then
\begin{equation}
\frac{1}{\sqrt{N}}\sum_{i\in\mathcal{F}_k}Q_i^*R_i^*=\sqrt{\frac{n}{N}}\left\{\frac{1}{\sqrt{n}}\sum_{i\in\mathcal{F}_k}Q_iR_i-\sqrt{n}\,\overline{Q}_k\overline{R}_{k'}-\sqrt{n}\,\overline{Q}_{k'}\overline{R}_k+\sqrt{n}\,\overline{Q}_{k'}\overline{R}_{k'}\right\}.
\label{eq:lem1-generic-demeaning}
\end{equation}
Suppose, conditionally on $\mathcal{G}_k$, that $\E_k(Q_i)=0$, $\E_k(Q_iR_i)=0$, $\E_k(Q_i^2)=O_p(1)$, and $\Var_k(R_i)=O_p(1)$. Let $\mathcal{H}_k$ additionally contain the auxiliary-fold observations. Conditional on $\mathcal{H}_k$, the target-fold products $Q_i^*R_i^*$ are i.i.d. and
\[
\E(Q_i^*R_i^*\mid\mathcal{H}_k)=\overline{Q}_{k'}\{\overline{R}_{k'}-\E_k(R_i)\}.
\]
Moreover, conditional Chebyshev's inequality gives $\overline{Q}_{k'}=O_p(n^{-1/2})$ and $\overline{R}_{k'}-\E_k(R_i)=O_p(n^{-1/2})$. Hence
\[
\frac{n}{\sqrt{N}}\E(Q_i^*R_i^*\mid\mathcal{H}_k)=o_p(1).
\]
If, in addition, $\E\{(Q_i^*R_i^*)^2\mid\mathcal{H}_k\}=o_p(1)$, then
\[
\E\left[\left\{\frac{1}{\sqrt{N}}\sum_{i\in\mathcal{F}_k}\big[Q_i^*R_i^*-\E(Q_i^*R_i^*\mid\mathcal{H}_k)\big]\right\}^2\bigg|\mathcal{H}_k\right]\leqslant\frac{n}{N}\E\{(Q_i^*R_i^*)^2\mid\mathcal{H}_k\}=o_p(1).
\]
It follows that the left-hand side of \eqref{eq:lem1-generic-demeaning} is $o_p(1)$. The argument applies componentwise to vector- and matrix-valued products.

For $\widehat{R}_{1ki,\ell}+\widehat{R}_{2ki,\ell}$, the factor $Q_i$ is a lagged treatment or mediator innovation difference and $R_i$ is an estimation error in $f_0$ or $g_0$. The full-covariate-history restrictions in \eqref{eq:assignment-models} and \eqref{eq:mediator-residualization-exogeneity} imply $\E_k(Q_i)=\E_k(Q_iR_i)=0$. For $\widehat{R}_{3ki,\ell}$, take $Q_i$ to be the later outcome or mediator innovation difference and $R_i$ to be the corresponding error in $h_0$ or $\rho_0$. Because $p<t$, these errors are functions of histories dated no later than $p$ and quantities fixed conditional on $\mathcal{G}_k$. Expanding the later differenced innovation into its two level innovations and applying the sequential conditional moment restriction at periods $t$ and $t-1$ gives again $\E_k(Q_i)=\E_k(Q_iR_i)=0$. In particular, the identity
\[
\Delta\widehat{\rho}_{0k,ip}-\Delta\rho_{0,ip}=\Delta\widehat{g}_{0k,ip}-\Delta g_{0,ip}+\Delta\tau_{M,ip}'(\widetilde{\gamma}_k-\gamma_0)
\]
shows that the same argument covers the preliminary coefficient $\widetilde{\gamma}_k$.

Assumption \ref{ass:mediation-l2}(i)--(iv) supplies the remaining second-moment and weighted second-moment conditions in the generic argument. Therefore,
\begin{equation}
\frac{1}{\sqrt{N}}\sum_{i\in\mathcal{F}_k}\big(\widehat{R}_{1ki}+\widehat{R}_{2ki}+\widehat{R}_{3ki}\big)=o_p(1).
\label{eq:lem1-first-three-short}
\end{equation}

It remains to control the interaction remainder. Suppose first that Assumption \ref{ass:mediation-product}(i) holds and let $\mathcal{H}_k$ be the sigma-field generated by $\mathcal{G}_k$, the auxiliary-fold observations, and the fold partition. Conditional on $\mathcal{H}_k$, the variables $\widehat{\delta}_{ik}(W_i)$, $i\in\mathcal{F}_k$, are i.i.d. Hence
\begin{align*}
\E\left[\left\|\frac{1}{\sqrt{N}}\sum_{i\in\mathcal{F}_k}\left\{\widehat{\delta}_{ik}(W_i)-\E(\widehat{\delta}_{ik}(W_i)\mid\mathcal{H}_k)\right\}\right\|^2\bigg|\mathcal{H}_k\right]&\leqslant\frac{n}{N}\int\|\widehat{\delta}_{ik}(w)\|^2F_0(\mathrm{d}w)=o_p(1).
\end{align*}
Moreover,
\[
\left\|\frac{1}{\sqrt{N}}\sum_{i\in\mathcal{F}_k}\E(\widehat{\delta}_{ik}(W_i)\mid\mathcal{H}_k)\right\|=\frac{n}{\sqrt{N}}\left\|\int\widehat{\delta}_{ik}(w)F_0(\mathrm{d}w)\right\|\leqslant\sqrt{N}\left\|\int\widehat{\delta}_{ik}(w)F_0(\mathrm{d}w)\right\|=o_p(1).
\]
Thus
\begin{equation}
\frac{1}{\sqrt{N}}\sum_{i\in\mathcal{F}_k}\widehat{\delta}_{ik}(W_i)=o_p(1).
\label{eq:lem1-delta-short}
\end{equation}
The same conclusion follows immediately from alternative (ii) by the triangle inequality and from alternative (iii) by assumption.

Combining \eqref{eq:lem1-decomposition-short}, \eqref{eq:lem1-first-three-short}, and \eqref{eq:lem1-delta-short}, and then summing over the fixed number of folds, yields
\[
\frac{1}{\sqrt{N}}\sum_{k=1}^K\sum_{i\in\mathcal{F}_k}\widehat{\psi}_i^k(\beta_0)=\frac{1}{\sqrt{N}}\sum_{k=1}^K\sum_{i\in\mathcal{F}_k}\psi(W_i;\eta_0,\beta_0)+o_p(1).
\]
This proves the lemma. Q.E.D.

\subsection{Proof of Lemma \ref{lem:mediation-jacobian}}
Let $n\coloneq N/K$. For $q\in\{Y,M\}$ and a fold $l$, define the blockwise fold means
\[
\overline{\boldsymbol{A}}_l^q\coloneq\frac{1}{n}\sum_{i\in\mathcal{F}_l}\boldsymbol{A}_i^q,\qquad \overline{\boldsymbol{\mathcal{T}}}_l^q\coloneq\frac{1}{n}\sum_{i\in\mathcal{F}_l}\boldsymbol{\mathcal{T}}_i^q,
\]
and write
\[
\boldsymbol{A}_{i,k}^{q*}\coloneq\boldsymbol{A}_i^q-\overline{\boldsymbol{A}}_{k'}^q,\qquad \boldsymbol{\mathcal{T}}_{i,k}^{q*}\coloneq\boldsymbol{\mathcal{T}}_i^q-\overline{\boldsymbol{\mathcal{T}}}_{k'}^q.
\]
Since the residual vectors are affine in $\theta$ and $\gamma$, respectively, for every $\beta=(\theta',\gamma')'$,
\begin{equation*}
\frac{\partial\widehat{\psi}_i^k(\beta)}{\partial\beta'}=-\begin{pmatrix}\widehat{\boldsymbol{A}}_{i,k}^Y\boldsymbol{\mathcal{T}}_{i,k}^{Y*}&\boldsymbol{0}\\ \boldsymbol{0}&\widehat{\boldsymbol{A}}_{i,k}^M\boldsymbol{\mathcal{T}}_{i,k}^{M*}\end{pmatrix}.
\end{equation*}
The right-hand side does not depend on $\beta$. Hence, for any possibly data-dependent $\overline{\beta}$,
\begin{equation}
\frac{\partial\widehat{\psi}_i^k(\overline{\beta})}{\partial\beta'}=\frac{\partial\widehat{\psi}_i^k(\beta_0)}{\partial\beta'}.
\label{eq:lem2-concise-beta}
\end{equation}

Let $\mathcal{H}_k$ be the sigma-field generated by the observations used to estimate $\widehat{\eta}_k$, the observations in the auxiliary fold $\mathcal{F}_{k'}$, and the fold partition. Conditional on $\mathcal{H}_k$, the observations $\{W_i:i\in\mathcal{F}_k\}$ are i.i.d. with distribution $F_0$. Therefore, by the conditional Markov inequality, the triangle inequality, and Assumption \ref{ass:mediation-jacobian}(ii),
\begin{equation}
\frac{1}{n}\sum_{i\in\mathcal{F}_k}\frac{\partial\psi(W_i;\widehat{\eta}_k,\beta_0)}{\partial\beta'}=\frac{1}{n}\sum_{i\in\mathcal{F}_k}\frac{\partial\psi(W_i;\eta_0,\beta_0)}{\partial\beta'}+o_p(1).
\label{eq:lem2-concise-oracle-reduction}
\end{equation}

At the true nuisance values, the instrument matrices in the score are $\boldsymbol{A}_{i,k}^{q*}$, $q\in\{Y,M\}$. The conditional moment restrictions imply
\begin{equation}
\E(\boldsymbol{A}_i^Y)=\boldsymbol{0},\qquad \E(\boldsymbol{A}_i^M)=\boldsymbol{0}.
\label{eq:lem2-concise-zero-mean}
\end{equation}
Since $T$ is fixed, Assumption \ref{ass:mediation-l2}(i), Assumption \ref{ass:mediation-jacobian}(i), and the cross-sectional i.i.d. condition yield, for $q\in\{Y,M\}$,
\begin{gather}
\overline{\boldsymbol{A}}_k^q=o_p(1),\qquad \overline{\boldsymbol{A}}_{k'}^q=o_p(1),\qquad \overline{\boldsymbol{\mathcal{T}}}_k^q=O_p(1),\qquad \overline{\boldsymbol{\mathcal{T}}}_{k'}^q=O_p(1),\notag\\
\frac{1}{n}\sum_{i\in\mathcal{F}_k}\boldsymbol{A}_i^q\boldsymbol{\mathcal{T}}_i^q=\E(\boldsymbol{A}_i^q\boldsymbol{\mathcal{T}}_i^q)+o_p(1).
\label{eq:lem2-concise-lln}
\end{gather}
Moreover,
\begin{align*}
\frac{1}{n}\sum_{i\in\mathcal{F}_k}\boldsymbol{A}_{i,k}^{q*}\boldsymbol{\mathcal{T}}_{i,k}^{q*}=&\frac{1}{n}\sum_{i\in\mathcal{F}_k}\boldsymbol{A}_i^q\boldsymbol{\mathcal{T}}_i^q-\overline{\boldsymbol{A}}_k^q\overline{\boldsymbol{\mathcal{T}}}_{k'}^q-\overline{\boldsymbol{A}}_{k'}^q\overline{\boldsymbol{\mathcal{T}}}_k^q+\overline{\boldsymbol{A}}_{k'}^q\overline{\boldsymbol{\mathcal{T}}}_{k'}^q\\
=&\E(\boldsymbol{A}_i^q\boldsymbol{\mathcal{T}}_i^q)+o_p(1),
\end{align*}
where the last equality follows from \eqref{eq:lem2-concise-zero-mean}--\eqref{eq:lem2-concise-lln}. Hence
\[
\frac{1}{n}\sum_{i\in\mathcal{F}_k}\frac{\partial\psi(W_i;\eta_0,\beta_0)}{\partial\beta'}=\boldsymbol{G}+o_p(1).
\]
Combining this result with \eqref{eq:lem2-concise-oracle-reduction}, using \eqref{eq:lem2-concise-beta}, and averaging over the fixed number of folds gives
\[
\frac{1}{N}\sum_{k=1}^K\sum_{i\in\mathcal{F}_k}\frac{\partial\widehat{\psi}_i^k(\overline{\beta})}{\partial\beta'}=\boldsymbol{G}+o_p(1),
\]
which proves the result. Q.E.D.

\subsection{Proof of Theorem \ref{th:mediation-beta-normality}}
Let $n\coloneq N/K$ and define
\[
\widehat{\boldsymbol{\psi}}_N(\beta)\coloneq\frac{1}{N}\sum_{k=1}^K\sum_{i\in\mathcal{F}_k}\widehat{\psi}_i^k(\beta),\qquad \widehat{\boldsymbol{G}}_N\coloneq\frac{1}{N}\sum_{k=1}^K\sum_{i\in\mathcal{F}_k}\frac{\partial\widehat{\psi}_i^k(\beta)}{\partial\beta'}.
\]
Since the residual vectors are affine in $\theta$ and $\gamma$, while the fold-specific instrument matrices do not depend on the final GMM candidate $\beta$, the Jacobian $\widehat{\boldsymbol{G}}_N$ does not depend on its evaluation point. Hence, for every $\beta\in\mathbb{R}^{s_Y+s_M}$,
\begin{equation}
\widehat{\boldsymbol{\psi}}_N(\beta)=\widehat{\boldsymbol{\psi}}_N(\beta_0)+\widehat{\boldsymbol{G}}_N(\beta-\beta_0).
\label{eq:proof-med-beta-exact-affine-short}
\end{equation}

By Lemma \ref{lem:mediation-jacobian} and $\widehat{\boldsymbol{\Omega}}\xrightarrow{p}\boldsymbol{\Omega}$,
\[
\widehat{\boldsymbol{G}}_N\xrightarrow{p}\boldsymbol{G},\qquad \widehat{\boldsymbol{G}}_N'\widehat{\boldsymbol{\Omega}}\widehat{\boldsymbol{G}}_N\xrightarrow{p}\boldsymbol{G}'\boldsymbol{\Omega}\boldsymbol{G}.
\]
Since $\boldsymbol{G}'\boldsymbol{\Omega}\boldsymbol{G}$ is nonsingular, $\widehat{\boldsymbol{G}}_N'\widehat{\boldsymbol{\Omega}}\widehat{\boldsymbol{G}}_N$ is nonsingular with probability approaching one. On this event, the first-order condition for the quadratic GMM criterion and \eqref{eq:proof-med-beta-exact-affine-short} give
\begin{align}
\sqrt{N}(\widehat{\beta}_0^D-\beta_0)&=-\widehat{\boldsymbol{C}}_N\sqrt{N}\widehat{\boldsymbol{\psi}}_N(\beta_0),
\label{eq:proof-med-beta-foc-short}\\
\widehat{\boldsymbol{C}}_N&\coloneq(\widehat{\boldsymbol{G}}_N'\widehat{\boldsymbol{\Omega}}\widehat{\boldsymbol{G}}_N)^{-1}\widehat{\boldsymbol{G}}_N'\widehat{\boldsymbol{\Omega}}\xrightarrow{p}\boldsymbol{C}\coloneq(\boldsymbol{G}'\boldsymbol{\Omega}\boldsymbol{G})^{-1}\boldsymbol{G}'\boldsymbol{\Omega}.\notag
\end{align}

We next obtain the limit of the score at $\beta_0$. For $q\in\{Y,M\}$, write
\[
\overline{e}_l^q\coloneq\frac{1}{n}\sum_{i\in\mathcal{F}_l}e_i^q,\qquad \overline{\boldsymbol{A}}_l^q\coloneq\frac{1}{n}\sum_{i\in\mathcal{F}_l}\boldsymbol{A}_i^q.
\]
At the true nuisance and parameter values, the auxiliary-fold transformed score satisfies, for $i\in\mathcal{F}_k$,
\begin{equation}
\psi(W_i;\eta_0,\beta_0)=\begin{pmatrix}(\boldsymbol{A}_i^Y-\overline{\boldsymbol{A}}_{k'}^Y)(e_i^Y-\overline{e}_{k'}^Y)\\ (\boldsymbol{A}_i^M-\overline{\boldsymbol{A}}_{k'}^M)(e_i^M-\overline{e}_{k'}^M)\end{pmatrix}.
\label{eq:proof-med-beta-oracle-decomposition-short}
\end{equation}
The conditional moment restrictions and Assumption \ref{ass:mediation-l2} imply, for $q\in\{Y,M\}$,
\[
\E(\boldsymbol{A}_i^q)=\boldsymbol{0},\qquad \E(e_i^q)=\boldsymbol{0},\qquad \E\|\boldsymbol{A}_i^q\|_F^2+\E\|e_i^q\|^2<\infty.
\]
Consequently,
\[
\overline{\boldsymbol{A}}_l^q=O_p(n^{-1/2}),\qquad \overline{e}_l^q=O_p(n^{-1/2})
\]
for every fold $l$. Expanding the products in \eqref{eq:proof-med-beta-oracle-decomposition-short} gives
\begin{align*}
&\frac{1}{\sqrt{N}}\sum_{i\in\mathcal{F}_k}\left\{(\boldsymbol{A}_i^q-\overline{\boldsymbol{A}}_{k'}^q)(e_i^q-\overline{e}_{k'}^q)-\boldsymbol{A}_i^qe_i^q\right\}\\
&\quad=-\frac{n}{\sqrt{N}}\left(\overline{\boldsymbol{A}}_k^q\overline{e}_{k'}^q+\overline{\boldsymbol{A}}_{k'}^q\overline{e}_k^q-\overline{\boldsymbol{A}}_{k'}^q\overline{e}_{k'}^q\right)=o_p(1).
\end{align*}
Since $K$ is fixed, summing over folds and using Lemma \ref{lem:mediation-score-reduction} yields
\begin{equation*}
\sqrt{N}\widehat{\boldsymbol{\psi}}_N(\beta_0)=\frac{1}{\sqrt{N}}\sum_{i=1}^N\widetilde{\psi}_i+o_p(1).
\end{equation*}

For $p<t$, each component of the lagged innovation instruments is orthogonal to the later outcome or mediator innovation difference under \eqref{eq:observed-mediation-model}, \eqref{eq:assignment-models}, and \eqref{eq:mediator-residualization-exogeneity}. Therefore $\E(\widetilde{\psi}_i)=\boldsymbol{0}$. Under \eqref{iid}, the vectors $\widetilde{\psi}_i$ are i.i.d., and Assumption \ref{ass:mediation-l2}(i) gives a finite second moment. The multivariate central limit theorem thus gives
\begin{equation}
\frac{1}{\sqrt{N}}\sum_{i=1}^N\widetilde{\psi}_i\xrightarrow{d}N(\boldsymbol{0},\boldsymbol{\Psi}).
\label{eq:proof-med-beta-clt-short}
\end{equation}
Combining \eqref{eq:proof-med-beta-foc-short}--\eqref{eq:proof-med-beta-clt-short} and applying Slutsky's theorem yields
\[
\sqrt{N}(\widehat{\beta}_0^D-\beta_0)=-(\boldsymbol{G}'\boldsymbol{\Omega}\boldsymbol{G})^{-1}\boldsymbol{G}'\boldsymbol{\Omega}\frac{1}{\sqrt{N}}\sum_{i=1}^N\widetilde{\psi}_i+o_p(1)\xrightarrow{d}N(\boldsymbol{0},\boldsymbol{V}_\beta),
\]
which proves the theorem. Q.E.D.

\subsection{Proof of Theorem \ref{prop:mediation-sandwich}}
Lemma \ref{lem:mediation-jacobian} gives $\widehat{\boldsymbol{G}}\xrightarrow{p}\boldsymbol{G}$. It remains to establish $\widehat{\boldsymbol{\Psi}}\xrightarrow{p}\boldsymbol{\Psi}$. For $i\in\mathcal{F}_k$, write
\begin{align*}
\widehat{\psi}_i-\widetilde{\psi}_i=&\{\psi(W_i;\widehat{\eta}_k,\widehat{\beta}_0^D)-\psi(W_i;\widehat{\eta}_k,\beta_0)\}\\
&+\{\widehat{\psi}_i^k(\beta_0)-\psi(W_i;\eta_0,\beta_0)\}+\{\psi(W_i;\eta_0,\beta_0)-\widetilde{\psi}_i\}.
\end{align*}
The first term converges to zero in empirical mean square by the additional condition in the proposition and the conditional law of large numbers. By the decomposition \eqref{eq:lem1-decomposition-short}, Assumption \ref{ass:mediation-l2}(ii)--(iv), and the second additional condition in the proposition, the same is true of the second term. For the third term, \eqref{eq:proof-med-beta-oracle-decomposition-short} and the bounds $\overline{\boldsymbol{A}}_{k'}^q=O_p(n^{-1/2})$ and $\overline{e}_{k'}^q=O_p(n^{-1/2})$ imply
\[
\frac{1}{N}\sum_{k=1}^K\sum_{i\in\mathcal{F}_k}\|\psi(W_i;\eta_0,\beta_0)-\widetilde{\psi}_i\|^2=o_p(1).
\]
Consequently,
\begin{equation}
\frac{1}{N}\sum_{k=1}^K\sum_{i\in\mathcal{F}_k}\|\widehat{\psi}_i-\widetilde{\psi}_i\|^2=o_p(1).
\label{eq:proof-score-l2-sandwich}
\end{equation}
Let $r_i\coloneq\widehat{\psi}_i-\widetilde{\psi}_i$. The identity $aa'-bb'=(a-b)a'+b(a-b)'$ and the Cauchy--Schwarz inequality yield
\begin{align*}
&\left\|\widehat{\boldsymbol{\Psi}}-\frac{1}{N}\sum_{i=1}^N\widetilde{\psi}_i\widetilde{\psi}_i'\right\|_F\notag\\
&\quad\leqslant\left(\frac{1}{N}\sum_{i=1}^N\|r_i\|^2\right)^{1/2}\left\{\left(\frac{1}{N}\sum_{i=1}^N\|\widehat{\psi}_i\|^2\right)^{1/2}+\left(\frac{1}{N}\sum_{i=1}^N\|\widetilde{\psi}_i\|^2\right)^{1/2}\right\}=o_p(1).
\end{align*}
where \eqref{eq:proof-score-l2-sandwich} and the law of large numbers imply that the bracketed term is $O_p(1)$. Since
\[
\frac{1}{N}\sum_{i=1}^N\widetilde{\psi}_i\widetilde{\psi}_i'\xrightarrow{p}\boldsymbol{\Psi},
\]
we obtain $\widehat{\boldsymbol{\Psi}}\xrightarrow{p}\boldsymbol{\Psi}$. Finally, $\widehat{\boldsymbol{\Omega}}\xrightarrow{p}\boldsymbol{\Omega}$, $\widehat{\boldsymbol{G}}\xrightarrow{p}\boldsymbol{G}$, and the nonsingularity of $\boldsymbol{G}'\boldsymbol{\Omega}\boldsymbol{G}$ imply $\widehat{\boldsymbol{V}}_\beta\xrightarrow{p}\boldsymbol{V}_\beta$ by the continuous mapping theorem. Q.E.D.

\end{appendices}

\spacingset{1.4}
\bibliography{reference}

@article{rubin1974estimating,
  title={Estimating causal effects of treatments in randomized and nonrandomized studies.},
  author={Rubin, Donald B},
  journal={Journal of educational Psychology},
  volume={66},
  number={5},
  pages={688},
  year={1974},
  publisher={American Psychological Association}
}

@book{imbens2015causal,
  title={Causal inference in statistics, social, and biomedical sciences},
  author={Imbens, Guido W and Rubin, Donald B},
  year={2015},
  publisher={Cambridge University Press}
}

@article{vanderweele2017mediation,
  title={Mediation analysis with time varying exposures and mediators},
  author={VanderWeele, Tyler J and Tchetgen Tchetgen, Eric J},
  journal={Journal of the Royal Statistical Society Series B: Statistical Methodology},
  volume={79},
  number={3},
  pages={917--938},
  year={2017},
  publisher={Oxford University Press}
}

@article{chernozhukov2018double,
  author = {Chernozhukov, Victor and Chetverikov, Denis and Demirer, Mert and Duflo, Esther and Hansen, Christian and Newey, Whitney and Robins, James},
  title = "{Double/debiased machine learning for treatment and structural parameters}",
  journal = {The Econometrics Journal},
  volume = {21},
  number = {1},
  pages = {C1-C68},
  year = {2018},
  month = {01},
  issn = {1368-4221},
  doi = {10.1111/ectj.12097},
}

@article{kapetanios2008bootstrap,
  title={A bootstrap procedure for panel data sets with many cross-sectional units},
  author={Kapetanios, George},
  journal={The Econometrics Journal},
  volume={11},
  number={2},
  pages={377--395},
  year={2008},
  publisher={Oxford University Press Oxford, UK}
}

@article{richardson2013single,
  title={Single world intervention graphs (SWIGs): A unification of the counterfactual and graphical approaches to causality},
  author={Richardson, Thomas S and Robins, James M},
  journal={Center for the Statistics and the Social Sciences, University of Washington Series. Working Paper},
  volume={128},
  number={30},
  year={2013},
  publisher={Citeseer}
}

@article{cerqua2026misuse,
  author  = {Cerqua, Augusto and Letta, Marco and Pinto, Gabriele},
  title   = {On the (Mis)Use of Machine Learning With Panel Data},
  journal = {Oxford Bulletin of Economics and Statistics},
  volume  = {88},
  number  = {3},
  pages   = {506--518},
  year    = {2026},
  doi     = {10.1111/obes.70019}
}

@article{blanchard1992regional,
  author  = {Blanchard, Olivier Jean and Katz, Lawrence F.},
  title   = {Regional Evolutions},
  journal = {Brookings Papers on Economic Activity},
  volume  = {1992},
  number  = {1},
  pages   = {1--75},
  year    = {1992}
}

@book{bartik1991benefits,
  author    = {Bartik, Timothy J.},
  title     = {Who Benefits from State and Local Economic Development Policies?},
  publisher = {W. E. Upjohn Institute for Employment Research},
  address   = {Kalamazoo, MI},
  year      = {1991}
}

@article{intro_robins1992,
  author  = {Robins, James M. and Greenland, Sander},
  title   = {Identifiability and Exchangeability for Direct and Indirect Effects},
  journal = {Epidemiology},
  year    = {1992},
  volume  = {3},
  number  = {2},
  pages   = {143--155},
  doi     = {10.1097/00001648-199203000-00013}
}

@article{intro_imai2010,
  author  = {Imai, Kosuke and Keele, Luke and Yamamoto, Teppei},
  title   = {Identification, Inference and Sensitivity Analysis for Causal Mediation Effects},
  journal = {Statistical Science},
  year    = {2010},
  volume  = {25},
  number  = {1},
  pages   = {51--71},
  doi     = {10.1214/10-STS321}
}

@article{intro_tchetgen2012,
  author  = {Tchetgen Tchetgen, Eric J. and Shpitser, Ilya},
  title   = {Semiparametric Theory for Causal Mediation Analysis: Efficiency Bounds, Multiple Robustness, and Sensitivity Analysis},
  journal = {The Annals of Statistics},
  year    = {2012},
  volume  = {40},
  number  = {3},
  pages   = {1816--1845},
  doi     = {10.1214/12-AOS990}
}

@article{intro_zheng2017,
  author  = {Zheng, Wenjing and van der Laan, Mark J.},
  title   = {Longitudinal Mediation Analysis with Time-Varying Mediators and Exposures, with Application to Survival Outcomes},
  journal = {Journal of Causal Inference},
  year    = {2017},
  volume  = {5},
  number  = {2},
  doi     = {10.1515/jci-2016-0006}
}

@article{intro_diaz2023,
  author  = {D{\'i}az, Iv{\'a}n and Williams, Nicholas and Rudolph, Kara E.},
  title   = {Efficient and Flexible Mediation Analysis with Time-Varying Mediators, Treatments, and Confounders},
  journal = {Journal of Causal Inference},
  year    = {2023},
  volume  = {11},
  number  = {1},
  doi     = {10.1515/jci-2022-0077}
}

@article{intro_farbmacher2022,
  author  = {Farbmacher, Helmut and Huber, Martin and Laff{\'e}rs, Luk{\'a}{\v{s}} and Langen, Henrika and Spindler, Martin},
  title   = {Causal Mediation Analysis with Double Machine Learning},
  journal = {The Econometrics Journal},
  year    = {2022},
  volume  = {25},
  number  = {2},
  pages   = {277--300},
  doi     = {10.1093/ectj/utac003}
}

@inproceedings{intro_zenati2025,
  author    = {Zenati, Houssam and Ab{\'e}cassis, Judith and Josse, Julie and Thirion, Bertrand},
  title     = {Double Debiased Machine Learning for Mediation Analysis with Continuous Treatments},
  booktitle = {Proceedings of the 28th International Conference on Artificial Intelligence and Statistics},
  series    = {Proceedings of Machine Learning Research},
  volume    = {258},
  pages     = {4150--4158},
  publisher = {PMLR},
  year      = {2025}
}

@article{intro_huber2026,
  author  = {Huber, Martin and Oberh{\"a}nsli, Sarina Joy},
  title   = {Difference-in-Differences for Mediation Analysis Using Double Machine Learning},
  journal = {arXiv preprint arXiv:2602.23877},
  year    = {2026}
}

@article{intro_deng2026,
  author  = {Deng, Yuhao and Wei, Haoyu and Ouyang, Zhongzhe},
  title   = {Difference-in-Differences with a Mediator},
  journal = {arXiv preprint arXiv:2604.24049},
  year    = {2026}
}

@inproceedings{intro_lewis2021,
  author    = {Lewis, Greg and Syrgkanis, Vasilis},
  title     = {Double/Debiased Machine Learning for Dynamic Treatment Effects},
  booktitle = {Advances in Neural Information Processing Systems},
  volume    = {34},
  year      = {2021}
}

@article{intro_semenova2023,
  author  = {Semenova, Vira and Goldman, Matt and Chernozhukov, Victor and Taddy, Matt},
  title   = {Inference on Heterogeneous Treatment Effects in High-Dimensional Dynamic Panels under Weak Dependence},
  journal = {Quantitative Economics},
  year    = {2023},
  volume  = {14},
  number  = {2},
  pages   = {471--510},
  doi     = {10.3982/QE1670}
}

@article{intro_clarke2026,
  author  = {Clarke, Paul S. and Polselli, Annalivia},
  title   = {Double Machine Learning for Static Panel Models with Fixed Effects},
  journal = {The Econometrics Journal},
  year    = {2026},
  volume  = {29},
  number  = {1},
  pages   = {69--86},
  doi     = {10.1093/ectj/utaf011}
}

@article{intro_wu2026,
  author  = {Wu, Peikai and Xiao, Zhiguo},
  title   = {Double/debiased machine learning for treatment effects in dynamic panels},
  journal = {Journal of the American Statistical Association},
  pages = {1--12},
  year  = {2026},
  note = {https://doi.org/10.1080/01621459.2026.2696607}
}

@article{wei2025debiased,
  title={Debiased machine learning for ultra-high dimensional mediation analysis},
  author={Wei, Kecheng and Liu, Yahang and Huang, Chen and Lin, Ruilang and Yu, Yongfu and Qin, Guoyou},
  journal={Bioinformatics},
  volume={41},
  number={6},
  pages={btaf282},
  year={2025},
  publisher={Oxford University Press}
}

\end{document}